\documentclass[showpacs,preprintnumbers,amsmath,amssymb,aps,nofootinbib]{revtex4}
\usepackage{graphicx}
\usepackage{dcolumn}
\usepackage{bm}
\newcommand{\ds}{\displaystyle}
\newcommand{\ident}{{\bf 1}} 
\newcommand{\nin}{\noindent}
\newcommand{\be}{\begin{equation}}
\newcommand{\ee}{\end{equation}}
\newcommand{\beq}{\begin{eqnarray}}
\newcommand{\eeq}{\end{eqnarray}}

\newcommand{\nn}{\nonumber\\}
\newcommand{\MD}{M_{\Delta}}

\newcommand{\BE}{\begin{equation}}
\def\EE{\end{equation}}
\def\BEA{\begin{eqnarray}}
\def\EEA{\end{eqnarray}}

\def\D{{\Delta}}

\newcounter{saveeqn}

\newcommand{\lsim}{\raisebox{-0.07cm}{$\;\stackrel{<}{{\scriptstyle \sim}}\; $} }
\usepackage{color}
\begin{document}

\title{Determination of the $\Delta(1232)$ axial and pseudoscalar form factors from lattice QCD}

\author{Constantia Alexandrou}
\affiliation{Department of Physics, University of Cyprus, P.O. Box 20537,
1678 Nicosia, and\\ The Cyprus Institute, P.O. Box 27546, 1645 Nicosia, Cyprus}

\author{Eric B. Gregory} 
\affiliation{Bergische Universit\"at Wuppertal, Gaussstr.\,20, D-42119 Wuppertal, Germany}

\author{Tomasz Korzec}
\affiliation{Institut f\"ur Physik, Humboldt Universit\"at zu Berlin, 
Newtonstrasse 15, 12489 Berlin,Germany}
\author{Giannis Koutsou}
\affiliation{Cyprus Institute, CaSToRC, 20 Kavafi Street, Nicosia 2121, Cyprus}

\author{John Negele}
  \affiliation{Center for Theoretical Physics, 
Laboratory for Nuclear Science and 
        Department of Physics, Massachusetts Institute of
        Technology, Cambridge, Massachusetts 02139, U.S.A.}
\author{Toru Sato}
  \affiliation{Department of Physics, Osaka University, Osaka 560-0043, Japan}
\author{Antonios Tsapalis}     
\affiliation{Hellenic Naval Academy, Hatzikyriakou Ave., Pireaus 18539, 
Greece}       
\affiliation{Department of Physics, National Technical University of
       Athens, Zografou Campus 15780, Athens, Greece}

\date{\today}
    
\begin{abstract}

We present a lattice QCD calculation of the $\Delta(1232)$  matrix elements of the  axial-vector and pseudoscalar currents. The
 decomposition of these matrix elements into the appropriate 
Lorentz invariant
form factors is carried out and  
the techniques to calculate the form factors are developed and tested
using quenched configurations.
Results are obtained for 
$2+1$  domain wall fermions  and within
a hybrid scheme with domain wall valence and staggered sea quarks.  Two Goldberger-Treiman type relations connecting
the axial to the pseudoscalar effective couplings are derived. These and further relations based on the  pion-pole dominance hypothesis are examined
 using the lattice QCD results, finding support for their validity.
Utilizing  lattice QCD results on the  axial charges of the nucleon and
the $\Delta$, 
as well as the nucleon-to-$\Delta$ transition coupling constant, we perform  a combined chiral fit to all three
quantities and study  their pion mass dependence 
as the chiral limit is approached.

\end{abstract}
   
\pacs{}

\maketitle

\section{INTRODUCTION} \label{se:INTRODUCTION}
Great progress has been made in lattice QCD  studies of hadron spectroscopy 
and   structure and lattice QCD results are beginning to provide input to phenomenology and experiment. 
Simulations with dynamical quarks near 
and at the physical pion 
mass~\cite{Jansen:2008vs,Durr:2008zz,Aoki:2008sm,Durr:2010aw} have been shown to  
produce the observed low-lying hadron 
spectrum~\cite{Durr:2008zz,Alexandrou:2009qu,Aoki:2009ix} and $\pi^+-\pi^+$ scattering lengths 
have been calculated to good accuracy~\cite{Beane:2011sc,Dudek:2012gj,Feng:2009ij,Yamazaki:2004qb}.

Whereas producing experimentally measured quantities from first principles provides
a powerful validation of the lattice QCD methodology, calculating quantities that are difficult to extract
or have impact in probing physics beyond the standard model is a much more challenging prospective.
Studying the structure of  the $\Delta$ resonance  is an example
of the input lattice QCD can provide to phenomenology that cannot be directly extracted from experiments.    
This is because the  $\Delta$ decays strongly with a  lifetime of 
$\sim 10^{-23}$ seconds ~\cite{Kotulla:2002cg,LopezCastro:2000cv} and
resists experimental probing. Measurements of the $\Delta^+$ 
magnetic moment exist albeit with a large experimental uncertainty. The $\Delta$, having width  
$\Gamma \sim 118$ MeV and lying  close to the $\pi N$ threshold, plays an important role 
in chiral expansions. In heavy baryon chiral perturbation theory it has been included 
as an explicit degree of freedom~\cite{Bernard:2005fy, Hemmert:1997ye,Jenkins:1991es,Fettes:2000bb}, where it is argued that it
improves  chiral expansions applied in the description of lattice QCD results such as the nucleon axial charge~\cite{Hemmert:2003cb}.
 Chiral langragians with $\Delta$ degrees of freedom involve additional coupling constants that are difficult
to measure. Therefore, one either 
 treats them as free parameters to be fitted
along other parameters 
using  lattice QCD results~\cite{Bernard:2005fy,Bernard:2007zu} and data extracted from
partial-wave analysis of scattering
measurements~\cite{Jenkins:1991es,Fettes:2000bb}
or estimates them based on phenomenology and symmetries.
For example, one can relate 
 the nucleon axial 
charge $g_A$, which is well measured, to  the $\Delta$  axial charge, in
 the large-$N_c$ limit
\cite{Dashen:1993jt} or using $SU(4)$ symmetry
\cite{Brown:1975di}.
The Goldberger-Treiman (GT) relation  is then used to get the effective 
$\pi \Delta \Delta$ coupling. 
Another framework to extract the $\pi\Delta\Delta$ coupling is via
sum rules~\cite{Choi:2010ty}. 

Lattice QCD provides a nice framework to study the $\Delta$ properties
and calculate the $\Delta$ coupling constants.
 In some of our recent work we developed the formalism to study  the  $N-\Delta$ transition 
form-factors within lattice QCD~\cite{Alexandrou:2007xj,Alexandrou:2010uk}, as well as the $\Delta$
electromagnetic form-factors~\cite{Alexandrou:2009hs}. The quadrupole electromagnetic form factor,
extracted for the first time, provided input for 
 the deformation of the $\Delta$ showing that in the infinite momentum frame the
$\Delta$ is prolate~\cite{Alexandrou:2012da}.

In this work, we present a detailed study of the axial-vector and pseudoscalar form factors of the $\Delta$.  
The theoretical framework and a subset of the results were given in Ref.~\cite{Alexandrou:2011py}.
Here we discuss in detail the lattice techniques  developed and 
utilized for the extraction of these form factors. In addition, we present 
an extended analysis of the momentum dependence of all the form factors
using  an additional ensemble of dynamical domain wall fermions.
We also include a study of the  pion-pole dominance predictions and compare them
to our lattice QCD results.

The outline of the paper is as follows: In Section II we present the 
decomposition of the $\Delta$ matrix elements of the axial-vector and pseudoscalar currents. 
In Section III we explain our lattice techniques and discuss the ensembles
utilized for the calculation. In Section IV we present the
lattice results on all form factors and examine several relations among them
and their phenomenological consequences.
In Section V we perform a combined
chiral fit using our results on the nucleon axial charge $g_A$~\cite{Alexandrou:2010hf}, the $\Delta$ axial charge
$G_{\D \D}$ calculated in this work and the dominant axial $N-\D$ transition form
factor, $C^A_5$, calculated  in previous work on the same sets of lattices~\cite{Alexandrou:2010uk}.
Finally, in Section VI we give a summary and conclusions. Technical details and our values on the form factors are
presented in the Appendices.

\section{The Axial and Pseudoscalar Matrix Element of the $\Delta$
} \label{se:DECOMPOSITION}

\nin
Lorentz invariance and spin-parity rules determine  the 
decomposition of the $\Delta^+$ 
matrix element of the isovector axial-vector current 
 in terms of four invariant 
functions of the momentum transfer squared, $q^2 = (p_f -p_i)^2$:
\be
\langle \Delta^+(p_f,s_f) | A^\mu(0)| \Delta^+(p_i,s_i)\rangle 
=\overline{u}^{\Delta}_{\sigma}(p_f,s_f)\left[{\mathcal O}^{\mu {\rm A }}\right]^
{ \sigma \tau} u^{\Delta}_\tau(p_i,s_i) \nonumber
\ee
\be
\label{decomp_eqs}
\left[{\mathcal O}^{{\rm \mu {\rm A}}}\right]^{\sigma\tau}= -\frac{1}{2}\left[
g^{\sigma\tau}
\left(g_1(q^2)\gamma^\mu\gamma^5 + g_3(q^2) \frac{q^\mu}{2M_\Delta}\gamma^5\right)
+\frac{\ds q^\sigma q^\tau}{\ds 4M_\Delta^2}
\left(h_1(q^2)\gamma^\mu\gamma^5 +h_3(q^2) \frac{q^\mu}{2M_\Delta}\gamma^5\right)
\right],
\ee
where $p_i (s_i)$ denotes the initial momentum (spin) of the $\Delta$  and 
$p_f (s_f)$ the final momentum (spin).
The flavor-isovector axial-vector current operator is defined as
\be
A^{\mu}(x) = \overline{\psi}(x) \gamma^\mu\gamma_5 \frac{\tau^3}{2} \psi(x)
\ee
where $\tau^3$ denotes the Pauli  matrix acting in flavor space 
and $\psi(x)$ is the isospin quark doublet.
The four axial form factors, $g_1, g_3, h_1$ 
and $h_3$ as defined in Eq.~(\ref{decomp_eqs}) are grouped into 
the familiar structure of the nucleon axial-vector vertex.

 In the description of spin-3/2 energy-momentum eigenstates, classical
solutions of the Rarita-Schwinger equation play a central role. Each
component of a vector-spinor $u_\sigma$, with $\sigma$  a Lorentz four-vector index solves the free Dirac equation
\begin{equation}
   \left[ \displaystyle{\not} {p} - M_\Delta \right] u^\Delta_\sigma(p,s) = 0\, .
\end{equation}
Implementing additionally the constraint equations,
\begin{equation}
   p^\sigma u^\Delta_\sigma(p,s) = 0 \qquad {\rm and} \qquad \gamma^\sigma u^\Delta_\sigma(p,s) = 0 \>,
\end{equation}
the unphysical components are eliminated and the remaining eight degrees of freedom describe a spin-3/2 (anti-)particle.
Rarita-Schwinger spinors satisfy the spin sum relation:
\begin{eqnarray}
\label{RS_spinsum_min}
\Lambda_{\sigma\tau} &\equiv& \sum^{3/2}_{s=-3/2} u^\Delta_\sigma(p,s)\overline{u}^\Delta_\tau(p,s)
\nonumber\\
&=&-\frac{\ds \displaystyle{\not} p + M_\Delta}{\ds 2M_\Delta}
\bigg(g_{\sigma\tau} - \frac{\ds\gamma_\sigma\gamma_\tau}{\ds 3}
-\frac{\ds 2p_\sigma p_\tau}{\ds 3 M_\Delta^2} 
+\frac{\ds p_\sigma\gamma_\tau -p_\tau \gamma_\sigma }{\ds 3 M_\Delta}\bigg),
\end{eqnarray}
where the normalization $\overline{u}^{\Delta\sigma} u^{\Delta}_\sigma = -1$ is assumed.

The zero momentum transfer limit of (\ref{decomp_eqs}) defines the axial
charge, $G_{\Delta\Delta}$ of the $\Delta$ multiplet. Ref~\cite{Jiang:2008we}
normalizes the axial charge via
\begin{equation}
\label{ax_charge_FF_relation}
\langle\Delta^{++}|A_\mu^3|\Delta^{++}\rangle -\langle\Delta^{-}|A_\mu^3|
\Delta^{-}\rangle = G_{\Delta\Delta}{\mathcal M}_\mu
\end{equation}
where ${\mathcal M}_\mu$ encodes the spin structure of the forward matrix
element
\be
{\mathcal M}_\mu = \overline{u}^{\Delta \sigma}(p) \gamma_\mu \gamma_5 u^\Delta_\sigma(p) \;.
\ee
Following the above normalization we establish via Eq.~(\ref{decomp_eqs}):
\be
G_{\Delta\Delta}=-3g_1(0) \;. 
\ee
We note that while the Lorentz decomposition of the axial current is naturally expressed 
via $g_1, g_3, h_1$ and $h_3$ as in Eq.(\ref{decomp_eqs}), a decomposition in terms of
multipoles is possible as for example in the case of the $\Delta$ electromagnetic 
transition ~\cite{Alexandrou:2009hs}.
Such a representation is more easily expressed in the Breit frame. This    
decomposition is performed in Appendix A in terms  of four multipoles, $L_1, L_3, E_1$ and $E_3$, and 
their relation to the form factors $g_1, g_3, h_1$ and $h_3$ is given.

The $\Delta^+$ matrix element of the pseudoscalar density operator 
\be
P(x) = \overline{\psi}(x) \gamma_5 \frac{\tau^3}{2} \psi(x)
\ee
is decomposed in terms of two Lorentz invariant form factors, denoted by
$\tilde{g}(q^2)$ and $\tilde{h}(q^2)$: 
\be
\langle \Delta^+(p_f,s_f) | P(0)| \Delta^+(p_i,s_i)\rangle 
=\overline{u}^\Delta_\sigma(p_f,s_f)\left[{\mathcal O}^{{\rm P}}\right]
^{ \sigma \tau} u^{\Delta}_\tau(p_i,s_i), \nonumber
\ee
\be
\label{ps_decomp}
\left[{\mathcal O}^{{\rm PS}}\right]^{\sigma\tau}=
- \frac{1}{2}\left[ g^{\sigma \tau}\left({\tilde g }\gamma^5 \right) 
+ \frac{\ds q^\sigma q^\tau}
{\ds 4M_\Delta^2} \left(\tilde{h }\gamma^5 \right) \right] \;.
\ee
\nin
While $\tilde{g}(q^2)$ and $\tilde{h}(q^2)$ are the directly computable form factors from the
three-point pseudoscalar correlator, they can be
related to the  phenomenologically more interesting pion-$\Delta$ vertex using
the partially conserved axial current hypothesis (PCAC).
Using PCAC on the hadronic level one can write
\be
\partial^\mu A_\mu^a=f_\pi m_\pi^2 \pi^a ~,
\label{PCAC}
\ee
with $\pi^a$ denoting the isotriplet pion field operator.
In the SU(2) symmetric limit of QCD with $m_q$ denoting 
the up/down mass, the pseudo-scalar density is related  
to the divergence of the axial-vector current through the
axial Ward-Takahashi identity (AWI) 
\be
 \partial^\mu A_\mu^a = 2 m_q P^a=f_\pi m_\pi^2 \pi^a ~,
\label{AWI}
\ee
with operators now defined as quark bilinears.
Using the relations of Eqs.~(\ref{PCAC}) and (\ref{AWI}) we identify the physically relevant
pion-$\Delta$-$\Delta$ from factor  $G_{\pi\Delta\Delta}(q^2)$, which
at $q^2=0$ gives the $\pi \Delta \Delta$ coupling, as well as a
 second form factor  $H_{\pi\Delta\Delta}(q^2)$,
by rewriting the pseudoscalar matrix element as
\be
\label{eff_couple_eq}
2 m_q \langle\Delta^+ (p_f,s_f) |P(0)| \Delta^+ (p_i,s_i) \rangle  \equiv 
\frac{\ds f_\pi m_\pi^2 }{\ds (q^2- m_\pi^2)}\times 
\overline{u}^\Delta_\sigma \left [g^{\sigma\tau} G_{\pi\Delta\Delta}(q^2)
+\frac{\ds q^\sigma q^\tau}{\ds 4M_\Delta^2} H_{\pi\Delta\Delta}(q^2)
\right] \gamma^5 u^\Delta_\tau ~,
\ee
where we effectively make the identification
\beq 
G_{\pi\Delta\Delta} (q^2)&\equiv&\frac{m_q (m_\pi^2-q^2)}{f_\pi m_\pi^2} \tilde{g}(q^2) 
\label{gpdd}
\\
H_{\pi\Delta\Delta} (q^2)&\equiv&\frac{ m_q (m_\pi^2-q^2)}{f_\pi m_\pi^2} \tilde{h}(q^2) ~.
\label{hpdd}
\eeq
At zero momentum transfer $q^2=0$ only $G_{\pi \Delta\Delta}$ can be extracted. This coupling
is analogous to the known $\pi-N$ pseudoscalar coupling constant $G_{\pi NN}$ defined for the nucleon. For the discussion presented in the next section
it is useful to recall the definition of the corresponding quantities in the nucleon sector~\cite{Alexandrou:2007xj}.
For the matrix elements of the axial-vector current we have 
\be 
\langle N (p_f,s_f) |A_\mu^3| N (p_i,s_i) \rangle  = i\frac{1}{2} \bar{u}_N \left[G_A(q^2)\gamma_\mu\gamma_5 + \frac{q_\mu \gamma_5}{2 m_N} G_p(q^2)\right]u_N
\label{gA}
\ee
and for the pseudoscalar density 
\be 
2 m_q \langle N (p_f,s_f) |P^3| N (p_i,s_i) \rangle  =
\frac{\ds f_\pi m_\pi^2 }{\ds (q^2- m_\pi^2)}\times
\overline{u}_N \left [G_{\pi NN}(q^2)
\right] i\gamma^5 u_N ~.
\label{gpnn}
\ee
Note that we have dropped for simplicity an overall kinematical factor arising from the normalization of lattice states, since it is of no relevance for our discussion here.

\subsection{Goldberger-Treiman Relations}

In this section we apply PCAC to derive GT relations for the $\Delta$.
We recall that PCAC   has been shown to 
apply satisfactorily in the nucleon case 
leading to the Goldberger-Treiman (GT) relation.
This can be derived from Eqs.~(\ref{gA}) and (\ref{gpnn}) related by AWI
and taking $q^2=0$ to obtain $G_{\pi NN}$ in terms of the nucleon
axial charge via the relation
\be
f_\pi G_{\pi NN}(0)=m_N G_A(0) .
\label{GT nucleon}
\ee
Assuming $G_{\pi NN}$ varies smoothly with $q^2$ so that $G_{\pi NN}(0) \sim G_{\pi NN}(m_\pi^2)\equiv g_{\pi NN}$ then
the GT relates the physical coupling constant $g_{\pi NN}$ with the nucleon axial charge $g_A$. 
At the chiral limit, using $\partial_\mu A_\mu=0$ one 
derives that  $G_p(q^2)=-\frac{4m_N^2}{q^2} G_A(q^2)$. Therefore,
$g_{\pi N N}$
measures the chiral symmetry breaking. 
PCAC dictates that the form factor 
$G_p(q^2)$ has a pion pole  given by $G_p(q^2)=\frac{4m_Nf_\pi}{m_\pi^2-q^2} G_{\pi NN}(q^2)$.
The validity of the GT relation and the momentum dependence
of $G_p(q^2)$ in the nucleon case has been studied
in Ref.~\cite{Alexandrou:2007xj}.
Similarly, a non-diagonal GT relation, applicable to the axial $N$-to-$\Delta$ transition
is formulated and relates the axial $N\Delta$ coupling $c_A$ to the 
$\pi N \Delta$ effective coupling. 
Lattice calculations examined the validity of the non-diagonal GT relation using the same ensembles as 
in this work~\cite{Alexandrou:2010uk}.

One can similarly derive GT relations for the $\Delta$  by taking the 
 matrix elements of the AWI  with $\Delta$ states,  
$ \langle\Delta| \partial_\mu A^\mu | \Delta \rangle = 2m_q\langle\Delta|P | 
\Delta \rangle $. 
Taking the dot-product of $q_\mu$ with  the  matrix element of the axial-vector current given in Eq.~(\ref{decomp_eqs})
we obtain

\beq
\label{GTmatrix}
m_\Delta \left [ g^{\sigma\tau} (g_1 - \tau g_3) 
+\frac{q^\sigma q^\tau}{4M_\Delta^2} (h_1 - \tau h_3) \right ]= 
\frac{f_\pi m_\pi^2 }{(m_\pi^2-q^2)}~
\left [g^{\sigma\tau} G_{\pi\Delta\Delta}
+\frac{q^\sigma q^\tau}{4M_\Delta^2} H_{\pi\Delta\Delta}
\right] ~,
\eeq
where $\tau=-q^2/(2M_\Delta)^2$.
 By considering
 $\sigma\ne\tau$ in Eq.~(\ref{GTmatrix}) we derive the relation
\be
m_\Delta\left(h_1  -\tau h_3\right)  = \frac{f_\pi m_\pi^2 
H_{\pi\Delta\Delta}(q^2) }{m_\pi^2-q^2},
\label{GT2}
\ee
which implies that 
\be
M_\Delta\left(g_1  -\tau g_3\right) = \frac{f_\pi m_\pi^2 
G_{\pi\Delta\Delta}(q^2) }{m_\pi^2-q^2}. 
\label{GT1}
\ee

One possible linear combination of Eqs.~(\ref{GT2}) and (\ref{GT1}) can be obtained by taking the dot product of Eq.~(\ref{GTmatrix}) with $q_\tau$ leading to 
\be
M_\Delta\left[(g_1  -\tau g_3) - \tau (h_1  -\tau h_3)\right] = 
\frac{f_\pi m_\pi^2 }{m_\pi^2-q^2}~\left[ G_{\pi\Delta\Delta}-\tau
H_{\pi\Delta\Delta} \right]\>,
\label{GT3}
\ee
which can be considered as  a generalized GT-type relation connecting all the six form factors.
 At $q^2=0$ and assuming all terms in Eq.~(\ref{GT3}) are finite  we obtain
\be
f_\pi G_{\pi \Delta \Delta}(0) = m_\Delta g_1(0).
\label{GT1 Delta}
\ee
 If  $G_{\pi \Delta \Delta}$ is a continuous  
slow varying function of $q^2$ as $q^2\rightarrow 0$ then $G_{\pi \Delta \Delta}(m_\pi^2)\sim G_{\pi \Delta \Delta}(0)$
and we thus derive a GT relation for the $\Delta$ analogous to the one for the nucleon case.

Using Eq.~\ref{GT2} and setting $q^2=0$ 
we obtain a second GT relation
\be
f_\pi H_{\pi \Delta \Delta}(0) = m_\Delta h_1(0).
\label{GT2 Delta}
\ee
If one invokes pion-pole dominance and noting that $g_1$ and $G_{\pi\Delta\Delta}$ are both finite at the origin, it follows from  Eqs.~(\ref{GT1}, \ref{GT2}) that as $q^2\rightarrow m_\pi^2$
$g_3$ and $h_3$ must have a pole at $q^2=m_\pi^2$.
We thus  arrive at
the relations
\be
g_1 = \frac{f_\pi}{M_\Delta} G_{\pi\Delta\Delta} \;\;\;\; , \;\;\;\;
g_3 = \frac{4 f_\pi M_\Delta }{m_\pi^2-q^2} G_{\pi\Delta\Delta} \;.
\label{gr1}
\ee
and 
\be
h_1 = \frac{f_\pi}{M_\Delta} H_{\pi\Delta\Delta} \;\;\;\; , \;\;\;\;
h_3 = \frac{4 f_\pi M_\Delta }{m_\pi^2-q^2} H_{\pi\Delta\Delta} \;.
\label{gr2}
\ee
It is thus interesting to note how the spin-3/2 nature of the $\Delta$ state combined with PCAC
leads to a {\it pair} of Goldberger-Treiman relations,  given by Eqs.~(\ref{GT1 Delta}) and (\ref{GT2 Delta}).
Let us examine further these relations at the chiral limit. From Eq.~(\ref{GT3}) we find that
\be
 h_1-\tau h_3=\frac{g_1-\tau g_3}{\tau},
\label{pole-dependence}
\ee
 which means  that in the limit $q^2\rightarrow 0$ the leadin
behavior of  $h_1\sim 1/q^2$,  $h_3\sim 1/(q^2)^2$ via Eq.~(\ref{GT2})
and  $H_{\pi \Delta \Delta}\sim 1/q^2$ via Eq.~(\ref{gr2}).
Therefore, the second GT-type relation given in Eq.~(\ref{GT2 Delta}) cannot 
be extrapolated to physical pion mass since the assumption that $h_1$ and $H_{\pi\Delta\Delta}$ are slowly varying functions of $q^2$ no longer holds.  However, since they both display a pion-pole behaviour one can factor it out on both sides and
thus the ratio $h_1/H_{\pi \Delta \Delta}$ can be extrapolated to the physical pion. In this sense, this constitutes a second GT relation. 

\section{LATTICE EVALUATION} \label{se:LAT_CALC}

\subsection{Euclidean Correlators and Form Factors}
\nin
Standard techniques are employed on the Euclidean space-time lattice for the
evaluation of hadronic form factors. The following two-point and 
three-point functions are required:
\begin{eqnarray}
\label{23pts}
G_{\sigma\tau}(\Gamma^\nu,\vec{p},t_f)&=& 
\sum_{\vec{x}_f}e^{-i\vec{x}_f\cdot\vec{p}}
\,\Gamma^\nu_{\alpha^\prime\alpha}
\langle\chi_{\sigma\alpha}(t_f,\vec{x}_f)\overline{\chi}_{\tau\alpha^\prime}(0,\vec{0})\rangle
\nonumber\\
G_{\sigma\mu\tau}^{\rm A}(\Gamma^\nu,\vec{q},t;t_f)&=& 
\sum_{\vec{x},\vec{x}_f}e^{+i\vec{x}\cdot\vec{q}}
\,\Gamma^\nu_{\alpha^\prime\alpha}
\langle\chi_{\sigma\alpha}(t_f,\vec{x}_f)A_\mu(t,\vec{x})
\overline{\chi}_{\tau\alpha^\prime}(0,\vec{0})\rangle
\nonumber\\
G_{\sigma\tau}^{\rm PS}(\Gamma^\nu,\vec{q},t;t_f)&=& 
\sum_{\vec{x},\vec{x}_f}e^{+i\vec{x}\cdot\vec{q}}
\,\Gamma^\nu_{\alpha^\prime\alpha}
\langle\chi_{\sigma\alpha}(t_f,\vec{x}_f)P(t,\vec{x})
\overline{\chi}_{\tau\alpha^\prime}(0,\vec{0})\rangle,
\end{eqnarray}
where $P(t,\vec{x})$ and $A_\mu(t,\vec{x})$ are the lattice pseudoscalar 
or axial 
current insertions, and $\chi$ is the standard lattice interpolating field 
with overlap with the $\Delta^+$ quantum numbers:

\begin{equation}
   {\bf\chi}^{\Delta^+}_{\sigma\alpha}(x) = \frac{1}{\sqrt{3}} \epsilon^{abc}\Bigl[2\left({\bf u}^{a\top}(x) C\gamma_\sigma {\bf d}^b(x)\right)
 {\bf u}_\alpha^c(x)  
+ \left({\bf u}^{a\top}(x) C\gamma_\sigma {\bf u}^b(x)\right) {\bf d}_\alpha^c(x)\Bigr]\, .
\end{equation}
The overlap of $\chi$ with the spin-3/2 $\Delta^+$ is
\begin{equation}
   \langle \Omega| \chi_{\sigma\alpha}(0) |\Delta(p,s)\rangle = Z\, u^\Delta_{\sigma\alpha}(p,s)\, , \qquad \qquad   
   \langle \Delta(p,s) |\bar \chi_{\sigma\alpha}(0) |\Omega\rangle = Z^*\, \bar u^\Delta_{\sigma\alpha}(p,s)\, .
\end{equation}
We will use the following $\Gamma$ matrices, which project onto positive parity for zero momentum, for our
calculation
\begin{equation}
   \Gamma^4 = \frac{1}{4}(\ident  + \gamma^4)\, , \qquad \qquad \Gamma^k =
   \frac{i}{4}(\ident  + \gamma^4)\gamma_5\gamma_k\, , \qquad k=1, 2, 3\, .
\end{equation}
The Fourier transforms in (\ref{23pts}) enforce a static $\Delta$ sink at the
timeslice $t_f$ and a momentum transfer $\vec{q} = -\vec{p}\,\,$ injected via the
operator insertion at an intermediate timeslice $t$.

\nin
We insert into these correlators complete sets of hadronic energy momentum 
eigenstates:
\begin{equation}
\sum_{n,p,\xi}\frac{\ds M_n}{\ds V E_{n(p)}}\left|n(p,\xi)\rangle
\langle n(p,\xi)\right|=\ident,
\end{equation}
where with $\xi$ we denote collectively all quantum numbers including spin.
For large Euclidean time separations $t$ and $t_f-t$ the ground state 
propagation dominates the correlator:
\begin{eqnarray}
G_{\sigma\tau}(\Gamma^\nu,\vec{p},t)&=& 
\frac{\ds M_\Delta}{\ds E_{\Delta(p)}}\left|Z\right|^2e^{-E_{\Delta(p)}t}
{\rm tr} \left[ \Gamma^\nu\Lambda^E_{\sigma\tau}(p)\right] 
+ {\rm excited\ states}
\nonumber\\
G_{\sigma\mu\tau}^{\rm A}(\Gamma^\nu,\vec{q},t;t_f)&=& 
\frac{\ds M_\Delta}{\ds E_{\Delta(p)}}\left|Z\right|^2e^{-M_{\Delta}(t_f-t)}e
^{-E_{\Delta(p)}t}
{\rm tr}\left[ \Gamma^\nu
\Lambda^E_{\sigma\sigma^\prime}(0)
{\mathcal O}^{E,A}_{\sigma^\prime\mu\tau^\prime}
\Lambda^E_{\tau\tau^\prime}(p)
\right]
\nonumber\\
&& + {\rm excited\ states}
\nonumber\\
G_{\sigma\tau}^{\rm PS}(\Gamma^\nu,\vec{q},t;t_f)&=& 
\frac{\ds M_\Delta}{\ds E_{\Delta(p)}}\left|Z\right|^2e^{-M_{\Delta}(t_f-t)}e
^{-E_{\Delta(p)}t}
{\rm tr}\left[ \Gamma^\nu
\Lambda^E_{\sigma\sigma^\prime}(0)
{\mathcal O}^{E,PS}_{\sigma^\prime\tau^\prime}
\Lambda^E_{\tau^\prime\tau}(p)
\right]
 \nonumber\\
&& + {\rm excited\ states}
\end{eqnarray}

\nin
The Wick-rotated axial and pseudoscalar operators take the form
\begin{equation}
\label{euc_ax_op}
{\mathcal O}_{\sigma\mu\tau}^{E,A}=
\frac{1}{2}\Biggl[\delta_{\sigma\tau}\left(g_1(Q^2)\gamma_\mu \gamma_5 - i\frac{g_3(Q^2)}{2M_\Delta}Q_\mu\gamma_5\right)
-\frac{Q_\sigma^E Q_\tau^E}{(2M_\Delta)^2}
\left(h_1(Q^2)\gamma_\mu\gamma_5  -i \frac{h_3(Q^2)}{2M_\Delta}Q_\mu \gamma_5\right)\Biggr],
\end{equation}


\begin{equation}
\label{euc_ps_op}
{\mathcal O}_{\sigma\tau}^{E,PS}=
\frac{1}{2}\biggl[\delta_{\sigma\tau}
\left(\tilde{g}(Q^2) \gamma_5\right) -
\frac{Q_\sigma Q_\tau}{(2M_\Delta)^2}
\left(\tilde{h}(Q^2) \gamma_5 \right)\Biggr],
\end{equation}
with the Euclidean four-momentum transfer $Q_\mu = 
\left(i(M_\Delta-E_{\Delta(p)}), -\vec{q} \right)$.
The Rarita-Schwinger spin-sum relation becomes
\begin{equation}
\Lambda_{\sigma\tau}^E =
-\frac{\ds -i\displaystyle{\not} p + M_\Delta}{\ds 2M_\Delta}
\bigg(\delta_{\sigma\tau} - \frac{\ds\gamma_\sigma\gamma_\tau}{\ds 3}
+\frac{\ds 2p_\sigma p_\tau}{\ds 3 M_\Delta^2} 
-i\frac{\ds p_\sigma\gamma_\tau -p_\tau \gamma_\sigma}
{\ds 3 M_\Delta}\bigg),
\end{equation}
where all the $\gamma$ matrices are in Euclidean space: $\gamma_0=\gamma_4$ and $\gamma^M_k=-i\gamma^E_{k}$.

\nin
Forming an appropriate ratio of the 3-point to the 2-point correlator
serves to  cancel out the unknown $Z$-factors and leading time-dependence. 
A particular product of 2-point correlators which minimizes the denominator noise-level
is utilized as it contains smaller time-extents.
The proposed ratios are:
\begin{equation}
 R_{\sigma\mu\tau}^{A}(\Gamma^\nu,\vec Q,t) = 
\frac{G_{\sigma\mu\tau}^{A}(\Gamma,\vec Q,t)}{G_{k k}(\Gamma^4,\vec 0, t_f)}
\sqrt{\frac{G_{kk}(\Gamma^4,\vec p_i, t_f-t)G_{kk}(\Gamma^4,\vec 0  ,t)
            G_{kk}(\Gamma^4,\vec 0,t_f)}
{G_{kk}(\Gamma^4,\vec 0, t_f-t)
G_{kk}(\Gamma^4,\vec p_i,t)G_{kk}(\Gamma^4,\vec p_i,t_f)}}\, 
\label{ax_ratio}
\end{equation}
and
\begin{equation}
        R_{\sigma\tau}^{PS}(\Gamma^\nu,\vec Q,t) = \frac{G_{\sigma\tau}^{PS}(\Gamma,
\vec Q,t)}{G_{k k}(\Gamma^4,\vec 0, t_f)}\ 
                                         \sqrt{\frac{G_{kk}(\Gamma^4,\vec p_i, t
_f-t)G_{kk}(\Gamma^4,\vec 0  ,t)G_{kk}(\Gamma^4,\vec 0,t_f)}
                                                    {G_{kk}(\Gamma^4,\vec 0, t_f
-t)G_{kk}(\Gamma^4,\vec p_i,t)G_{kk}(\Gamma^4,\vec p_i,t_f)}}\, ,
\label{ps_ratio}
\end{equation}
for the axial and pseudoscalar vertices.
Summation over $k=1,2,3$ is implicit in the 2-point correlators.
At large Euclidean time separations $t_f-t$  and $t$ these ratios become time-independent
(plateau region).
\begin{equation}
\label{generic_ratio}
R^X_{\sigma(\mu)\tau}(\Gamma^\nu,\vec{Q},t) \longrightarrow C\Pi_{\sigma(\mu)\tau}^X
=C{\rm tr} \left[\Gamma^\nu \Lambda_{\sigma\sigma^\prime}(0) {\mathcal O}
^X_{\sigma(\mu)\tau} \Lambda_{\tau^\prime\tau}(p)\right],
\end{equation}
where $X$ stands for the axial $(A_\mu)$ or pseudoscalar $(P)$ current.  
It is easy to show that the 2-point correlators are dominated by
\be
G_{kk}(\Gamma^4, \vec{p},t) = \left|Z\right|^2 e^{-E_{\Delta(p)}t} 
\frac{\ds E_{\Delta(p)} + M_\Delta}{ \ds E_{\Delta(p)}} 
\left(1 + \frac{ \vec{p}^{\,2}}{3 M_\Delta^2} \right)
\ee
and therefore the constant $C$ is determined as
\begin{equation}
C\equiv\sqrt{\frac{3}{2}}\left[\frac{2 E_{\Delta(p_i)}}{M_\Delta} 
                          +\frac{2 E^2_{\Delta(p_i)}}{M^2_\Delta} 
                          +\frac{  E^3_{\Delta(p_i)}}{M^3_\Delta} 
                          +\frac{  E^4_{\Delta(p_i)}}{M^4_\Delta} \right]^
{-\frac{1}{2}} \;.
\end{equation}

\nin
There are at most 256 available combinations of the Dirac and Lorentz indices 
in Equation (\ref{generic_ratio}), each one expressed as a linear 
combination of the axial (pseudoscalar) form factors times kinematical
tensor coefficients. Since we are interested in the momentum dependence of the
matrix elements, evaluation of the 3-point correlators is required for a 
large set of transition momenta $\vec{q}\,\,$ for both $A_\mu$ and $P$ operators.
In order to perform this economically we utilize the 
{\it sequential inversion through the sink} technique~\cite{Alexandrou:2009hs}
by fixing the sink timeslice $t_f$ and performing a backward sequential 
inversion through the sink. The sequential vector is coupled with
a forward quark propagator and the Fourier transformed 
insertion operator at all intermediate time-slices $0 \le t \le t_f$ 
at a small computational cost, obtaining thus the full momentum dependence of the 
amplitude. A drawback in this approach is the fact that the quantum numbers
of the source and sink interpolators --which correspond to the Lorentz indices
$\sigma, \tau$ and $\Gamma^\nu$ are now {\it fixed per sequential inversion}. 
Within the space of 64 available 3-point correlators corresponding to choices 
of $\sigma, \tau$ and $\nu$ we perform an {\it optimization by forming
appropriate linear combinations} such as the degree of rotational symmetry of 
the summed correlator is maximal and consequently all transition momentum vectors $\vec{q}$
that correspond to a fixed virtuality $q^2$ will contribute to the form factor
measurement in a rotationally symmetric fashion.
This optimization technique has proved extremely useful in obtaining high
accuracy results in the Nucleon elastic, 
Nucleon-to-$\Delta$ electromagnetic~\cite{Alexandrou:2010uk},
axial and pseudoscalar transitions~\cite{Alexandrou:2007xj}
as well as the $\Delta$ electromagnetic form factors~\cite{Alexandrou:2009hs}.
We evaluate the Dirac traces in Eq.~(\ref{generic_ratio}) using symbolic software
such as {\tt form} \cite{form_ref} and Mathematica. 

\nin
We construct the following two optimal linear combinations, which we refer to as Type-I and Type-II. 
\beq
{\rm Type-I:} \hspace{0.5cm}&& \Pi^{I{\rm A}}_\mu(q)\equiv 
\sum_{i=1}^3 \sum_{\sigma,\tau=1}^3
\delta_{\sigma\tau}{\rm tr}\left[\Gamma^i 
\Lambda_{\sigma\sigma^\prime}(0) 
{\mathcal O}_{\sigma^\prime \mu \tau^\prime}^{E,A}
\Lambda_{\tau^\prime\tau}(p)\right] \\
{\rm Type-II:} \hspace{0.5cm} &&\Pi^{II{\rm A}}_\mu(q)\equiv 
\sum_{\sigma,\tau =1}^3 T_{\sigma\tau}{\rm tr}\left[
\Gamma^4 \Lambda_{\sigma\sigma^\prime}(0) 
{\mathcal O}_{\sigma^\prime \mu \tau^\prime}^{E,A}
\Lambda_{\tau^\prime\tau}(p)\right]
\label{types}
\eeq
with the matrix $T$: 
\be
T_{\sigma\tau}=\left[\begin{array}{rrr}
0&1&-1\\
-1&0&1\\
1&-1&0\end{array}
\right].
\ee
Detailed expressions for the decomposition of the above combinations to the four 
axial form factors
are provided in Appendix~B. The above types are in addition utilized for the extraction of the
two pseudoscalar couplings (Appendix~B). 
A large number ($O(10^3))$ correlators of axial $(A_\mu ,\mu=1,2,3,4)$  and pseudoscalar $(P)$ 
insertion momenta $\vec{q}$ are combined  
for momentum transfers ranging up to $\sim 3$ GeV$^2$ per ensemble. 
We stress that {\it only two} sequential inversions through sink --one for each Type above--
are required in order to disentangle completely all six form factors from the relevant
3-point functions.

Correlators corresponding to a fixed momentum transfer $q^2$ are analyzed simultaneously in an
overconstrained system analysis for the extraction of the form factors. Typically $O(20-50)$
plateau averages for the optimal ratios given in Eq.~(\ref{generic_ratio}) will contribute to the
determination of the 
form factor for each $Q^2$ value.
A global $\chi^2$-minimization amounts technically to the singular value decomposition
of an $N \times M$ over-complete linear system, with $M$ unknowns 
(4 for the axial or 2 for the pseudoscalar) and $N$ input data 
(the O(20-50) plateau averages). Further details on this kind of analysis can be found 
in Ref~\cite{Tsapalis:2006kn}. Jackknife estimates are utilized for all levels of variance 
extraction on observables.

\subsection{Ensembles and Parameters}

\nin
In Table~\ref{Table:params} we summarize the parameters and number of configurations for the 
ensembles used in this work. As can be seen, three sets are employed. These are
the same as the ones we used previously for the study of the 
nucleon axial form factors as well as the  nucleon-to-$\Delta$ axial
transition form factors. Therefore, these ensembles provide a complete
calculation of the nucleon/$\Delta$ sector, allowing a direct extraction of low energy
couplings from a combined fit.


\begin{table}[t]
\small
\begin{center}
\begin{tabular}{ccccccc}
\hline\hline 
V& stat. & $m_\pi$ (GeV) & $m_N$ (GeV) & $m_\Delta$ (GeV) & $\kappa$\\ 
\hline
\multicolumn{5}{c}{Quenched Wilson fermions}\\
\multicolumn{5}{c}{$\beta=6.0,~~a^{-1}=2.14(6)$~GeV} \\
\hline
$32^3\times 64$& 200  &0.563(4)& 1.267(11) & 1.470(15)& 0.1554\\
$32^3\times 64$& 200  &0.490(4)& 1.190(13) & 1.425(16)& 0.1558\\
$32^3\times 64$& 200  &0.411(4)& 1.109(13) & 1.382(19)&0.1562\\
\hline
\hline
\multicolumn{5}{c}{Mixed action, $a^{-1} = 1.58(3)$~GeV} \\
\multicolumn{5}{c}{
Asqtad ($am_{\mbox{\tiny u,d/s}} = 0.02/0.05$),  
DWF ($ am_{\mbox{\tiny u,d}} = 0.0313$)}\\ 
\hline
$20^3\times 64$ &264 &   0.498(3) & 1.261(17)& 1.589(35)\\
\multicolumn{5}{c}{
Asqtad ($am_{\mbox{\tiny u,d/s}} = 0.01/0.05$),  
DWF ($ am_{\mbox{\tiny u,d}} = 0.0138$)}\\ \hline
$28^3\times 64$ &550 &   0.353(2) & 1.191(19)& 1.533(27)\\
\hline\hline
\multicolumn{5}{c}{Domain Wall Fermions (DWF)} \\
\multicolumn{5}{c}{
 $ m_{\mbox{\tiny u,d}}/m_s=0.004/0.03$, $a^{-1} =2.34(3)$~GeV}\\\hline
$32^3\times 64$ &1428 &   0.297(5) &1.27(9) &1.455(17) \\
\hline
\hline
\end{tabular}
\end{center}
\caption{Ensembles and parameters used in this work.  We give in the first 
column  the lattice size, in the second the statistics, in the third, fourth 
and fifth the pion, nucleon and $\Delta$ mass in GeV respectively.
We did not do a full form-factor analysis on the $20^3\times 64$ mixed-action
ensemble.  Rather we merely determined the axial matrix element at $q^2=0$ 
(a much cheaper computation) for our axial charge chiral fits.}
\label{Table:params}
\end{table}

The gauge configurations used in the analysis include a set of quenched configurations
 on a $32^3 \times 64$, at $\beta = 6.0$, corresponding to a 
lattice spacing  $a=0.092$ fm  with pion masses
560~MeV, 490~MeV and 411~MeV. The low  statistical noise makes this 
ensemble appropriate for checking our lattice 
methodology and some of the phenomenological relations.
We apply Gaussian smearing at the source and sink in order to minimize the excited state
contamination on the baryon correlators. The parameters $\alpha=4.0$ and $n=50$ have been
tuned to provide optimal overlap to a nucleon state~\cite{Alexandrou:2007xj}. 
The source-sink separation
is set at $\Delta T = 12a = 1.1$ fm. In our previous studies involving the $\Delta$
such a time separation was found sufficient  for
ground state dominance.
The second set consists of two  ensembles that use two degenerate light and one strange ($N_f=2+1$) Asqtad-improved dynamical staggered fermions
generated by the MILC collaboration~\cite{Bernard:2001av}. The strange quark mass is fixed to its physical value, the lattice spacing is set to 0.124~fm and the lowest pion mass is 353~MeV.
Our calculation  employs Domain Wall (DW) valence quarks 
with light quark mass  tuned so as the pion mass  matches the lowest pion mass obtained using staggered fermions.
The extent of the fifth dimension of the domain wall action is set to $L_5 = 16a$, which was
 demonstrated to
provide minimal violations to the chiral symmetry properties of the domain wall fermion (DWF) operator. 
 The source-sink separation is set to $\Delta T = 8a = 1.0$ fm and Gaussian smearing is applied at the source
and sink with APE smearing on gauge links that enter the smearing function applied on the interpolating fields. The parameters are given in Ref.~\cite{Alexandrou:2007xj}.  
Finally, the third set  
is an $N_f=2+1$ ensemble
of DWF  generated by the 
RBC-UKQCD collaborations~\cite{Aoki:2010dy} with a lattice spacing  $a=0.084$~fm and the physical volume of
 $(2.7\,{\rm fm})^3$  
and  pion mass of 0.297~MeV. The
extent of the fifth dimension is also here $L_5 = 16a$. It turns out that the residual quark mass introduced
via the chiral symmetry breaking effects is $a m_{\rm res} = 0.000665(3)$, or  $17\%$ 
of the bare quark mass. The smearing parameters for the interpolating fields are given in Ref.~\cite{Alexandrou:2010uk}. The sink-source time  separation 
is set at $\Delta T = 12a = 1.01$ fm.
In order to increase the statistics at this lowest pion mass  we use
the {\it coherent sink technique}, employed in our study of the nucleon to $\Delta$ transition using the same ensemble~\cite{Alexandrou:2010uk}. The four quark
sources are placed at timeslice $t_i = (i-1) 16, i=1,\cdots, 4$ for each configuration. Four forward  
propagators must be computed -- each with a  source at one of the time-slices.
 The $\Delta$ sinks are constructed  at all four equally spaced time-slices 
$t_f(i) = t_i + 12$ and  one sequential inversion is performed in order to  construct 
 the three-point correlator. 
Gauge invariance ensures that combining the sequential vector with each one of 
the forward quark propagators generated at $t_i$, projects the appropriate $\Delta$ matrix element
 between $t_i$ and $t_i + 12$ as the other cross-terms will average to zero. It has been shown
in Ref.~\cite{Alexandrou:2010uk} that while statistics is thus multiplied by 4, the noise level is
not raised above what expected from the four completely independent correlators which 
participate in the coherent sink. This means that we can reduce
the error by a factor of two at the cost of one sequential inversion.
Therefore the 1428 statistics given in the table correspond to  357 coherent sequential inversions per each type
of combinations (see
Eq.~(\ref{types})).

\section{Results}

\subsection{Axial-vector and Pseudoscalar Form Factors}

In this section we present results on the $\Delta$ axial-vector and pseudoscalar form factors from
the ensembles utilized in this work.
The axial current is renormalized multiplicatively in all ensembles. 
Values for the renormalization constant $Z_A$ are provided in Table \ref{tableZ}.

\begin{figure}
\begin{center}
\scalebox{0.4}{\includegraphics{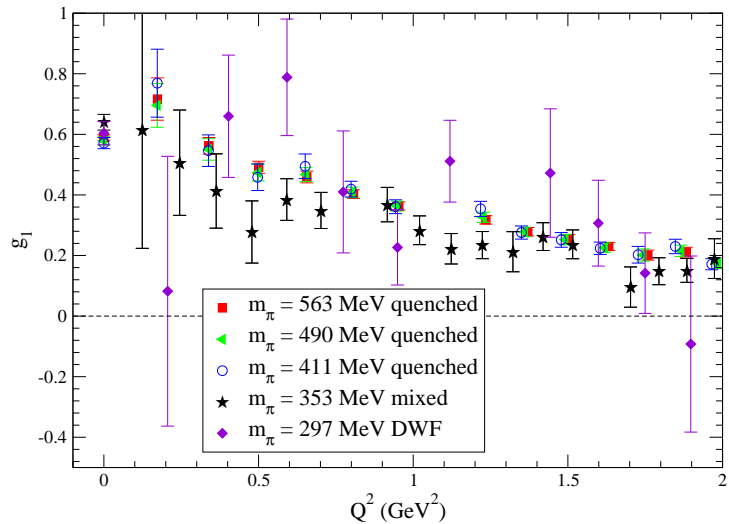}}
\end{center}
\caption{\label{g1_FFS_fig} Lattice QCD results for the $g_1$ axial form-factor.}
\end{figure}

\begin{figure}
\begin{center}
\scalebox{0.4}{\includegraphics{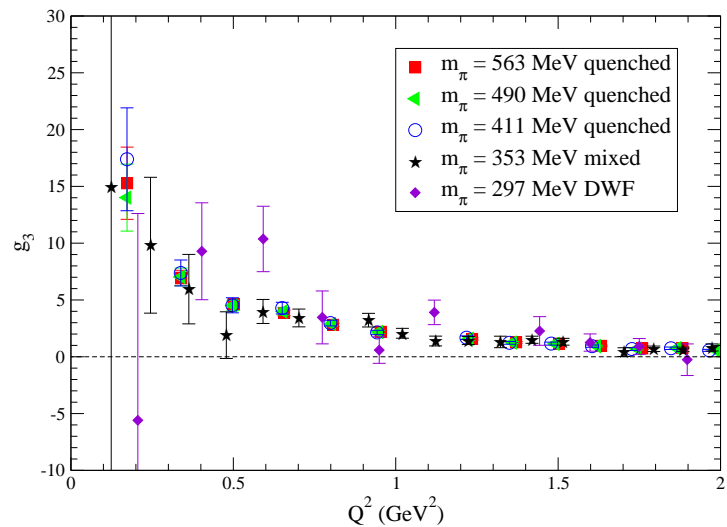}}
\end{center}
\caption{\label{g3_FFS_fig} Lattice QCD results for the $g_3$ axial form-factor.}
\end{figure}

\begin{figure}
\begin{center}
\scalebox{0.4}{\includegraphics{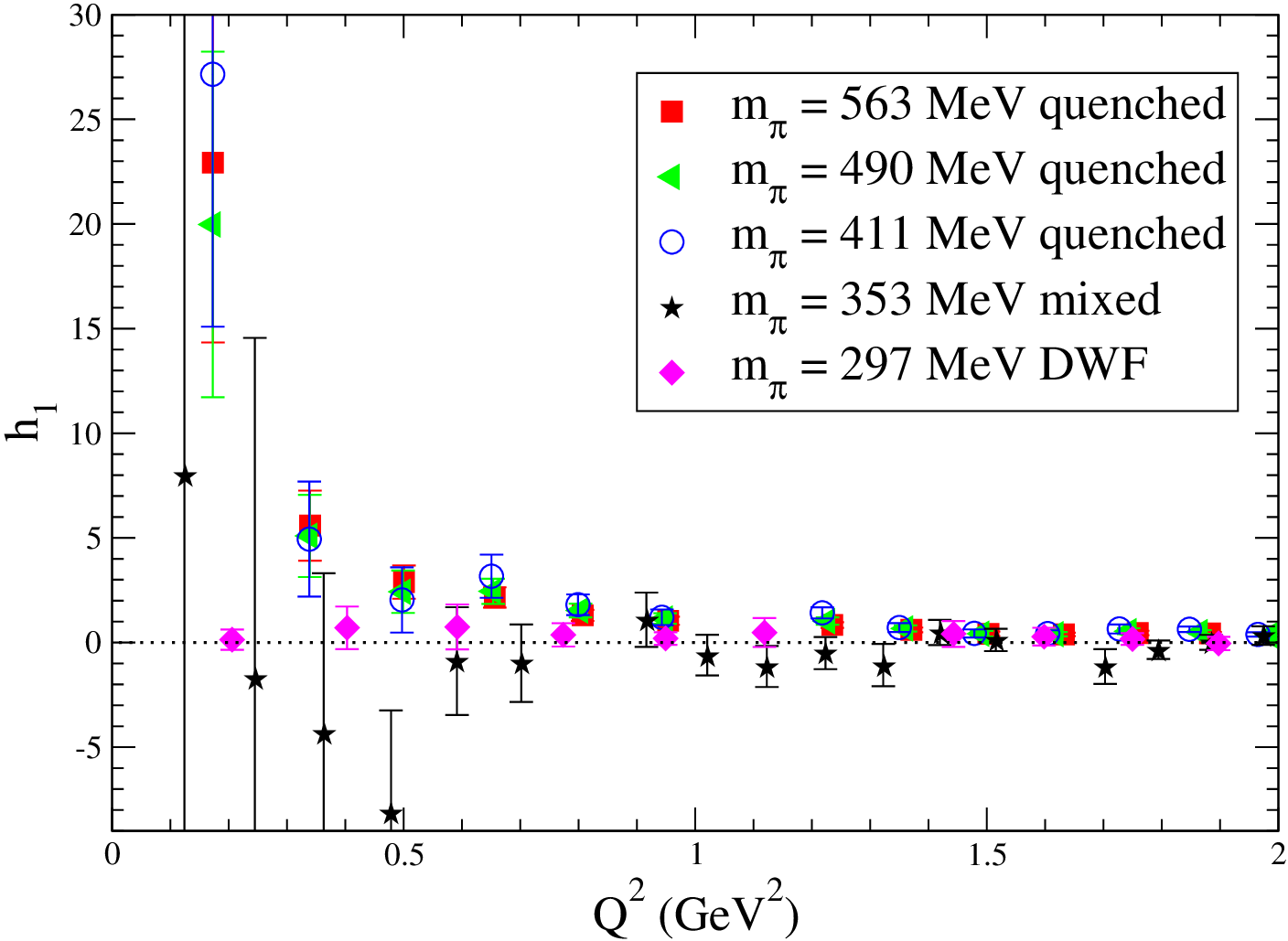}}
\end{center}
\caption{\label{h1_FFS_fig} Lattice QCD results for the $h_1$ axial form-factor.}
\end{figure}

\begin{figure}
\begin{center}
\scalebox{0.4}{\includegraphics{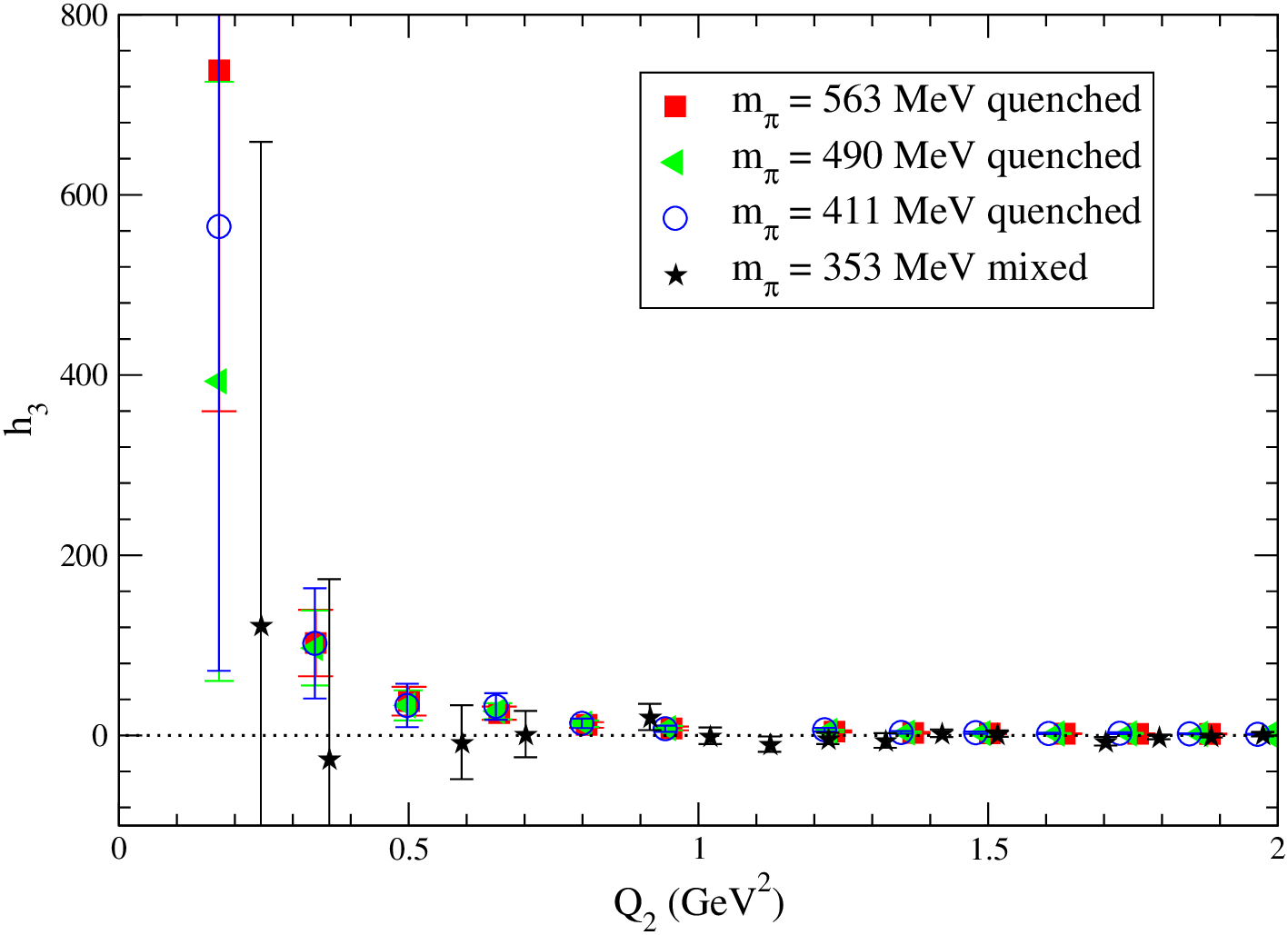}}
\end{center}
\caption{\label{h3_FFS_fig} Lattice QCD results for the $h_3$ axial form-factor.}
\end{figure}

In Figs.~\ref{g1_FFS_fig}, \ref{g3_FFS_fig}, \ref{h1_FFS_fig} and 
\ref{h3_FFS_fig} we show the results for the four
axial form factors, $g_1$, $g_3$, $h_1$ and $h_3$, respectively.
All the results on these form factors are provided in Appendix~C.
The form factor $g_1$ is the dominant axial-vector form factor and the only one that can be extracted directly from the matrix element 
 at $Q^2 = 0$, determining the axial charge of the $\Delta$.  
Based on PCAC and pion pole dominance we expect 
 $g_1$ to be a smooth function of $Q^2$, whereas
$h_1$ and $g_3$  to have a pion-pole and $h_3$ a double  pion-pole  behavior.
Given that $g_1$ and  $h_1$ are multiplied by $Q^2$, whereas $h_3$ is multiplied by $Q^4$ it is
 increasingly more difficult to resolve these form factors via the
simultaneous overconstrained analysis of the
measured  matrix element of the axial-vector current,
especially at small $Q^2$ -- a fact that is clearly reflected on the
statistical error of the form factors shown in the figures. 
The results from the quenched ensemble, although based on the analysis of 200 configurations, have the 
lowest statistical noise and this is the primary reason for using them in this
first calculation of the form factors.
The statistical noise is more severe for the DWF ensemble at $m_\pi=297$~MeV
for which results on $h_3$ are too noisy to be useful  and are omitted 
from plots. 
We do, however, include these
numbers in the tables in the Appendix~C for completeness.

\begin{figure}
\begin{center}
\scalebox{0.4}{\includegraphics{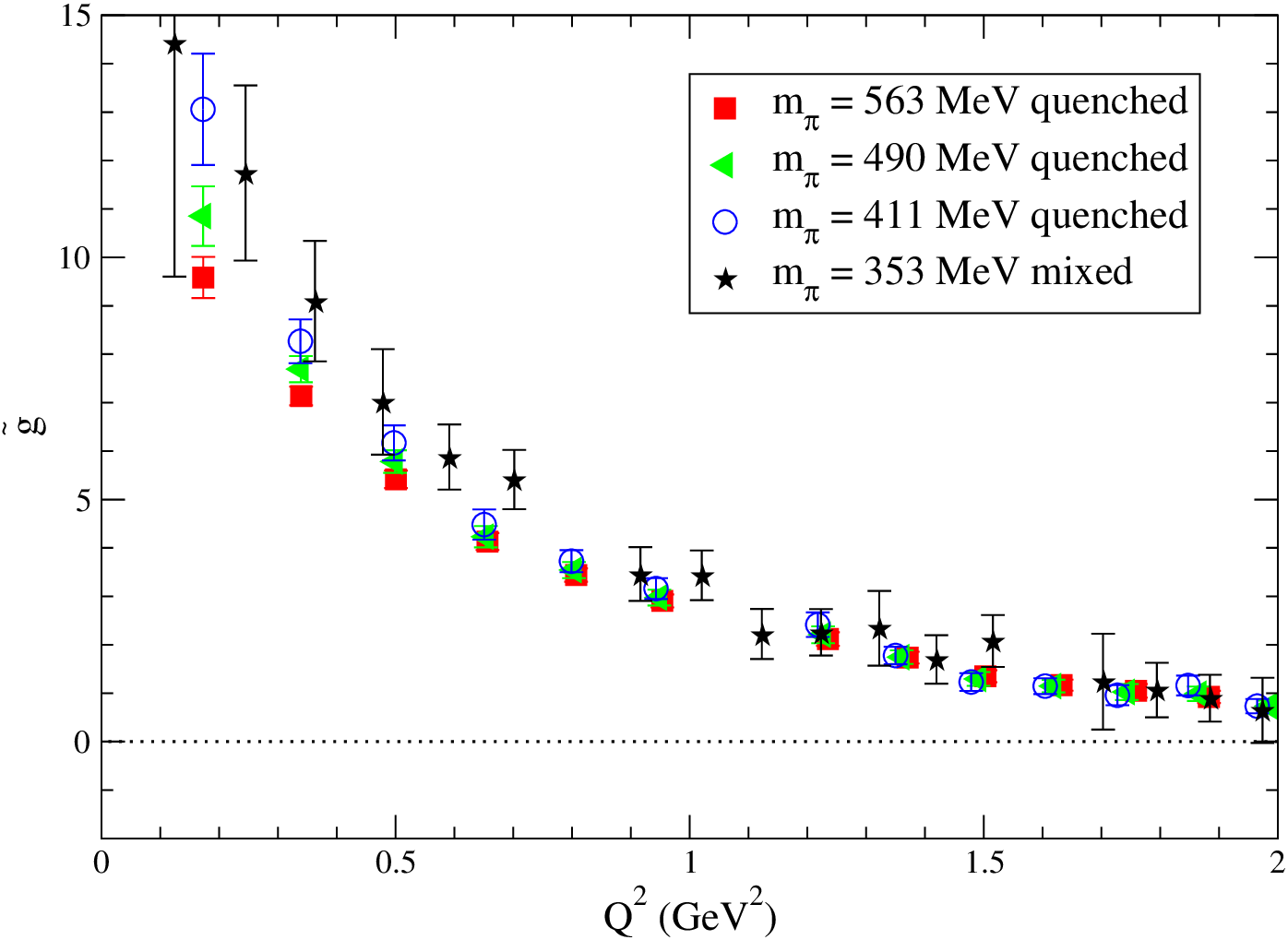}}
\end{center}
\caption{\label{g_PS_FFS_fig} Lattice QCD results for the $\tilde{g}$ pseudoscalar form-factor.}
\end{figure}

\begin{figure}
\begin{center}
\scalebox{0.4}{\includegraphics{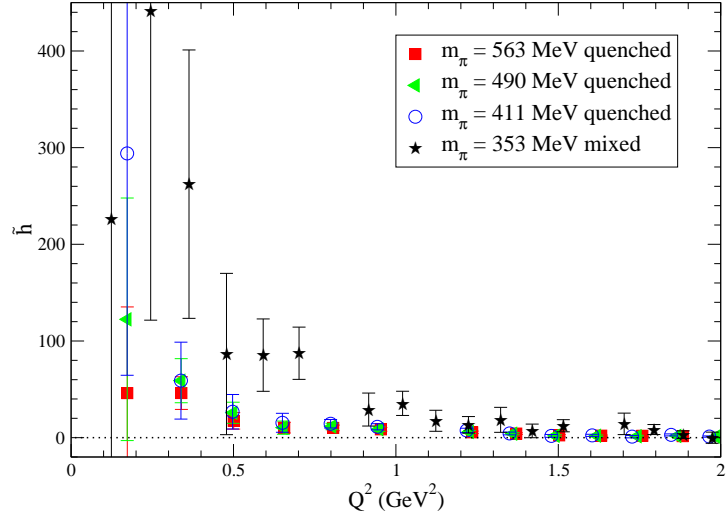}}
\end{center}
\caption{\label{h_PS_FFS_fig} Lattice QCD results for the $\tilde{h}$ pseudoscalar form-factor.}
\end{figure}

\begin{figure}
\begin{center}
\scalebox{0.4}{\includegraphics{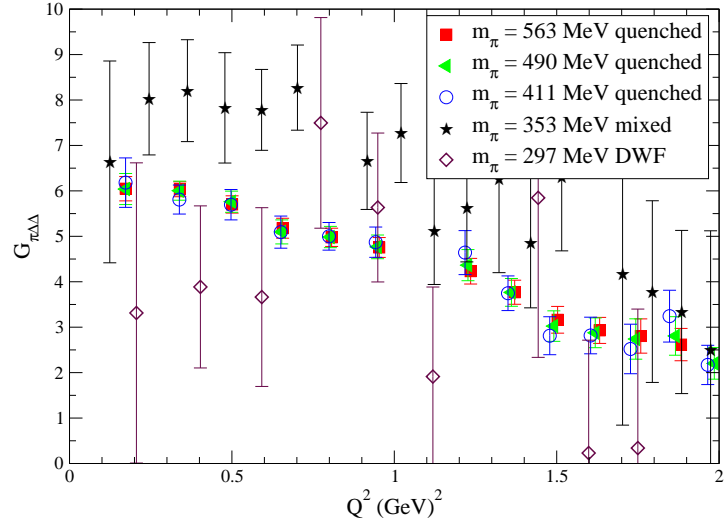}}
\end{center}
\caption{\label{GpDD_fig} Lattice QCD results for the primary $\pi\Delta\Delta$ coupling, $G_{\pi\Delta\Delta}$.}
\end{figure}

\begin{figure}
\begin{center}
\scalebox{0.4}{\includegraphics{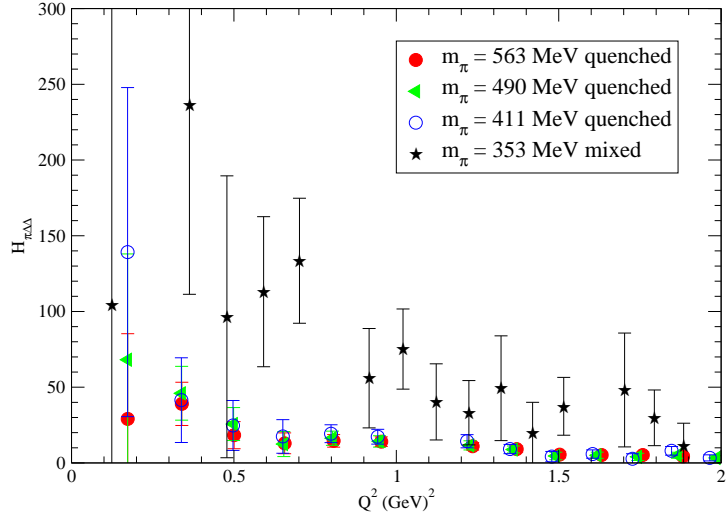}}
\end{center}
\caption{\label{HpDD_fig} Lattice QCD results for the secondary $\pi\Delta\Delta$ coupling, $H_{\pi\Delta\Delta}$.}
\end{figure}

Figs.~\ref{g_PS_FFS_fig} and \ref{h_PS_FFS_fig} show the pseudoscalar 
form factors $\tilde{g}$ and $\tilde{h}$, respectively, as defined in 
Eq.~(\ref{ps_decomp}), where the pion pole is explicitly written. The numerical  values of these form factors are provided in Appendix~C. 
 As confirmed by the numerical results,  $\tilde{g}$ is the dominant
pseudoscalar form factor showing a pion-pole dependence, whereas  the subdominant form factor $\tilde{h}$ shows a stronger $Q^2$-dependence consistent with 
a double pion-pole.
In section II we already defined the physically relevant pion-$\Delta$ coupling
$G_{\pi\Delta\Delta}(m_\pi^2)$ 
factoring out the pion-pole and fixing coefficients via PCAC through Eq.~\ref{gpdd}. 
$G_{\pi\Delta\Delta}(q^2)$  has a 
finite value at the origin, as can be seen in Fig.~\ref{GpDD_fig} where
 numerical results are depicted. 
 This value in fact defines the traditional strong coupling
$g_{\pi\Delta\Delta}$ of the pion to the $\Delta$ state via
\be
g_{\pi\Delta\Delta} = G_{\pi\Delta\Delta}(m_\pi^2) \, .
\ee
The secondary momentum-dependent coupling, $H_{\pi\Delta\Delta}(Q^2)$, is plotted
in Fig. \ref{HpDD_fig}. The numerical results are consisted with a pion-pole divergence
at small $Q^2$ as expected from the analysis given in the previous section. 
The
statistical error on this coupling is larger in particular at small $Q^2$
since in
the combined analysis  the pseudoscalar matrix element is multiplied by a factor of $Q^2$.

\begin{table}[h]
\begin{tabular}{cccc}
 \hline 
$\kappa$ or $am_l$&  $am_q$ & $af_\pi/Z_A$ &  $Z_A$  \\ 
\hline
\multicolumn{4}{c}{Quenched Wilson fermions} \\
\hline
0.1554 &  0.0403(4) & 0.0611(14) & 0.808(7)  \\
0.1558 &  0.0307(4) & 0.0587(16) & 0.808(7)  \\
0.1562 &  0.0213(4) & 0.0563(17) & 0.808(7)  \\
\hline
\multicolumn{4}{c}{Hybrid or mixed action} \\
\hline
0.02  &  0.0324(4) & 0.0648(8)   & 1.0994(4) \\
0.01  &  0.0159(2) & 0.0636(6)   & 1.0847(6) \\
\hline
\multicolumn{4}{c}{$N_F=2+1$ DWF} \\
\hline
0.004  &  0.004665(3) & 0.06575(12) & 0.74521(2) \\
\hline
\end{tabular}
\caption{The first column gives the hopping parameter $\kappa$ for
Wilson fermions or the lattice mass of the domain wall fermion, 
the second the renormalized quark mass, 
the third the unrenormalized pion decay constant $f_\pi/Z_A$ in lattice units, 
 and the fourth the axial current renormalization constant $Z_A$.}
\label{tableZ}
\end{table}

Notice that the extraction of $G_{\pi \Delta \Delta}$ and $H_{\pi \Delta \Delta}$ 
from Eqs.~(\ref{gpdd}) and (\ref{hpdd}) requires  knowledge of the light quark mass
$m_q$ and the pion decay constant, $f_{\pi}$, on each of the ensembles.
Calculation of $f_{\pi}$ requires the two-point functions of the axial-vector current $A_4^3$ with local-smeared (LS) 
and smeared-smeared (SS) quark sources,
\beq
C^{A}_{LS}(t) 
 = \sum_{{\bf x}}  \; \langle \Omega
|\;T\;\left( A_4^3 ({\bf x},t)
 \tilde{A}^3_4 ({\bf 0},0)\right) \; |  \Omega\;\rangle 
\eeq
(and similarly for $C^A_{SS}$), where $A_4^3 ({\bf x},t)$ denotes the local operator and $\tilde{A}^3_4 ({\bf x},t)$ the 
smeared operator. The pion decay constant
 $f_{\pi}$ is obtained from the pion-to-vacuum matrix element 
\be 
\langle 0|A_\mu^a(0)|\pi^b(p)\rangle = i f_\pi p_\mu \delta^{ab}
\label{pion decay}
\ee
extracted from the ratio of the two-point functions $C^A_{LS}$ and $C^A_{SS}$ and
\beq
f_\pi^{\rm eff}(t) = Z_A \sqrt{\frac{2}{m_\pi}}
\frac{C^{A}_{LS} (t)}{\sqrt{C^{A}_{SS}(t)}}
\; e^{m_\pi t/2} \,.
\label{fpi}
\eeq
in the large Euclidean time limit.

The renormalized quark mass $m_q$ is determined from AWI, via two-point functions of the pseudoscalar density with either
local ($P^3$) or smeared ($\tilde{P}^3$) quark fields,
\beq
C^{P}_{LS}(t) 
 = \sum_{{\bf x}}  \; \langle \Omega
|\;T\;\left( P^3 ({\bf x},t)
 \tilde{P}^3 ({\bf 0},0)\right) \; |  \Omega\;\rangle ~,
\eeq
(and similarly for $C^P_{SS}$).
The effective quark mass is defined by
\beq
m_{\rm eff}^{\rm AWI}(t) =\frac{m_\pi}{2}\frac{Z_A}{Z_P}
\frac{C^{A}_{LS} (t)}{C^{P}_{LS}(t)}
\sqrt{\frac{C^{P}_{SS} (t)}{C^{A}_{SS}(t)}} ~.
\label{meff}
\eeq
and its plateau value yields $m_q$.
Note that $Z_P$ will be needed only if ones wants $m_q$ alone. 
Since $Z_P$ enters also Eq.~(\ref{eff_couple_eq}), it  cancels --as does
 $Z_A$ since it comes with $f_{\pi}$--  and 
therefore $G_{\pi \Delta \Delta}$ and $H_{\pi \Delta \Delta}$ are  extracted directly 
from ratios of lattice three- and
two-point functions without prior knowledge of either $Z_A$ or $Z_P$. 
We also note that the quark mass computed 
through~(\ref{meff}) includes the
effects of residual chiral symmetry breaking from the finite extent $L_5$ 
of the fifth dimension. These effects are of the order of $17\%$ for the 
DWF ensemble and $15 \%$ for the hybrid  ensemble (also referred to as mixed scheme).
Chiral symmetry breaking affects the PCAC relations and 
therefore the value of both strong couplings $G_{\pi \Delta \Delta}$ and $H_{\pi \Delta \Delta}$ 
through Eq.~(\ref{eff_couple_eq}).

\subsection{Testing Pion-Pole Dominance in the Axial and Pseudoscalar Matrix Element}

\begin{figure}
\begin{center}
\scalebox{0.4}{\includegraphics{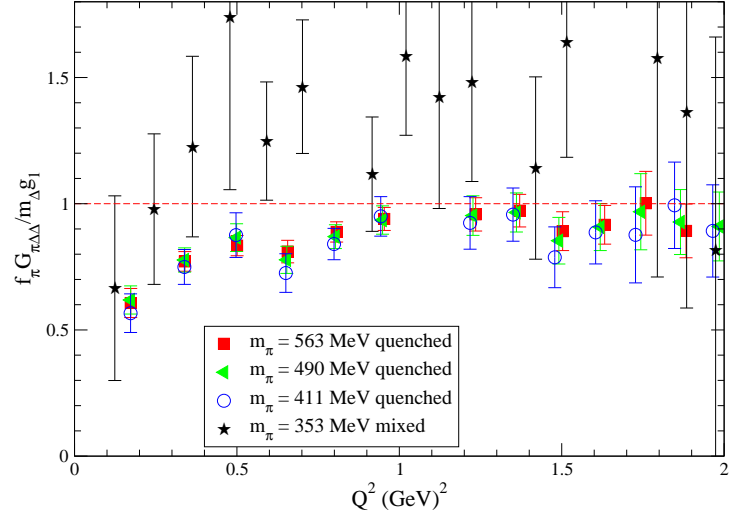}}
\end{center}
\caption{\label{GTR_GpDD_fig} Ratio test of the Goldberger-Treiman Relation for
$G_{\pi\Delta\Delta}$.}
\end{figure}

\begin{figure}
\begin{center}
\scalebox{0.4}{\includegraphics{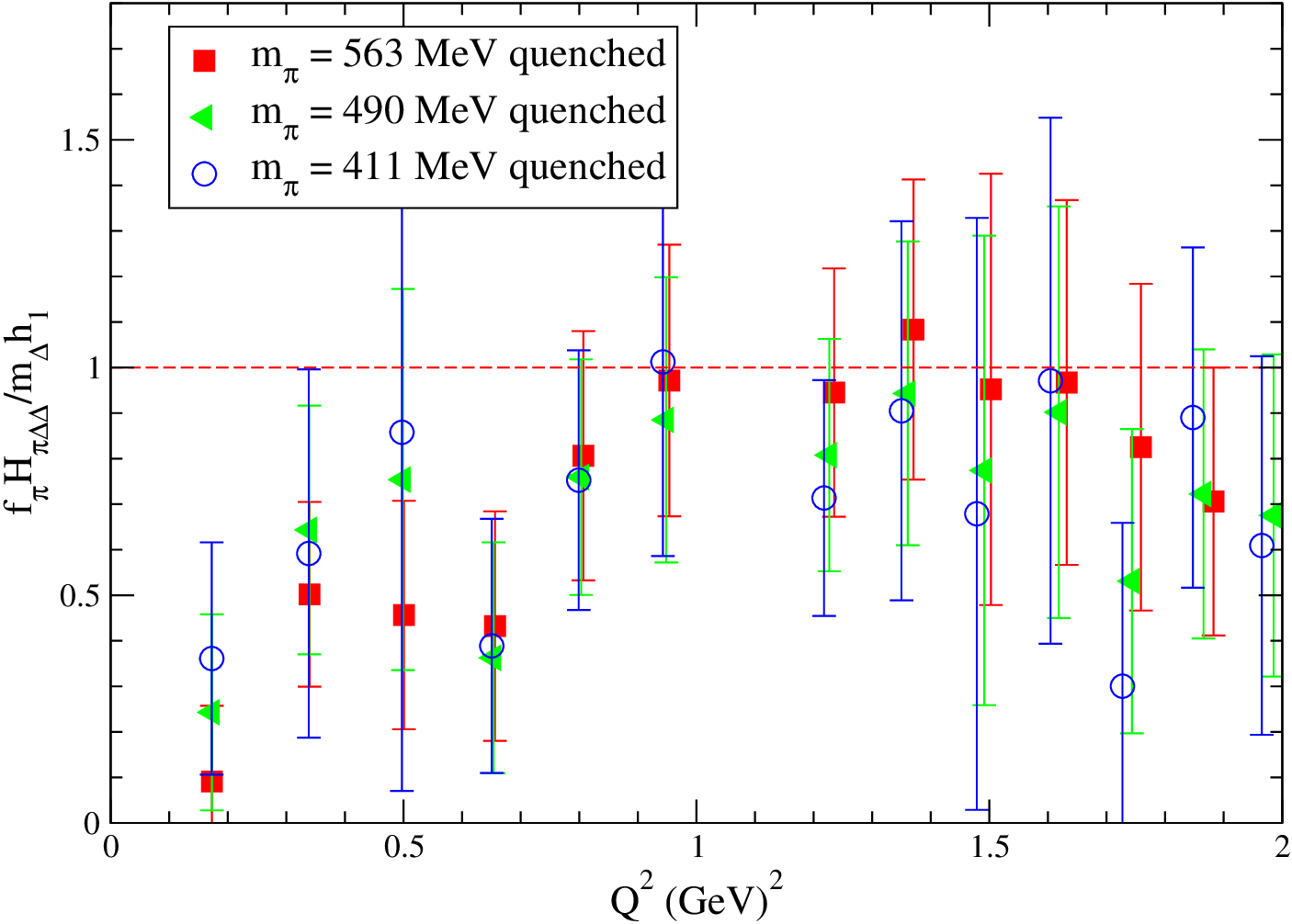}}
\end{center}
\caption{\label{GTR_HpDD_fig} Ratio test of the Goldberger-Treiman Relation for $H_{\pi\Delta\Delta}$.}
\end{figure}

In this section we  examine in detail the pion-pole dependence expected
for 
the $\Delta$ form factors by performing fits to the results
obtained.
First, we test the validity of the Goldberger-Treiman relations of Eqs.~(\ref{GT1 Delta}, \ref{GT2 Delta}) 
by evaluating the ratios 
\begin{equation}
\label{G-type_ratio}
\frac{f_\pi G_{\pi\Delta\Delta}(q^2)}{M_\Delta g_1(q^2)}
\end{equation}
and
\begin{equation}
\label{H-type_ratio}
\frac{f_\pi H_{\pi\Delta\Delta}(q^2)}{M_\Delta h_1(q^2)}.
\end{equation}
These relations are expected to hold at low $Q^2$.
We show the results in Figs.~\ref{GTR_GpDD_fig} and \ref{GTR_HpDD_fig}.
 The first ratio, given in Eq.~(\ref{G-type_ratio}), carries moderate 
statistical error. It is consistent with unity for $Q^2\stackrel{>}{\sim}0.8$ GeV$^2$ 
for the quenched ensembles while it is underestimated at smaller $Q^2$ values.
This discrepancy at smaller $Q^2$ can be 
attributed to chiral effects on $G_{\pi\Delta\Delta}$, which is expected to be
 more seriously affected by  pion cloud effects than $g_1$. The results using the hybrid dynamical ensemble, on the other
hand, are consistently higher  than unity for $Q^2>0.5$~GeV$^2$.
The large statistical errors carried by these data 
make it difficult to draw definite conclusions.
 The behavior of this ratio is very similar to  the
behavior shown by the corresponding ratio for the nucleon GT relation
as well as the nucleon-to-$\Delta$ axial transition~\cite{Alexandrou:2007xj}.

The second GT-type relation, given in Eq.~(\ref{H-type_ratio}),
is statistically consistent with unity for the quenched results
and $Q^2>0.8$~GeV$^2$. The results from the dynamical ensembles are
 plagued by too large statistical noise to be able to meaningfully
display  them on the plot.
We therefore have omitted these 
data from Fig.~\ref{GTR_HpDD_fig}. 
A very similar and consistent behavior with the first ratio is observed for the quenched data.
We remind the reader that it is the first 
Goldberger-Treiman relation that is more significant for phenomenology, 
as it is this relation  that connects the axial charge (from $g_1$ at $Q^2=0$)  
to the $G_{\pi\Delta\Delta}$ coupling.

\begin{figure}
\begin{center}
\scalebox{0.67}{\includegraphics{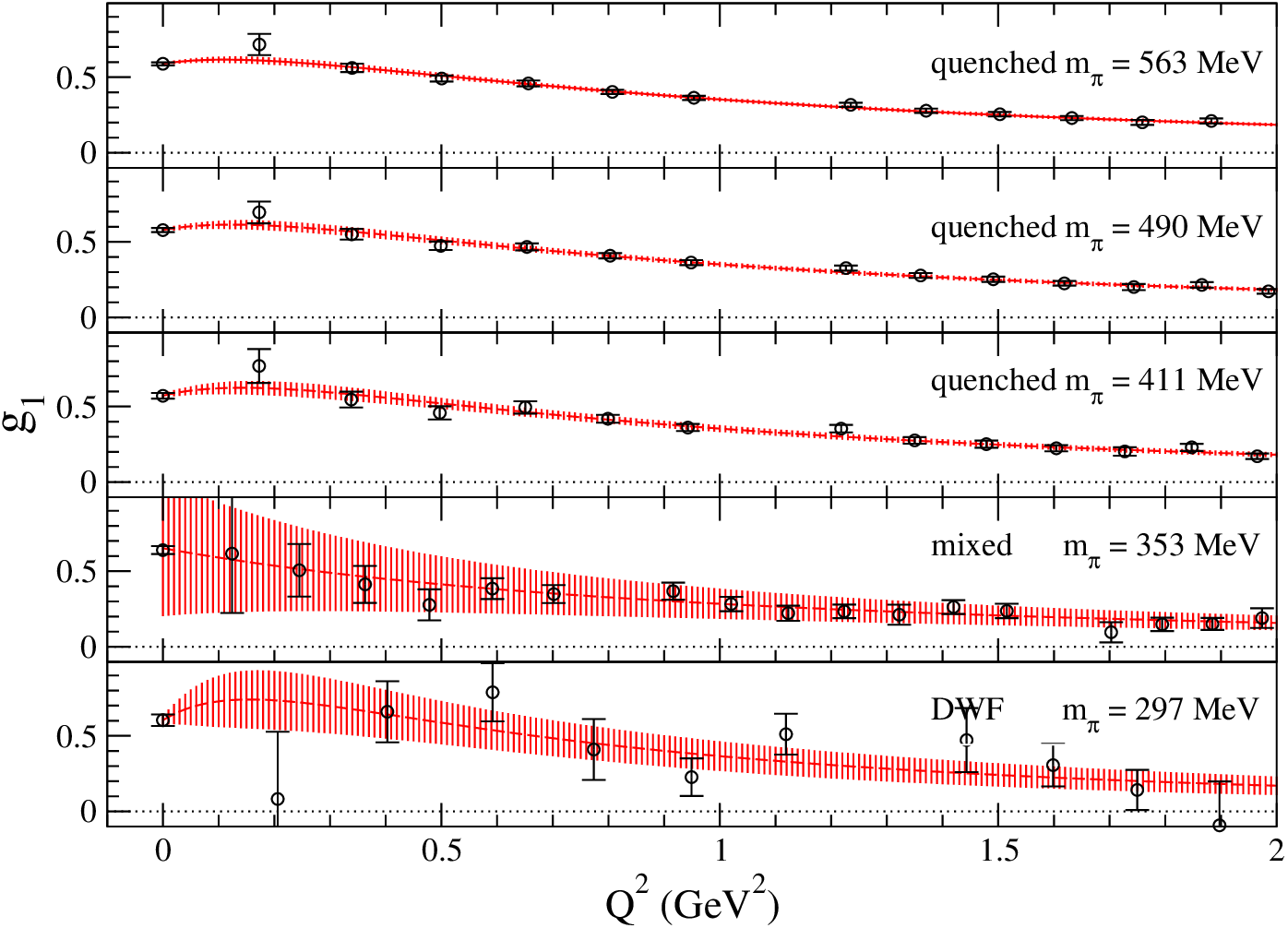}}
\end{center}
\caption{\label{g1_wtp_fits} Fits to the data for the $g_1$ form-factor  using the 
 form given in Eq. \ref{g1_fit_form}.}
\end{figure}

\begin{figure}
\begin{center}
\scalebox{0.67}{\includegraphics{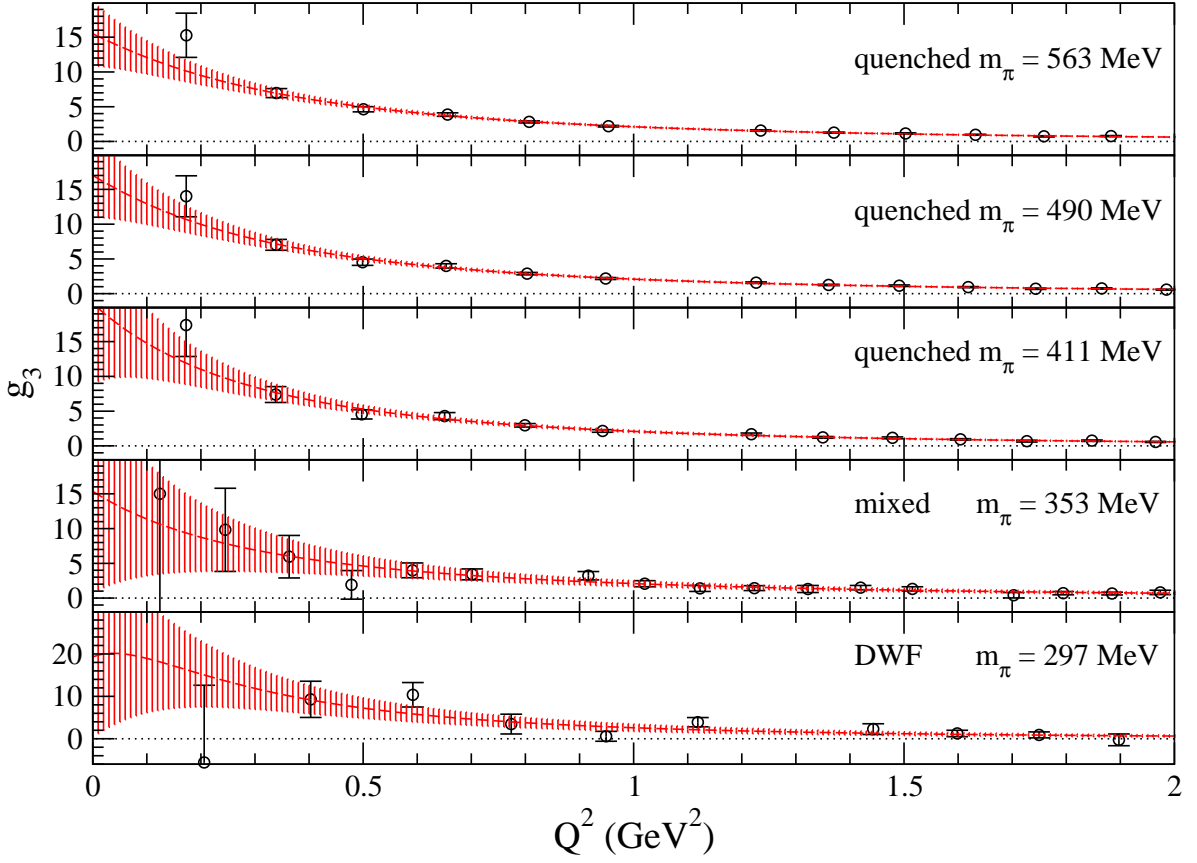}}
\end{center}
\caption{\label{g3_wtp_mono_fits} Fits to the data for the $g_3$ form-factor using the 
form given in Eq. \ref{g3_fit_form}.}
\end{figure}

\begin{figure}
\begin{center}
\scalebox{0.67}{\includegraphics{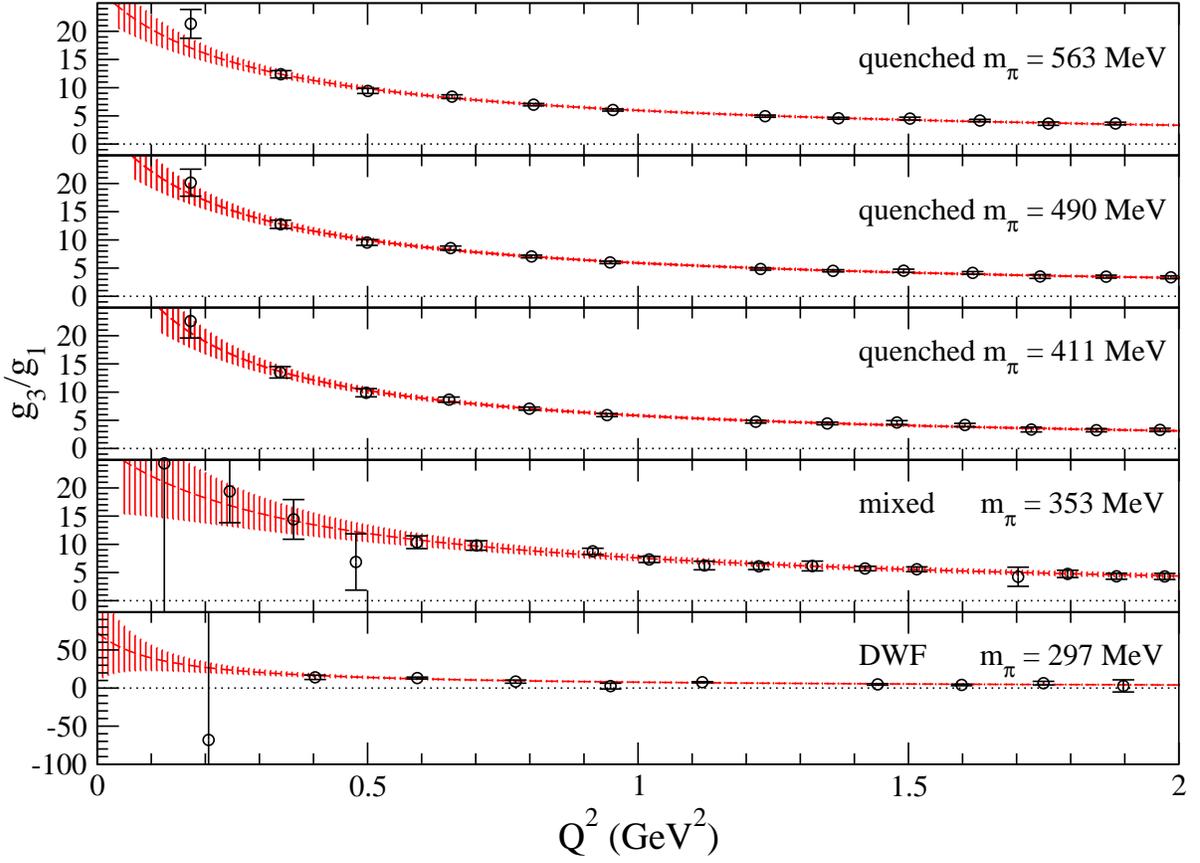}}
\end{center}
\caption{\label{g3g1_rat_mono_fits} Monopole fits as given by Eq. \ref{monopole_fit_form}
to the ratio $g_3/g_1$.}
\end{figure}

\begin{figure}
\begin{center}
\scalebox{0.67}{\includegraphics{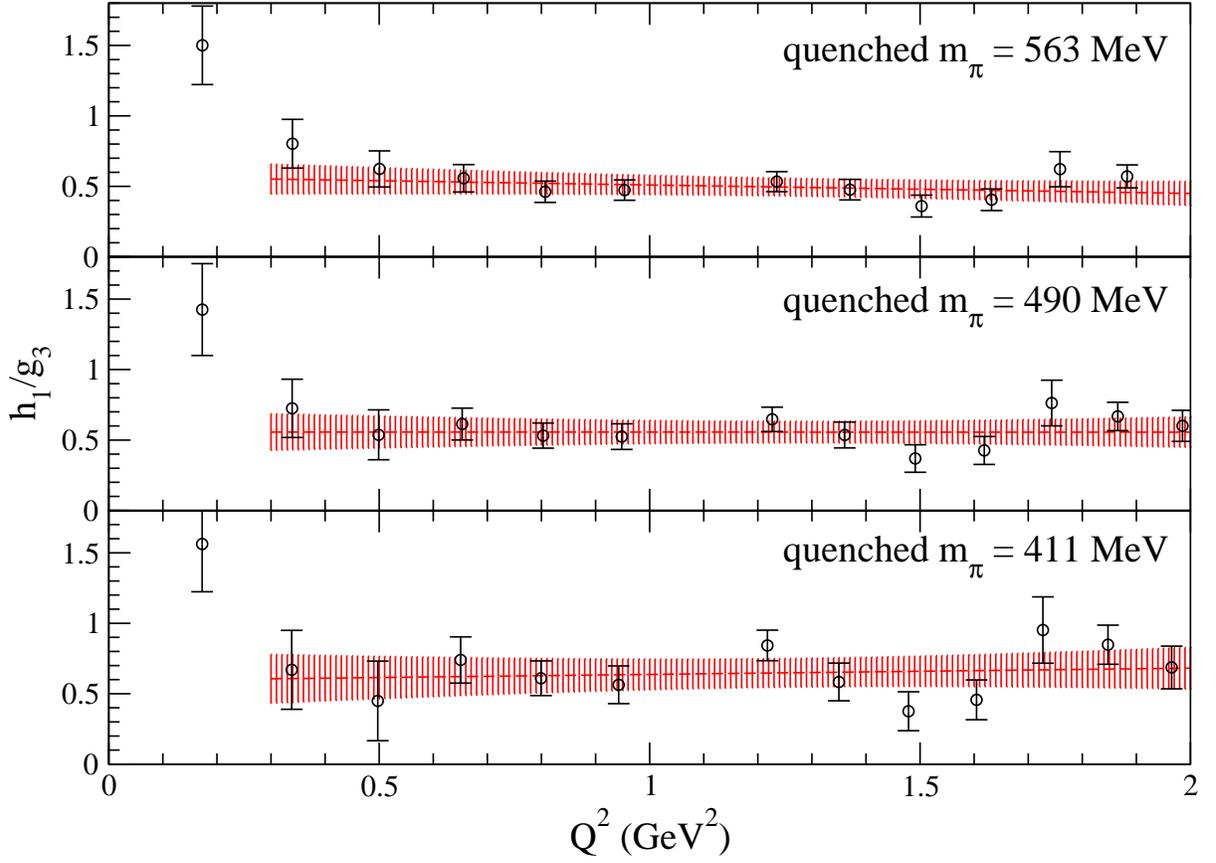}}
\end{center}
\caption{\label{h1g3_rat_fig}The ratio $h_1/g_3$ as a function of $Q^2$, with unity marked with a red line.}
\end{figure}

\begin{figure}
\begin{center}
\scalebox{0.67}{\includegraphics{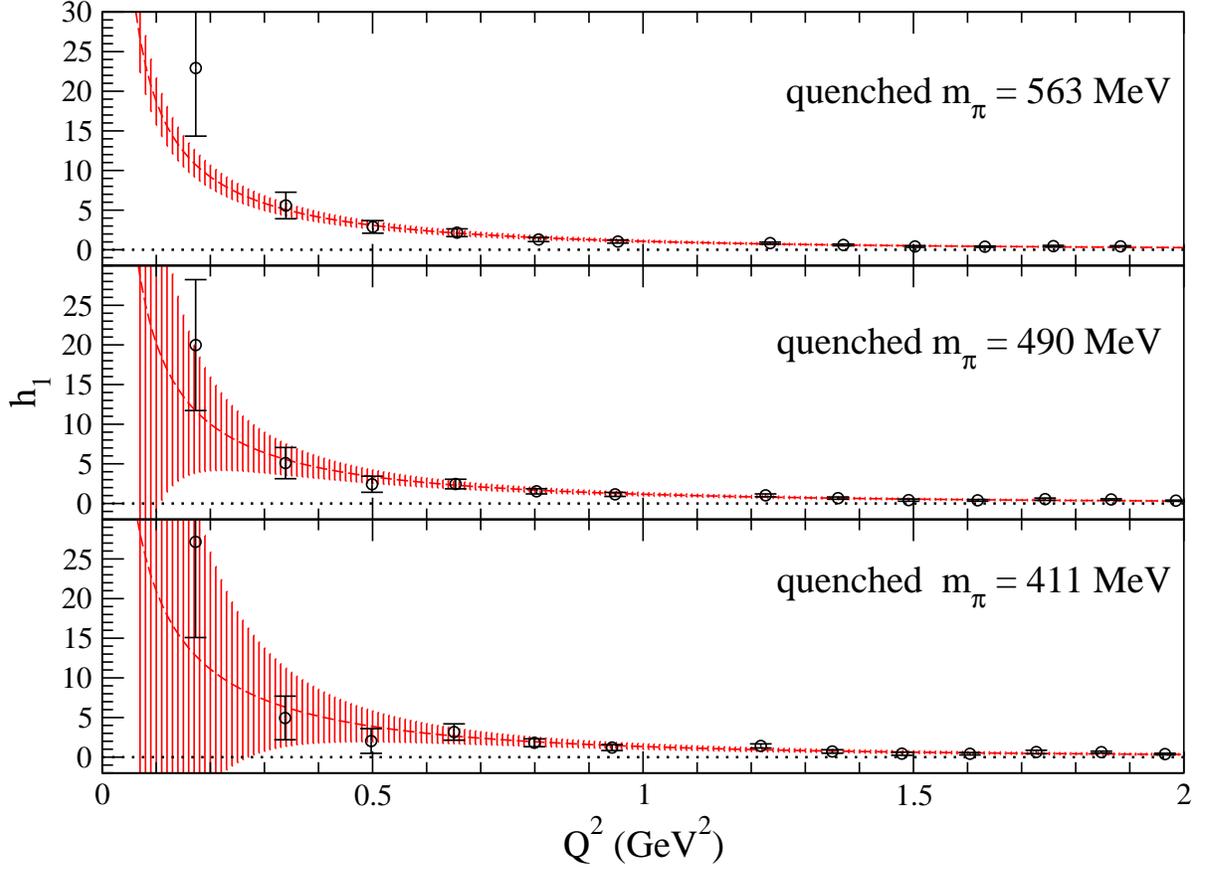}}
\end{center}
\caption{\label{h1_wtp_mono_fits} Fits to the data for the $h_1$ form-factor  using the 
form given in Eq. \ref{g3_fit_form}.}
\end{figure}

\begin{figure}
\begin{center}
\scalebox{0.67}{\includegraphics{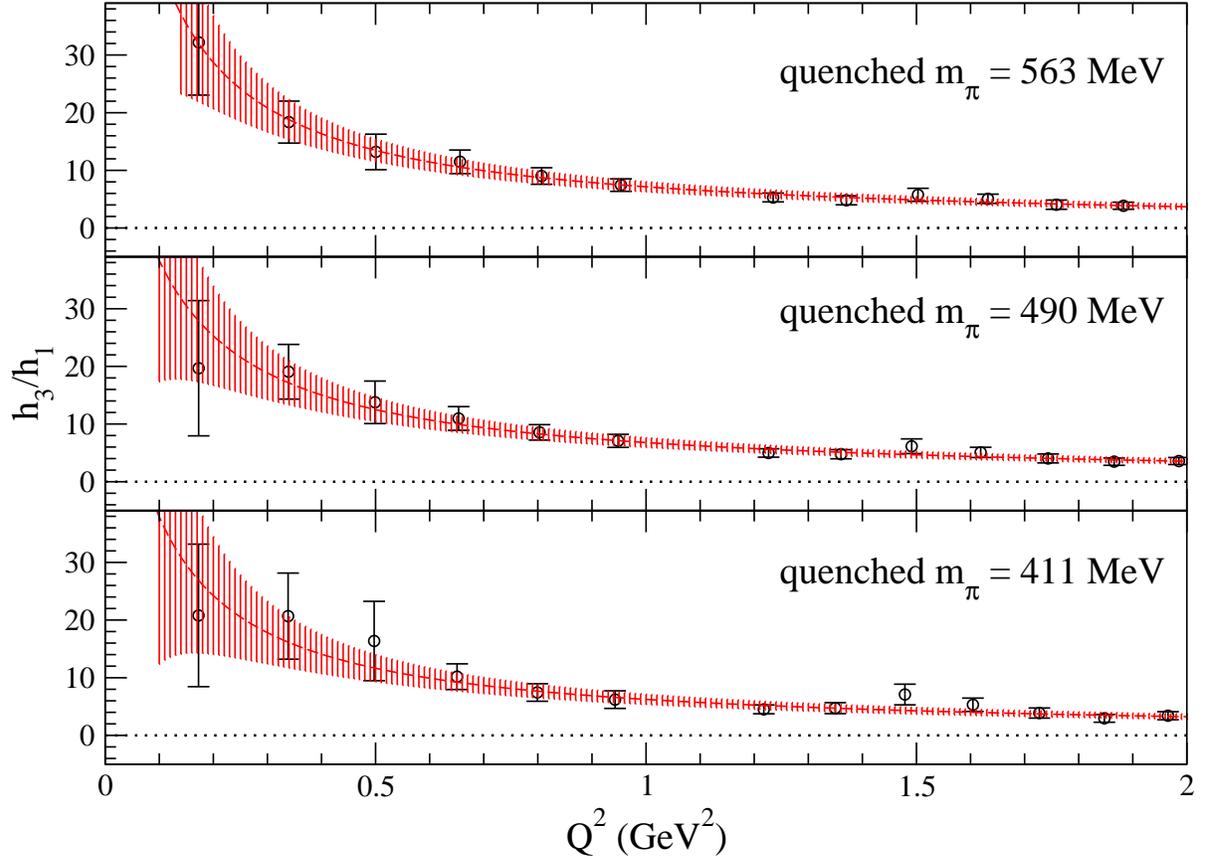}}
\end{center}
\caption{\label{h3h1_rat_mono_fits} Monopole fits to the ratio 
$h_3/h_1$ as described by Eq. \ref{monopole_fit_form}.}
\end{figure}

\begin{figure}
\begin{center}
\scalebox{0.67}{\includegraphics{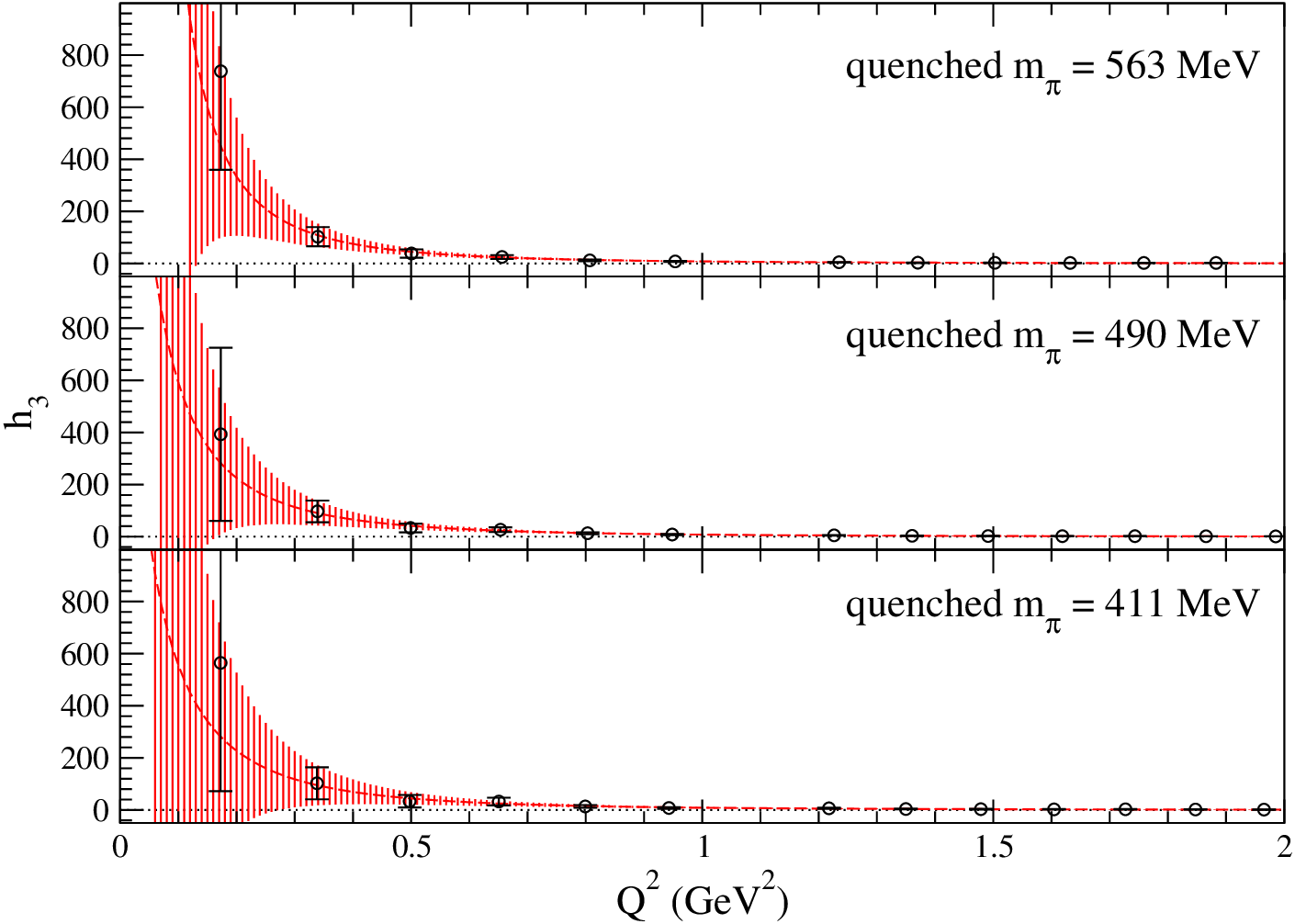}}
\end{center}
\caption{\label{h3_wtp_dip_fits} Fits to the data for the $h_3$ form-factor  using the 
 form given in Eq. \ref{h3_fit_form}.}
\end{figure}


To further probe the pion pole assumptions entering into our derivation of the 
GT relations we perform a set of fits to our form factor data. We have no
{\em a priori} theoretical expectation for the functional form of $g_1(q^2)$,
although typically a dipole form seems to accommodate well the nucleon axial form factor $G_A$
as well as the leading axial $N-\Delta$ transition form factor $C^A_5$.    
We note however that there seems to be a  small dip in the $g_1$ at $q^2=0$ 
for the quenched ensembles. To accommodate this we fit the data to:
\beq
\label{g1_fit_form}
g_1(Q^2) = 
\frac{a+ bQ^2}{\left(Q^2+ m_1^2\right)^3}
\eeq
The resulting fits are shown in Fig. \ref{g1_wtp_fits}. The values for the fitted parameters 
are given in Table~\ref{g1_fit_params}. We note that the mass parameter $m_1$ 
determining  the slope as $Q^2\rightarrow 0$  is around 1~GeV, a scale typical for  axial dipole 
masses
controlling the dependence of nucleon $G_A$ and the dominant axial $N-\Delta$ $C^A_5$ form factor.

We consider the form
\beq
\label{g3_fit_form}
\left[
\frac{a+ bQ^2}{\left(Q^2 + m_1^2\right)^3}
\right]
\frac{c}{\left(Q^2 + m_2^2\right)}
\eeq
for $g_3$ based on the pion pole dominance prediction given in Eq.(\ref{gr1}). 
The parameters $a$, $b$, and $m_1$ are fixed to the values arising from the
fit of the $g_1$ data using the Ansatz given in Eq. (\ref{g1_fit_form}). The fits are shown in 
Fig.~\ref{g3_wtp_mono_fits}. The fitted parameters are given in Table~\ref{g3_fit_params}. 
Note that the value of $m_2$ is considerably smaller compared to $m_1$, as in fact is
expected since this is detected from the presence of the pion-pole. 
This is especially verified by the quenched data
where $m_2$ is close to the actual pion mass $m_\pi$ of the ensemble.

Pion pole dominance fixes completely the ratio $ g_3/g_1$  
\be
\frac{g_3}{g_1} = \frac{4 M_\Delta^2 }{m_\pi^2-q^2} ~.
\label{monoform}
\ee
We form the ratio $g_3/g_1$ from our data and fit separately to a monopole form:
\beq
\label{monopole_fit_form}
\frac{c}{\left(Q^2 + m_2^2\right)}~.
\eeq
This fit is displayed in Fig.~\ref{g3g1_rat_mono_fits}. 
Using a ratio eliminates any need to know the theoretical form for 
$g_1(q^2)$ alone. The fitted parameters $c$ and $m_2$ are given in  
Table~\ref{g3g1_fit_params}. The verification of the predicted 
form given in Eq.~(\ref{monoform}) is very good, with the pole mass $m_2$ consistent with the 
pion  mass and the constant $c$ reasonably close to $4 M_\Delta^2$.

The form factor $h_1$ is similar to $g_3$ having a pion-pole dependence. We  
display the ratio $h_1/g_3$ in Fig. \ref{h1g3_rat_fig} for the quenched QCD ensembles.
This ratio is notably constant over the whole $Q^2$ range above 0.4 GeV$^2$, with the
constant $\sim 0.5$.  
Based on this observation, we use the Ansatz given in Eq. (\ref{g3_fit_form}) also for
$h_1$. The fit is shown in Fig.~\ref{h1_wtp_mono_fits} and the fitted parameters 
are given in Table~\ref{h1_fit_params}. Again, $m_2$ is considerably smaller 
compared to $m_1$, in accordance to the presence of a light (pion) mode.

From Eq.~(\ref{gr2}), the ratio $h_3/h_1$ is completely fixed:
\be
\frac{h_3}{h_1} = \frac{4 M_\Delta^2 }{m_\pi^2-q^2} ~.
\label{monoform2}
\ee
We plot this ratio in Fig.~\ref{h3h1_rat_mono_fits}. 
Fitting the data to the monopole form of Eq.~(\ref{monopole_fit_form}), we get parameters 
$m_2$ and $c$ within the range of the expected value~(Eq.~\ref{monoform2}) 
-- see Table~\ref{h3h1_fit_params} indicating that
 the subdominant form factor diverges with a double pion-pole- 
dependence.  

In Fig.~\ref{h3_wtp_dip_fits} we present the fit of $h_3$ to the Ansatz
\beq
\label{h3_fit_form}
\left[
\frac{a+ bQ^2}{\left(Q^2 + m_1^2\right)^3}
\right]
\frac{d}{\left(Q^2 + m_2^2 \right)^2},
\eeq
with $a$, $b$, and $m_1$ fixed to the values extracted from the fit of
$g_1$. The fitted  parameters  $d$ and $m_2$ are given in 
Table~\ref{h3_fit_params}, in accordance to the $h_3/h_1$ fit (Table~\ref{h3h1_fit_params}).

In the pseudoscalar sector, one expects a monopole dependence also for the ratio
$\tilde{h}/\tilde{g}$.
Fitting the data to the monopole form of Eq.~(\ref{monopole_fit_form}), we get the parameters
provided in Table~\ref{hg_PS_fit_params}. Indeed, an agreement of the $m_2$ pole mass to 
the pion mass
within the allowed by statistical noise regime is seen.

The overall conclusion from the fits in this section is that all form factors satisfy
qualitatively the pion-pole dependence predicted by PCAC. 
This is most clearly exemplified
in the case of quenched QCD where the level of statistical noise allows such detailed analysis.
In all cases the data fit these forms to good confidence levels, i.e., 
$\chi^2/{\rm dof} \lsim 1$.
Enhanced statistical noise for the dynamical ensembles limits the verification 
to the dominant form factors only, as the subdominant ones are beyond reach, but this still
is a useful result as it shows the consistency between quenched and dynamical
results. This  corroborates other baryon studies that show 
small effects due to dynamical quark for pion masses larger than about 300~MeV.

\begin{table}[H]
\begin{tabular}{lcccc}
\hline\hline 
 $m_\pi$ (GeV) & $a$ & $b$ & $m_1$  (GeV)&$\chi^2/{\rm dof}$\\
\hline
\multicolumn{5}{c}{quenched Wilson fermions}\\
0.563 & 0.53(18) & 2.15(31) & 0.98(5) & 0.82\\
0.490 & 0.47(18) & 2.08(33) & 0.99(6) & 1.07\\
0.411 & 0.40(19) & 1.98(38) & 0.94(8) & 1.48\\
\multicolumn{5}{c}{mixed action} \\
0.353 & 3.0(22.0) &2.4(1.9) & 1.3(1.2) & 0.44\\ 
\multicolumn{5}{c}{domain wall fermions} \\
0.297 & 0.19(24) & 1.5(9) & 0.82(18) & 1.1\\
\hline
\end{tabular}
\caption{Fit parameters for $g_1(Q^2)$  using Eq. \ref{g1_fit_form}.}
\label{g1_fit_params}
\end{table}

\begin{table}[H]
\small
\begin{tabular}{lccc}
\hline\hline 
 $m_\pi$ (GeV) & $c$ &$m_2$  (GeV)& $\chi^2/{\rm dof}$\\
\hline\hline
\multicolumn{4}{c}{quenched Wilson fermions}\\
0.563 & 7.77(88) & 0.54(11) & 0.61 \\
0.490 & 7.45(89) & 0.50(11) & 0.55 \\
0.411 & 7.1(1.0) & 0.44(15) & 0.65 \\
\multicolumn{4}{c}{mixed action} \\
0.353 & 10.7(4.6) &  0.67(43) & 0.49 \\
\multicolumn{4}{c}{domain wall fermions} \\
0.297 & 10.0(5.3) & 0.56(37) & 1.26\\
\end{tabular}
\caption{Fit parameters for $g_3(Q^2)$ using Eq. \ref{g3_fit_form}. Parameters $a$, $b$
and $m_1$ are fixed with the results of $g_1$ fits in Table \ref{g1_fit_params}.
\label{g3_fit_params}}
\end{table}

\begin{table}[H]
\small
\begin{tabular}{lcccc}
\hline\hline 
 $m_\pi$ (GeV) & $m_2$ (GeV) & $c$ & $4M_{\Delta}^2$ (GeV$^2$)  &$\chi^2/{\rm dof}$\\
\hline\hline
\multicolumn{5}{c}{quenched Wilson fermions}\\
0.563 &  0.523(64) &  7.60(52) &  8.64(18)  & 0.67 \\
0.490 &  0.477(63) & 7.25(49) &  8.12(18)  & 0.54 \\
0.411 &  0.396(75) & 6.76(48) &  7.64(21)  & 0.60 \\
\multicolumn{5}{c}{mixed action} \\
0.353 & 0.61(18) & 10.4(1.6) &  9.40(33) &  0.34 \\
\multicolumn{5}{c}{domain wall fermions}    \\
0.297 & 0.34(17) & 8.6(1.3)&   5.82(20)  & 0.66\\
\end{tabular}
\caption{Fit parameters for $g_3(Q^2)/g_1(Q^2)$ using the monopole form of 
Eq. \ref{monopole_fit_form}.
\label{g3g1_fit_params}}
\end{table}

\begin{table}[H]
\small
\begin{tabular}{lcccc}
\hline\hline 
 $m_\pi$ (GeV) & $c$ &$m_2$ (GeV)& $\chi^2/{\rm dof}$\\
\hline\hline
\multicolumn{4}{c}{quenched Wilson fermions}\\
0.563 & 3.04(48) & $2\times 10^{-8}(8\times 10^{-7})$ & 1.32\\
0.490 & 3.32(84) & 0.03(81) & 1.37\\
0.411 & 3.8(1.8) & 0.1(1.4) & 1.52 \\
\end{tabular}
\caption{Fit parameters for $h_1(Q^2)$ using Eq. \ref{g3_fit_form}. Parameters $a$, $b$
and $m_1$ are fixed with the results of $g_1$ fits in Table \ref{g1_fit_params}.
\label{h1_fit_params}}
\end{table}

\begin{table}[H]
\small
\begin{tabular}{lccccc}
\hline\hline 
 $m_\pi$ (GeV) & $m_2$ (GeV) & $c$ & $4M_{\Delta}^2$ (GeV$^4$)  &$\chi^2/{\rm dof}$\\
\hline\hline
\multicolumn{5}{c}{quenched Wilson fermions}\\
0.563 & 0.26(17) & 7.6(1.1) &  8.64(18) & 0.21 \\
0.490 & 0.31(19) & 7.4(1.1) &  8.12(18) & 0.38 \\
0.411 & 0.28(25) & 6.7(1.1) &  7.64(21) & 0.58 \\
\end{tabular}
\caption{Fit parameters for $h_3(Q^2)/h_1(Q^2)$ using the monopole form of 
Eq. \ref{monopole_fit_form}.
\label{h3h1_fit_params}}
\end{table}

\begin{table}[H]
\small
\begin{tabular}{lcccc}
\hline\hline 
 $m_\pi$ (GeV) & $d$ &$m_2$ (GeV)& $\chi^2/{\rm dof}$\\
\hline\hline
\multicolumn{4}{c}{quenched Wilson fermions}\\
0.563 & 22(6)  & 0.03(58) & 0.46\\
0.490 & 26(10) & 0.25(28) & 0.35\\
0.411 & 28(15) & 0.28(43) & 0.33\\
\end{tabular}
\caption{Fit parameters for $h_3(Q^2)$ using Eq. \ref{h3_fit_form}. Parameters $a$, $b$
and $m_1$ are fixed with the results of $g_1$ fits in Table \ref{g1_fit_params}.
The dynamical fermion data sets contain too much noise for the fits to be useful.
\label{h3_fit_params}}
\end{table}



\begin{table}[H]
\small
\begin{tabular}{lcccc}
\hline\hline 
 $m_\pi$ (GeV) & $m_2$ (GeV) & $c$ &$\chi^2/{\rm dof}$\\
\hline\hline
\multicolumn{4}{c}{quenched Wilson fermions}\\
0.563 & 0.73(39) & 4.2(1.6) & 0.23 \\
0.490 & 0.42(43) & 3.3(1.2) & 0.22 \\
0.411 & 0.45(70) & 3.9(2.0) & 0.32 \\
\end{tabular}
\caption{Fit parameters for $\tilde{h}(Q^2)/\tilde{g}(Q^2)$ using the monopole form of Eq. \ref{monopole_fit_form}.
\label{hg_PS_fit_params}}
\end{table}

\clearpage


\section{Phenomenological Couplings of the $\Delta$ and Combined Chiral Fit}

 Crucial parameters
in  heavy baryon chiral effective theories (HB$\chi$PT) with explicit $\Delta$ degrees of freedom are the axial
couplings of the nucleon, $g_A$, the axial $N-\Delta$ transition coupling, 
$c_A$, and the axial charge of the $\Delta$, $g_{\Delta \Delta}$.
Assuming PCAC, these can be related via GT relations to the effective 
$\pi NN$, $\pi N\Delta$ and $\pi \Delta\Delta$ strong couplings:
\be
g_A = \frac{f_\pi}{M_N}g_{\pi NN}\;\;\;,\;\;\;
c_A = \frac{f_\pi}{M_N}g_{\pi N \Delta}\;\;\;,\;\;\;
g_{\Delta \Delta} = \frac{f_\pi}{M_\Delta}g_{\pi \Delta \Delta}
\label{gteqs3}
\ee
We note that alternative notation and normalization factors exist in the literature 
in the definition of the
effective strong couplings for $\pi N \Delta$ and $\pi \Delta \Delta$. 
In addition, note that in such schemes Eqs.~(\ref{gteqs3}) are 
actually {\it defining} relations for the strong couplings.
$g_A$ is very well known experimentally and a variety of lattice and
theoretical calculations offer precise estimates. $c_A$  is much less-well 
determined, via the parity violating N-to-$\Delta$ amplitude which connects it to the dominant axial transition form factor $C_5^A(q^2)$.  
$g_{\Delta \Delta}$
remains 
undetermined from experiment and is typically treated 
--as is also the case for $c_A$-- as a fit parameter to be determined 
from fits to experimental or lattice data. 

\begin{figure}[h!]
\begin{center}
\includegraphics[width=0.8\linewidth,height=1\linewidth]{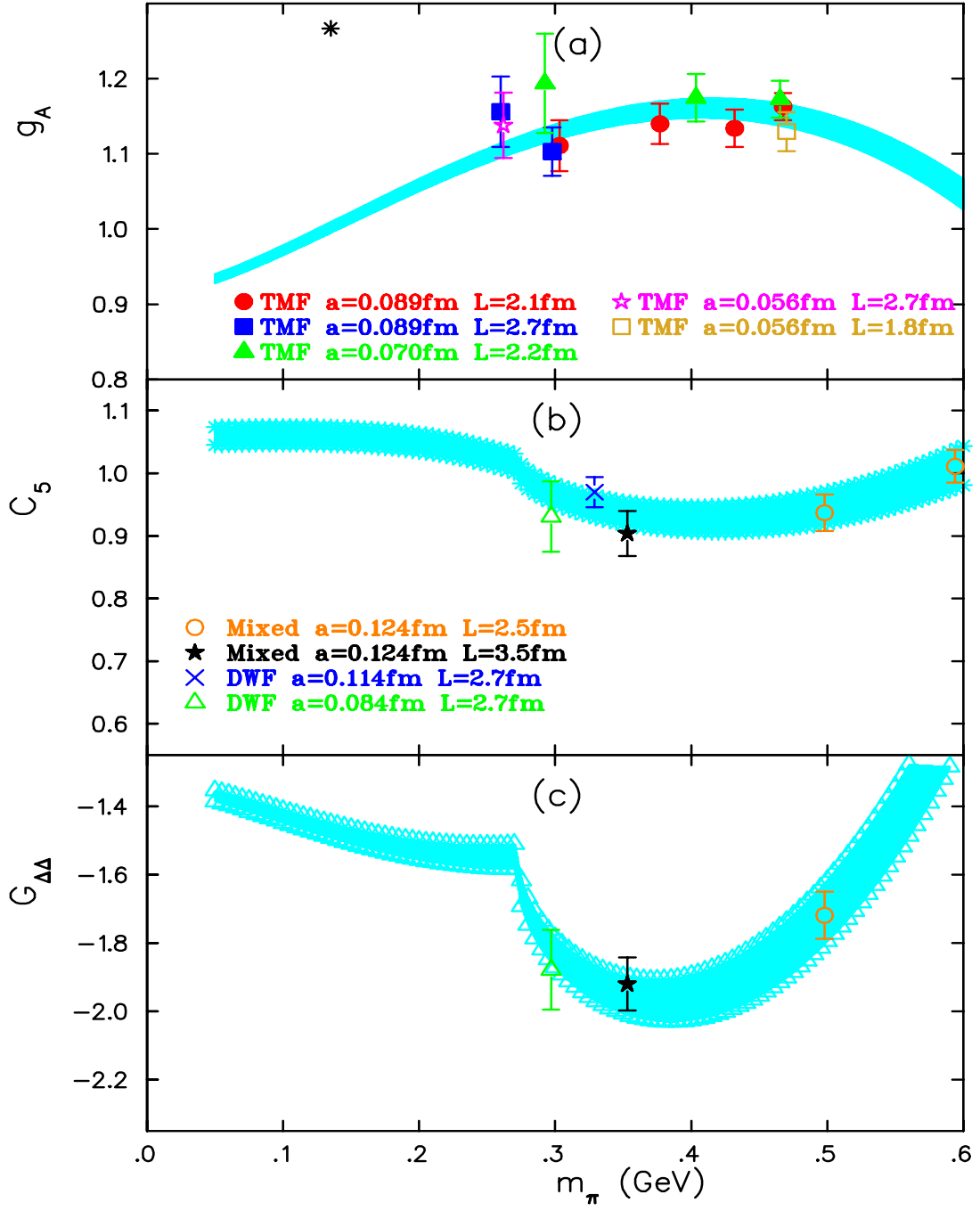}
\end{center}
\caption{\label{ax_chg_fit_fig} Combined chiral fit: (a) Nucleon axial charge, $g_A$, fitted to
 lattice data obtained with $N_f=2$ twisted mass fermions 
(TMF)~\cite{Alexandrou:2010hf}. The physical value is shown by the asterisk;
(filled circles: a=0.089~fm, L=2.1, filled squares: a=0.089~fm, L=2.8~fm, 
filled triangles: a=0.070~fm, L=2.2~fm, open square: a=0.056~fm, L=1.8~fm, 
star: a=0.056~fm, L=2.7~fm); 
(b) Real part of axial N to $\Delta$ transition coupling 
$C^A_5(0)$~\cite{Alexandrou:2010uk};
(open circles: a=0.124~fm, L=2.5~fm, filled square: a=0.124~fm, L=3.5~fm) and
 dynamical domain wall fermions (cross: a=0.114~fm, L=2.7~fm, open triangle: 
a=0.084~fm, 2.7~fm); 
(c) Real part of $\Delta$ axial charge $G_{\Delta \Delta}=-3g_1(0)$.} 
\label{chifit}
\end{figure}

There have been several sum-rules calculations of the effective 
$\pi\Delta\Delta$ coupling~\cite{Belyaev:1984ib,Zhu:2000zd,erkol_thesis}. 
In Ref.~\cite{Jido:1999hd} symmetry arguments in a quartet scheme 
where $N^*_+$, $N^*_-$, $\Delta_{+}$ and $\Delta_{-}$ form a chiral multiplet, 
lead to the
conclusion
 that $\pi\Delta_{\pm}\Delta_{\pm}$ 
couplings (with like-charged $\Delta$s) are forbidden at tree-level. 
Quark-model arguments~\cite{Brown:1975di} suggest that the 
$g_{\pi\Delta\Delta}=(4/5)g_{\pi NN}$.

\begin{table}[t]
\small
\begin{center}
\begin{tabular}{cc}
\hline
\hline
$m_\pi$ (GeV)& $g_1(0)$\\
\hline
\multicolumn{2}{c}{Quenched}\\
0.563(4) & 0.589(10)\\
0.490(4) & 0.578(13)\\
0.411(4) & 0.571(18) \\
\hline
\multicolumn{2}{c}{Mixed action}\\
0.498(3) & 0.573(23)\\
0.353(2) & 0.640(26)\\
\hline
\multicolumn{2}{c}{DWF}\\
0.297(5) & 0.604(38)\\
\hline
\hline
\end{tabular}
\end{center}
\caption{Numerical values for the dominant form factor $g_1(0)$ on each of the ensembles. $G_{\D \D}=-3 g_1(0)$ with our normalization. }
\label{g1_0_vals}
\end{table}

Lattice calculations for the nucleon axial charge $g_A$ are available on 
a variety of ensembles and pion masses~\cite{Alexandrou:2010hf}.
In addition, results on the axial $N-\Delta$
transition from factor $C^A_5$~\cite{Alexandrou:2010uk} have been obtained 
on most of the ensembles used also in this work. 
We are therefore in position to perform a {\it combined chiral fit} using 
small scale expansion (SSE) within (HB$\chi$PT) 
~\cite{Hemmert:2003cb,Procura:2008ze,Jiang:2008we}.
for  $g_A$, $C_5^A(q^2)$ and the $\Delta$ axial 
charge $G_{\Delta\Delta}$ as functions of the pion mass $m_\pi$.

The one-loop SSE expression for $C_5^A$ has been worked by 
Procura \cite{Procura:2008ze}. The expression for $C_5^A(q^2)$ 
as a function of $m_\pi$ is:
\begin{equation}
C^A_5=a_1+a_2\,m_\pi^2+a_3\,q^2 + {\rm loop}_{5}(m_\pi),
\label{c5sse}
\end{equation}
where the loop integral contribution is:
\begin{eqnarray}
{\rm loop}_5(m_\pi)&=&\frac{c_A}{15552 \,\pi^2 f_\pi^2}
\left\{\frac{1}{\Delta}\,\Big
[\frac{5}{4} g_{\Delta\Delta}^2\,(40 \pi\, m_\pi^3 +101 \Delta\, m_\pi^2+ 24 \Delta^3)+\frac{1170}{2} \,g_A\, g_{\Delta\Delta} m_\pi^2\, \Delta \right. \nonumber \\ 
&& \left.-\,12 \Delta\, c_A^2(162\, m_\pi^2-83 \Delta^2)-27\, g_A^2\,(24 \pi\, 
m_\pi^3+75 \Delta\, m_\pi^2 - 40 \Delta^
3)\Big] \nonumber \right. \\ 
&& \left. +\,\frac{72}{\Delta} \sqrt{m_\pi^2-\Delta
^2}\, \Big(m_\pi^2\, c_A^2 -28 \Delta^2\, c_A^2+18 g_A^2\, (m_\pi^2 - \Delta^2)
\Big) \arccos{\left(-\frac{\Delta}{m_\pi} \right)} \nonumber \right. \\ 
&& \left.
 -\,\frac{8}{\Delta} \sqrt{m_\pi^2-\Delta^2}\,\Big(9 \,m_\pi^2\, c_A^2+963\, 
\Delta^2 \,c_A^2+\frac{50}{4} g_{\Delta\Delta}^2\,(m_\pi^2-\Delta^2)\Big) 
\arccos{\left(\frac{\Delta}{m_\pi} \right)} \nonumber \right. \\ 
&& \left. - \,\Big[3\, m_\pi^2\, (900\, c_A^2
-\frac{425}{4} g_{\Delta\Delta}^2-\frac{450}{2} g_{\Delta\Delta}\, g_A +81\, g_A^2 +648)\right. \nonumber \\ 
&& \left.+\,8\Delta^2(-711 \,c_A^2+\frac{50}{4}\, g_{\Delta\Delta}^2+162\, g_A^2)\Big]\,
\ln{\left(\frac{m_\pi}{\lambda}\right)} \right\} 
\label{chC5}
\end{eqnarray}
Here $\Delta = M_\Delta - M_N$, $f_\pi = 92.4$ MeV, $a_1, a_2, a_3$ are 
unknown parameters and $\lambda$ is a cutoff scale set to $\lambda = 1$ GeV. 

We use the SSE expression for the nucleon axial charge is presented in Ref.~\cite{Procura:2006gq}
\begin{eqnarray}
g_A^{\rm SSE}(m_\pi^2)=g^0_A&-\frac{{g^0_A}^3\,m_\pi^2}{16\pi^2f_\pi^2}
                     +4\left[C^{\rm SSE}(\lambda)
                     +\frac{c_A^2}{4\pi^2 f_\pi^2}\left(\frac{155}{1944}\, g_{\Delta\Delta}
                     -\frac{17}{36}\, g^0_A\right)+\gamma^{\rm SSE}
                     \ln{\frac{m_\pi}{\lambda}}\right]m_\pi^2\nonumber\\
                  & +\frac{4c_A^2\, 
                     g^0_A}{27 \pi f_\pi^2\, \Delta}\,m_\pi^3
                     +\frac{8}{27\pi^2 f_\pi^2}\;c_A^2\, g^0_A\, m_\pi^2\,
                     \sqrt{1-\frac{m_\pi^2}{\Delta^2}}\,\ln{R}\nonumber\\
                  & +\frac{c_A^2\Delta^2}{81\pi^2 f_\pi^2}\left(\frac{25}{2}g_{\Delta\Delta}-
                     57g_A^0\right)\left(\ln\frac{2\Delta}{m_\pi}
                     -\sqrt{1-\frac{m_\pi^2}{\Delta^2}}\,\ln R\right)
                     +{\cal O}(\epsilon^4),
\label{gasse}
\end{eqnarray}
with
\begin{eqnarray}
\gamma^{\rm SSE}&=&\frac{-1}{16\pi^2 f_\pi^2}\left[g_A^0\left(\frac{1}{2}+{g_A^0}^2\right) + \frac{2}{9}\,c_A^2\left(g_A^0-\frac{25}{18}
g_{\Delta \Delta}\right)\right]\;,\nonumber\\
R&=&\frac{\Delta}{m_\pi }+\sqrt{\frac{\Delta^2}{m_\pi^2}-1}.
\label{R}
\end{eqnarray}
$g_A^0$ in the above expressions denotes the chiral limit value of the axial 
charge, i.e. corresponds to $g_A$ in~(\ref{chC5}). 

Finally, from Jiang and Tiburzi~\cite{Jiang:2008we} we obtain the chiral 
expansion for the axial charge of the $\Delta$:
\begin{eqnarray}
G_{\Delta\Delta} (m_\pi^2)
&=& 
g_{\Delta\Delta}Z_{\Delta} 
-
\frac{1}{(4 \pi f_\pi)^2} 
\Bigg[
 g_{\Delta\Delta} 
{\cal L}(m_\pi,\mu)
\Bigg( 1 + \frac{121}{324} g_{\D \D}^2 \Bigg)
\nonumber \\
&+& 
c_A^2
\Bigg(
\frac{8}{9} 
g_{\Delta\Delta}
{\cal K}(m_{\pi},-\Delta,\mu)
- 
g_A   {\cal J}(m_{\pi},-\Delta,\mu) 
\Bigg) \Bigg] \nonumber \\
&+&
A\, m^2_{\pi}  
\, .
\label{gddsse}
\end{eqnarray}
The $\Delta$ field renormalization is 
\be
Z_\D = 1 -\frac{1}{32 \pi^2 f_\pi^2} \left[\frac{25}{18} g_{\D \D}^2
{\cal L}(m_\pi,\mu) + 2 c_A^2 {\cal J}(m_{\pi},-\Delta,\mu) \right].
\ee
and the loop integrals from~(\cite{Jiang:2008aqa}) evaluated at the
scale $\mu = 1$ GeV: 
\begin{eqnarray}
{\cal L}(m,\mu) &=& m^{2}\log\Big(\frac{m^{2}}{\mu^2}\Big) \,, \nonumber \\
{\cal K}(m,\Delta,\mu) & = & \Big(m^2-\frac{2}{3}\Delta^2\Big)\log\Big(\frac{m^2}{\mu^2}\Big) \nonumber \\
&& \quad\,+\,  \frac{2}{3}\Delta \sqrt{\Delta^2-m^2}
\log\Big(\frac{\Delta-\sqrt{\Delta^2-m^2+ i \epsilon}}{
\Delta+\sqrt{\Delta^2-m^2+ i \epsilon}}\Big)
\nonumber\\
& & \quad \, +\, \frac{2}{3} \frac{m^2}{\Delta} \Big(\ \pi m - 
\sqrt{\Delta^2-m^2}
\log\Big(\frac{\Delta-\sqrt{\Delta^2-m^2+ i \epsilon}}{
\Delta+\sqrt{\Delta^2-m^2+ i \epsilon}}\Big)
\Big)\,,\nonumber\\
{\cal J}(m,\Delta,\mu) & = & \Big(m^2-2\Delta^2\Big)\log\Big(\frac{m^2}{\mu^2}\Big) \quad \qquad \nonumber \\
&&\quad \,+\,2\Delta\sqrt{\Delta^2-m^2}
\log\Big(\frac{\Delta-\sqrt{\Delta^2-m^2+ i \epsilon}}{
\Delta+\sqrt{\Delta^2-m^2 + i \epsilon}}\Big)\,,
\label{eq:decfun}
\end{eqnarray}
\nin
From the available lattice data on $C^A_5(q^2=0;m_\pi^2)$, $g_A(m_\pi^2)$ and
$G_{\D \D}(m_\pi^2)$  we perform a simultaneous 7-parameter fit to expressions
~(\ref{c5sse}),(\ref{gasse}) and (\ref{gddsse}) fitting the unknown constants
$a_1$, $a_2$, $A$, $C^{SSE}$ as well as the common chiral
couplings $g_A$, $c_A$ and $g_{\Delta\Delta}$.
We note that $C^{SSE}$ is independent of 
$m_\pi$; at a fixed value of $\lambda$ it can be fitted as a constant.

The lattice nucleon axial charge values $g_A(m_\pi^2)$ are taken
from twisted mass simulations~\cite{Alexandrou:2010hf}. 
Lattice values for the real part
of the axial $N-\Delta$ couplings $C^A_5(0)$ are taken 
from ~\cite{Alexandrou:2010uk} via a dipole extrapolation.
The values of the real part of the axial charge of the $\Delta$,
 $G_{\Delta\Delta}(m_\pi^2)$ are related to the dominant axial form factor 
$g_1$ at zero momentum transfer via  $G_{\Delta\Delta} = -3 g_1(0)$.
For an additional lattice point to assist the fit we 
computed the zero-momentum $g_1$ 
values (only) on the $20^3\times 64$ mixed-action ensemble with 
$m_\pi = 498$ MeV. Values are provided in Table~\ref{g1_0_vals}. 

In Figure~\ref{chifit} the combined fit is presented. The available 
lattice data for all three observables vary mildly in the pion mass regime
considered. $g_A$ remains underestimated with respect to the experimental 
value and the inclusion of $C^A_5$ and $G_{\D \D}$ into the SSE fit does
not improve this systematically observed behavior. Strong chiral effects are 
expected at lighter pion mass values, especially below the $\Delta$ decay
threshold, as is evident from the 1-loop trend of $C^A_5$ and $G_{\D \D}$.

\section{Conclusions}  \label{se:conclusions}

A detailed study of the axial structure of the $\D(1232)$
 has been presented, complementing  recent and ongoing 
studies of the axial structure of the nucleon as well as the axial 
$N-$ to $-\D$ transition.
The matrix element of the $\D$ state with the axial 
current has been parameterized via four Lorentz invariant form factors, 
$g_1, g_3, h_1$ and $h_3$,
and with  two, denoted $\tilde{g}$ and $\tilde{h}$,
 the pseudoscalar
matrix element, generalizing  the familiar nucleon axial structure. 
We detailed the lattice techniques required for the extraction of 
all six form factors for a complete $q^2-$ dependent evaluation via 
specially designed three-point functions. In fact,
the calculation is optimized such that only two sequential propagators
are needed for the numerical evaluation of the optimal correlators.
PCAC constrains strongly the nucleon matrix elements of the axial-vector and pseudoscalar currents 
as is manifestly evident by the phenomenological validity of
the Goldberger-Treiman relation. Lattice QCD provides a check of this
relation, which  is a result of
 chiral symmetry breaking present in the QCD Lagrangian,
confirming that the $q^2$-dependence of the  axial and pseudoscalar form factors 
 is in agreement  with the PCAC predictions. 
Furthermore, the recent studies of
the axial $N-$ to $-\D$ transition have shown that PCAC constrains strongly
also the transition from factors and  measurements of the dominant form 
factor $C_5^A$ and $G_{\pi N \Delta}$  provided a check of the non-diagonal GT relation. This work, examines extensions of similar relations
for the $\D$. The main result of the current work is that PCAC
plays a major role also in the relation between 
 the $\D$  matrix elements of the axial-vector and pseudoscalar currents,
  connecting the strength of the $\pi-\D-\D$ 
vertex to the $\D$ axial charge, $G_{\D\D}$. Actually, 
two independent pseudoscalar form factors, $G_{\pi\D\D}$ and $H_{\pi\D\D}$ are present and pion-pole dominance establishes relations among 
all six form factors. These predictions are qualitatively verified using results
obtained in the 
quenched QCD study, which carries  the smallest statistical noise.
Results from two dynamical ensembles 
are consistent with these findings, albeit within
large statistical errors. 
 Having obtained an evaluation of $g_A$ and $C_5^A$ from previous studies
and using the results of this work for $G_{\D\D}$ on similar lattice ensembles, we performed a 
simultaneous chiral fit for all three utilizing one-loop chiral effective
theory predictions in the SSE scheme which include a dynamical $\D$ field.  
The seven-parameter fit does not drive the prediction near the experimentally
known $g_A$ value, and this is not  surprising as it has become recently 
clear that  the correct value of $g_A$ is not reproduced even 
with pion masses very close to the physical one.
A careful isolation of excited state effects~\cite{Dinter:2011sg,Alexandrou:2011aa}
at pion mass of about 400~MeV failed to reveal excited state contamination
in the lattice extraction of $g_A$.
Resolving such discrepancies is important for sharpening the predictive power
of lattice QCD, which can yield
 phenomenologically important
quantities not accessible  experimentally.
  Fully chiral 2+1 domain wall flavour simulations 
are available now  below the 300 MeV pion mass utilized in this work, and
this leaves open the perspective for further investigations in the future 
which will elaborate on the relations studied in this
work and on the values of the major couplings that dominate
the low energy hadron interactions. However, as shown here, the gauge noise is
large and noise-reduction techniques will be needed in order to
extract useful results using ensembles with close to physical pion masses.

\section*{ACKNOWLEDGEMENTS}

We are grateful to Brian Tiburzi and K. S. Choi for helpful discussions.  
EBG was supported by Cyprus Research Promotion Foundation grant
$\Delta IE\Theta NH\Sigma/\Sigma TOXO\Sigma/0308/07$ and JWN in part by funds 
provided by the U.S. Department of Energy (DOE) under cooperative research agreement DE-FG02-94ER40818. 
Computer resources were provided by the National Energy Research Scientific 
Computing Center supported by the Office
 of Science of the DOE under Contract No. DE-AC02-05CH11231 and by the J\"ulich 
Supercomputing Center, awarded under the DEISA Extreme Computing Initiative, 
co-funded through the EU FP6 project RI-031513 and the FP7 project RI-222919.
This research was in part supported by the Research Executive Agency of the 
European Union under Grant Agreement number PITN-GA-2009-238353 (ITN STRONGnet)
and the 
Cyprus Research Promotion Foundation under contracts
KY-$\Gamma$A/0310/02 and  NEA Y$\Pi$O$\Delta$OMH/$\Sigma$TPATH/0308/31 (infrastructure project Cy-Tera  co-funded by the European Regional Development Fund and the Republic of Cyprus through the Research Promotion Foundation).

\newpage
\appendix
\section{Multipole Form Factors}\label{appendix_A}

\nin
The axial vector transition between $\Delta$ states can be parameterized via a multipole
expansion. This is most naturally performed on the Breit frame, where $\vec{p}_f = -\vec{p}_i =
\vec{q}/2$.
Let us denote the matrix element as 
\begin{equation}
\langle \Delta(\vec{q}/2,s_f) | \vec{A}\cdot \vec{\epsilon}_\lambda
| \Delta(-\vec{q}/2,s_i)\rangle = M(s_f,s_i,\lambda)\,.
\end{equation}
Generically four different transitions will occur  parameterized 
via
\begin{eqnarray}
M(\frac{1}{2},\frac{1}{2},0) &=& L_1 + 3L_3\nn
M(\frac{3}{2},\frac{3}{2},0) &=& 3L_1 - L_3\nn
M(\frac{1}{2},-\frac{1}{2},1) &=& -2E_1 - \sqrt{6}E_3\nn
M(\frac{3}{2},\frac{1}{2},1) &=& -\sqrt{3}E_1 + \sqrt{2}E_3\nn
\end{eqnarray}
with $L_J$, $E_J$ the longitudinal and electric multipole amplitudes of rank $J$.
The polarization vector $\vec{\epsilon}_\lambda$ has components
$\vec{\epsilon}_{+} = -(\hat{x}+i\hat{y})/\sqrt{2}$, 
$\vec{\epsilon}_{-} = -\vec{\epsilon}_{+}^{\,*} $,
$\vec{\epsilon}_0 = \hat{z}$.

\begin{figure}[h!]
\begin{center}
\scalebox{0.4}{\includegraphics{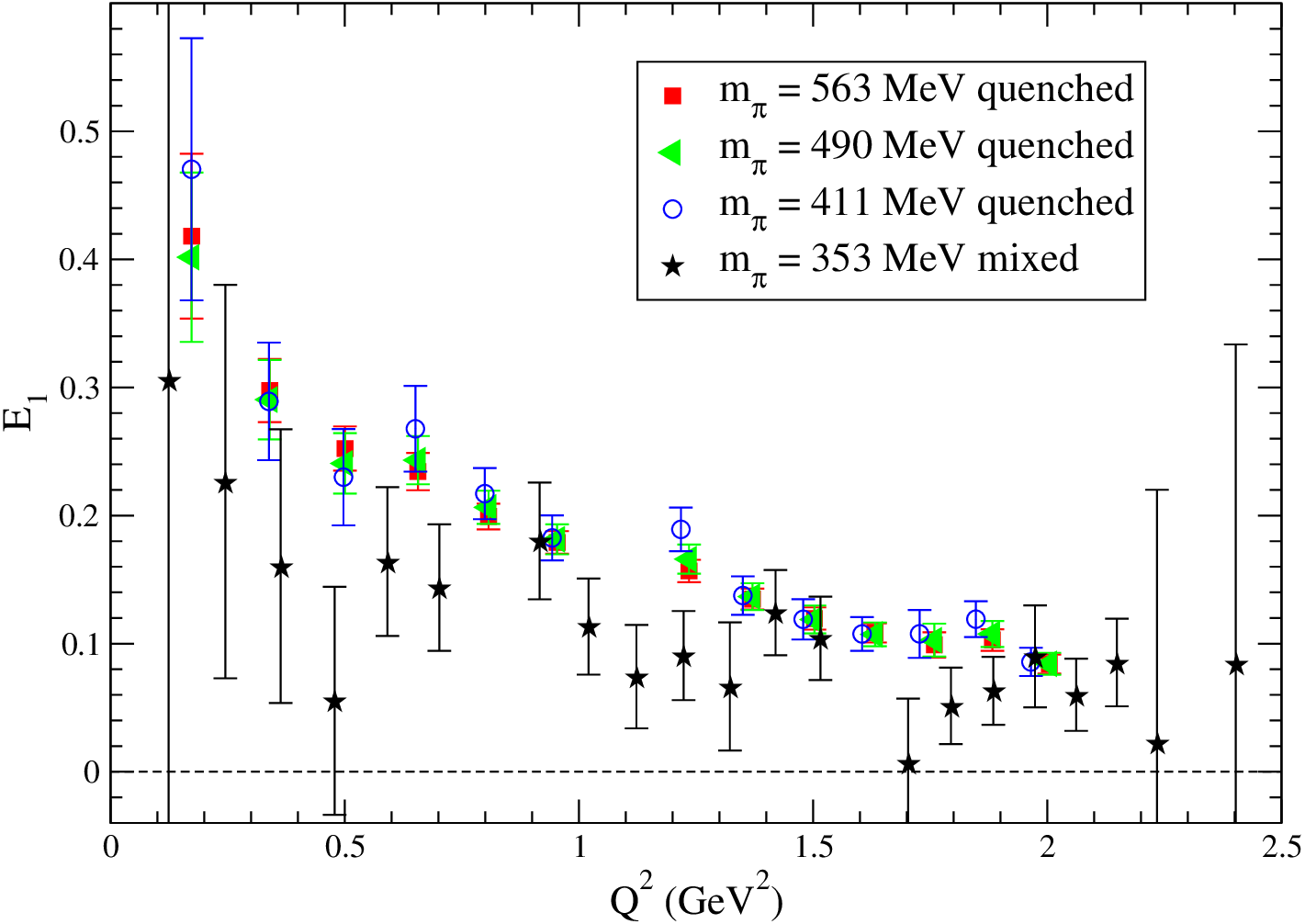}}
\end{center}
\caption{\label{E1_MP_FFS_fig} Lattice results for the $E_1$ multipole axial form-factor.}
\end{figure}

\begin{figure}[h!]
\begin{center}
\scalebox{0.4}{\includegraphics{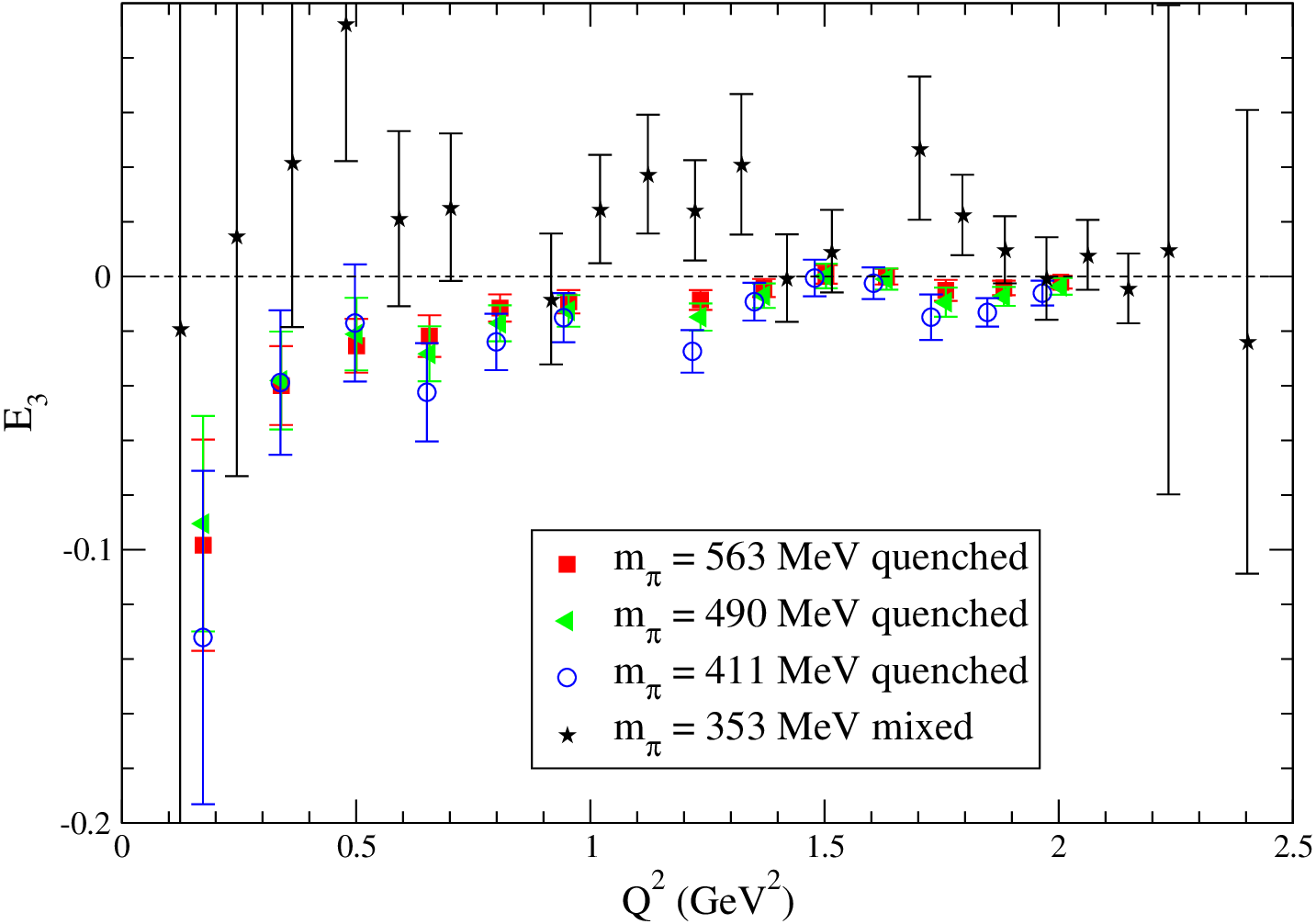}}
\end{center}
\caption{\label{E3_MP_FFS_fig} Lattice results for the $E_3$ multipole axial form-factor.}
\end{figure}

\begin{figure}[h!]
\begin{center}
\scalebox{0.4}{\includegraphics{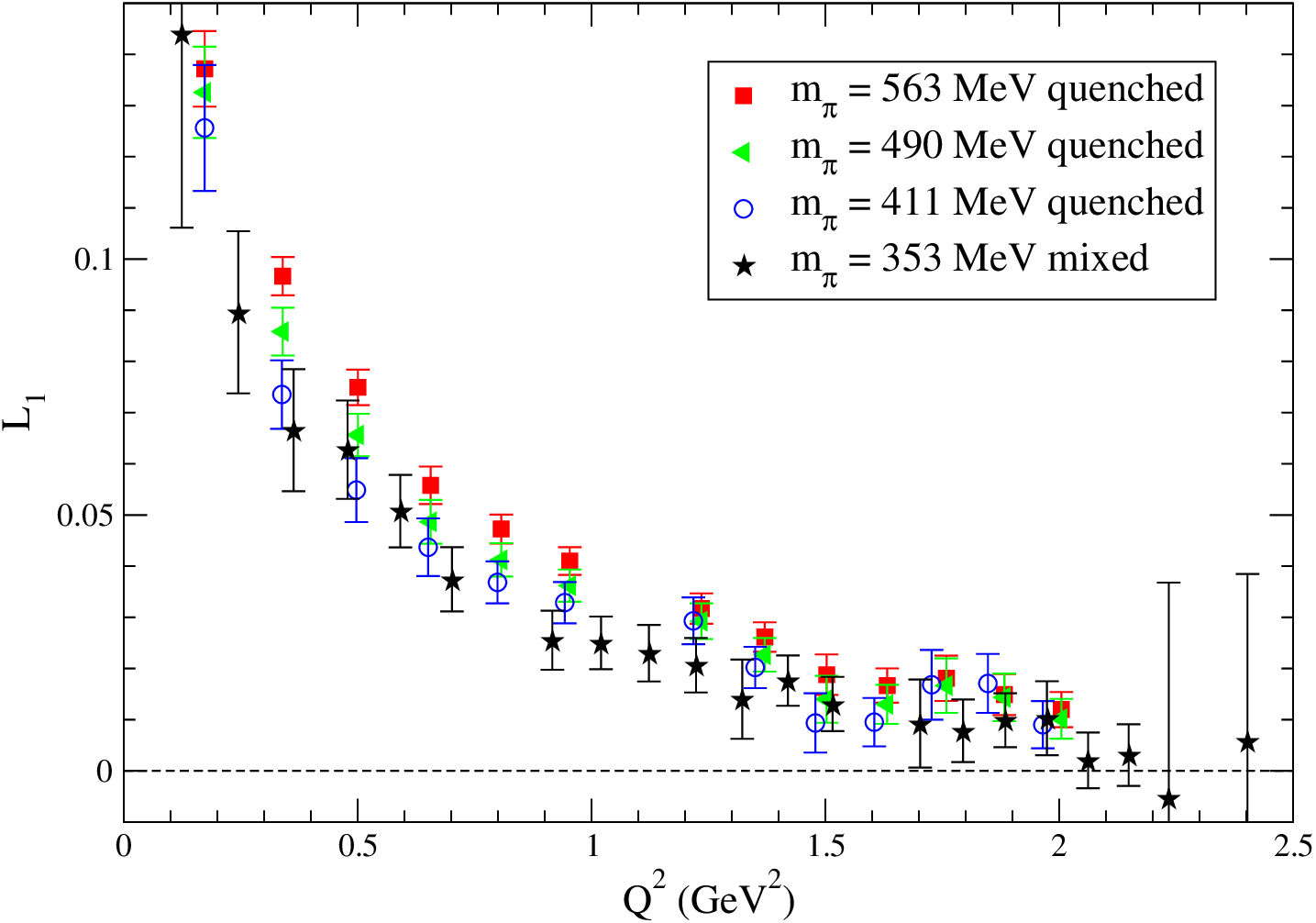}}
\end{center}
\caption{\label{L1_MP_FFS_fig} Lattice results for the $L_1$ multipole axial form-factor.}
\end{figure}

\begin{figure}[h!]
\begin{center}
\scalebox{0.4}{\includegraphics{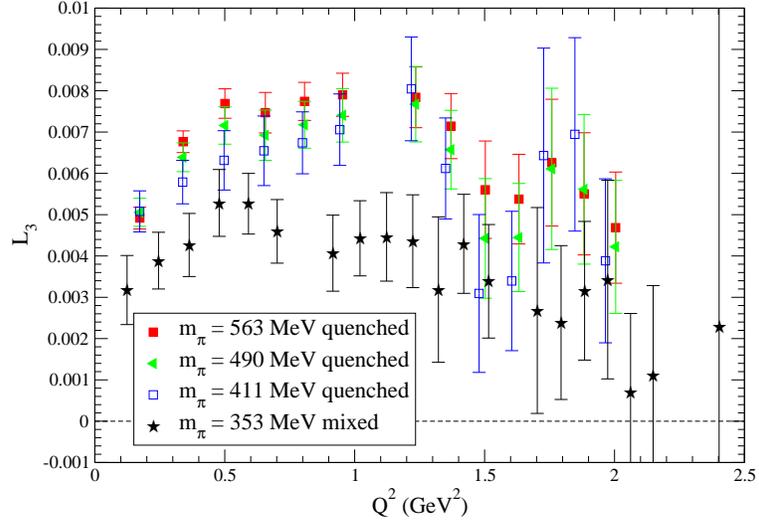}}
\end{center}
\caption{\label{L3_MP_FFS_fig} Lattice results for the $L_3$ multipole axial form-factor.}
\end{figure}

\begin{figure}[h!]
\begin{center}
\scalebox{0.4}{\includegraphics{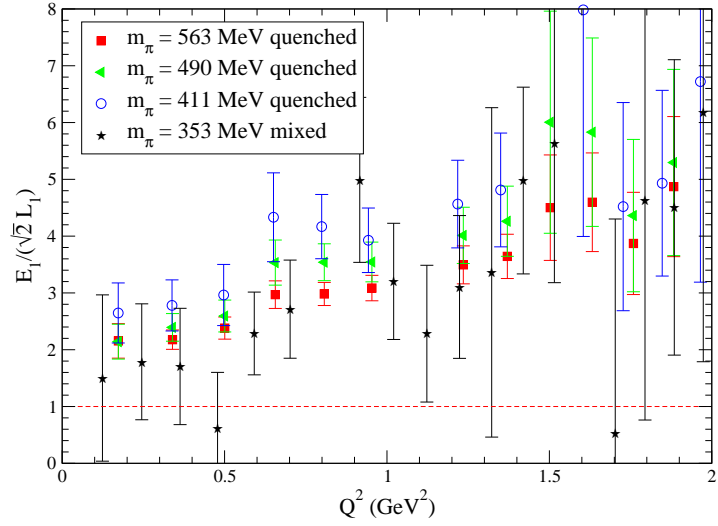}}
\end{center}
\caption{\label{e1vsl1} Lattice results for the ratio of the 
$E_1$ to $\sqrt{2} L_1$ multipole axial form-factors for the quenched ensembles.}
\end{figure}

\nin
We can relate the form factors $g_1$, $g_3$, $h_1$ and $h_3$ to the
multipole form factors $E_1$, $E_3$, $L_1$ and $L_3$, which have physical 
relevance in the multipole expansion.

\begin{eqnarray}
g_1 &=& \frac{3}{\sqrt{2}}E_1 + \sqrt{3}E_3\nn
\tau(1+\tau)h_1 &=& -3\sqrt{2}\tau E_1 + \frac{5+4\tau}{2} \sqrt{3}E_3\nn
(g_1 -\tau g_3) &=& \sqrt{1+\tau}(3L_1 - L_3)\nn
 \tau(1+\tau)(h_1 -\tau h_3) &=& \sqrt{1+\tau}(-6\tau L_1 + (5+2\tau)L_3)
\end{eqnarray}

from which the reverse relations can be verified:

\begin{eqnarray}
E_1&=& \frac{\sqrt{2} g_1}{3} - \frac{2\sqrt{2}\tau\left(2g_1 + h_1(1 + \tau)\right )}{3(5 + 8\tau)}\label{e1} \\
E_3 &=& \frac{2 \tau\left(2g_1 + h_1\left(1 + \tau\right)\right)}{ \sqrt{3}\left(5 + 8\tau\right)}\label{e3} \\
L_1 &=& \frac{\left(5+2\tau\right)\left( g_1-\tau g_3\right) +\tau\left(1+\tau\right)\left(h_1-\tau h_3\right)}{15\sqrt{1 +\tau}}\label{l1}\\
L_3 &=&\tau\frac{2\left( g_1-\tau g_3 \right) + \left(1+\tau\right)\left(h_1-\tau h_3\right) \label{l3}
}{5\sqrt{1+\tau}}\,.
\end{eqnarray}

Utilizing the above relations, we present results on the four multipole axial form-factors, 
$E_1$, $E_3$, $L_1$ and $L_3$, in Figures 
\ref{E1_MP_FFS_fig}, \ref{E3_MP_FFS_fig}, \ref{L1_MP_FFS_fig} and \ref{L3_MP_FFS_fig}, 
respectively.

In the low momentum transfer limit, $\tau \sim {\cal O}(\vec{q}^2/M_\Delta^2) << 1$ and 
from relations
(\ref{e3}, \ref{l3}) we deduce that $E_3, L_3 << 1$. 
On the other hand, $E_1$ and $L_1$ remain finite, as from relation (\ref{e1})
$E_1 \sim \sqrt{2} g_1/3$ and from (\ref{l1}) $L_1 \sim g_1/3$.
 Thus at the low momentum transfer limit we expect that 
$E_1 = \sqrt{2} L_1 + {\cal O}(\vec{q}^2)$.

We test these predictions explicitly in Figure~\ref{e1vsl1} where the ratio 
$E_1/\sqrt{2}L_1$ is plotted. We observe a behavior consistent with a constant in the 
low energy $(< 0.5 {\rm GeV}^2) $ regime although the numeric value of the constant is
largely overestimated by the quenched lattice data. 
In addition, this constancy is 
in accordance to the pion-pole dependence of both $E_1$ and $L_1$ 
which is evident from the quenched lattice data 
plotted in Figs.~(\ref{E1_MP_FFS_fig}) and~(\ref{L1_MP_FFS_fig}).
Despite the large statistical 
uncertainties, $E_3$ and $L_3$  are consistent with small values  
at small momentum transfers.

\section{Trace algebra for 3-point correlators}\label{appendix_B}

\subsection{Axial current correlator}

We define Type I as 
\begin{equation}
\Pi^{I{\rm A}}_\mu(q)\equiv
\sum_{i=1}^3 \sum_{\sigma,\tau=1}^3
\delta_{\sigma\tau}{\rm tr}\left[\Gamma^i 
\Lambda_{\sigma\sigma^\prime}(p_f) 
{\mathcal O}_{\sigma^\prime \mu \tau^\prime}^A
\Lambda_{\tau^\prime\tau}(p_i)\right]\,.
\end{equation}

After evaluating the Dirac traces we find two distinct cases, $\mu=4$ and $\mu=1,2,3$.
The kinematical frame is set to $\vec{p}_f = 0$ and $\vec{p}_i = -\vec{q}$. We note by
$E = (\vec{p}^{\,2} + \MD^2)^{1/2}$.

For $\mu=4$ we find
\begin{eqnarray}
\label{T1_mu_4_eq}
\Pi^{I{\rm A}}_{\mu=4}(q)
&=\frac{\ds -1}{\ds 36 \MD^3}&\bigg[2\left(2 E^2  - 2 E \MD + 5 \MD^2 \right)
\left(g_1-\tau g_3\right)\nonumber\\
&&-\tau\left(2 E - \MD\right) \left(E + \MD\right)\left( h_1-\tau h_3\right)
\bigg]
\left(p_1 + p_2 + p_3\right),
\end{eqnarray}
using
\begin{equation}
\tau \equiv \frac{(E-\MD)}{2\MD} = \frac{Q^2}{(2\MD)^2}.
\end{equation}
For $\mu=i$ we find
\begin{eqnarray}
\label{T1_mu_i_eq}
\Pi^{I{\rm A}}_{\mu=i}(q)&=&\frac{i(E+\MD)}{18 \MD^3}
\bigg[(2E^2+3\MD^2)g_1 - \tau E(E+\MD) h_1\bigg]\nonumber\\
&-&\frac{i}{72 \MD^4}\bigg[8 \MD^2 g_1 + 2\left(2E^2 - 2E\MD + 5\MD^2\right) g_3
\nonumber\\
&& - \MD(E+\MD) h_1 -\tau (E+\MD)(2E-\MD) h_3 \bigg]
p_i\left(p_1+ p_2 + p_3\right).
\end{eqnarray}

We define Type II as:
\begin{equation}
\Pi^{II{\rm A}}_\mu(q)\equiv
\sum_{\sigma,\tau =1}^3 T_{\sigma\tau}{\rm tr}\left[
\Gamma^4 \Lambda_{\sigma\sigma^\prime}(p_f) 
{\mathcal O}_{\sigma^\prime \mu \tau^\prime}^{A}
\Lambda_{\tau^\prime\tau}(p_i)\right]
\end{equation}

\begin{equation}
T_{\sigma\tau}=\left[\begin{array}{rrr}
0&1&-1\\
-1&0&1\\
1&-1&0\end{array}
\right].
\end{equation}
For $\mu=4$ we find
\begin{eqnarray}
\label{T2_mu_4_eq}
\Pi^{II{\rm A}}_{\mu=4}(q)&=
\frac{\ds i}{\ds 18\MD^2}
\bigg[&\left(E +4\MD\right)\left(g_1-\tau g_3\right)
\nonumber\\
&&-\frac{\tau}{2}\left(E + \MD\right)\left(h_1-\tau h_3\right)\bigg]
\left(p_1 + p_2 + p_3\right).
\end{eqnarray}
For $\mu=i$ we find
\begin{eqnarray}
\label{T2_mu_i_eq}
\Pi^{II{\rm A}}_{\mu=i}(q)&=&\frac{(E+\MD)^2}{36 \MD^3}
\bigg[(2E+3\MD)g_1 - \tau (E+\MD) h_1\bigg]\nonumber\\
&-&\frac{1}{36 \MD^3}\bigg[(2E+5\MD) g_1 + (E+4\MD) g_3
\nonumber\\
&& - \frac{E}{2 m}(E+\MD) h_1 -\frac{\tau}{2} (E+\MD) h_3 \bigg]
p_i\left(p_1+ p_2 + p_3\right).
\end{eqnarray}

\subsection{Pseudoscalar density correlator}
In a similar way we evaluate the trace algebra for the pseudoscalar vertices. The index summation types are defined in the same way as in the axial case.  Pseudoscalar Type I is
\begin{equation}
\label{PS_T1_def}
\Pi^I_{\rm PS}(q)\equiv
\sum_{i=1}^3 \sum_{\sigma,\tau=1}^3
\delta_{\sigma \tau}
{\rm tr}\left[\Gamma^i \Lambda_{\sigma\sigma^\prime}(p_f) 
{\mathcal O}_{\sigma^\prime\tau^\prime}^{\rm PS}
\Lambda_{\tau^\prime\tau}(p_i)\right]\,.
\end{equation}
After the trace evaluation we find:
\begin{eqnarray}
\Pi^I_{\rm PS}(q)&=&
 \frac{\ds -1}{\ds 18\MD^3}\Bigg[\tilde{g}\left(2E^2 -2E\MD + 5\MD^2\right)
\nonumber\\
&&-\tilde{h}\frac{\tau}{2}\left(2E-\MD \right)\left(E+\MD\right)\Bigg]
\left(p_1+p_2+p_3\right)
\end{eqnarray}
Pseudoscalar Type II  is
\begin{equation}
\label{PS_T2_def}
\Pi^{II}_{\rm PS}(q)\equiv
\sum_{\sigma,\tau=1}^3T_{\sigma\tau}
{\rm tr}\left[\Gamma^4 
\Lambda_{\sigma\sigma^\prime}(p_f) 
{\mathcal O}_{\sigma^\prime \tau^\prime}^{\rm PS}
\Lambda_{\tau^\prime\tau}(p_i)\right],
\end{equation}
giving us
\begin{equation}
\Pi^{II}_{\rm PS}(q)=\frac{i}{18 \MD^2}\left[\left(E+4\MD\right)\tilde{g} 
- \frac{\tau}{2} \left(E+\MD\right)\tilde{h}\right]
\left(p_1 + p_2 + p_3\right)
\end{equation}
after we evaluate the trace.

\section{Form Factor results}\label{appendix_C}

\begin{table}[H]
\small
\begin{tabular}{c|c|cccc|cc}
\hline\hline 
& & \multicolumn{4}{c}{Axial} &\multicolumn{2}{c}{Pseudoscalar} \\
& $Q^2$ (GeV$^2$) & $g_1$ & $g_3$ & $h_1$ & $h_3$ & $\tilde{g}$ &$\tilde{h}$\\
\hline
& 0.000000  & 0.5887(98)&    ---    &   ---     &    ---   &   ---   & --- \\
& 0.1730731 & 0.717(70) & 15.3(3.2) & 22.9(8.6) & 740(380) & 9.58(43) &46(89)\\
& 0.3396915 & 0.562(27) & 6.96(65)  & 5.6(1.7)  & 103(37)  & 7.14(19) &46(17)\\
& 0.5005274 & 0.491(20) & 4.63(38)  & 2.89(79)  & 38(16)   & 5.42(18) &17.4(8.5)\\
& 0.6561444 & 0.459(19) & 3.86(25)  & 2.15(48)  & 24.7(7.4)& 4.13(17) &10.3(5.5)\\
& 0.8070197 & 0.403(14) & 2.81(15)  & 1.30(26)  & 11.7(3.1)& 3.44(14) &10.1(2.9)\\
& 0.9535618 & 0.364(14) & 2.20(11)  & 1.04(19)  & 7.7(2.1) & 2.91(13) &8.6(1.9)\\
& 1.235014  & 0.318(14) & 1.57(10)  & 0.84(14)  & 4.4(1.1) & 2.12(14) &5.5(1.3)\\
$m_\pi = 563(4)$ MeV
& 1.370502  & 0.279(14) & 1.272(81) & 0.61(11)  & 2.91(87) & 1.73(12) &4.21(92)\\
& 1.502826  & 0.255(14) & 1.151(89) & 0.41(11)  & 2.38(79) & 1.35(13) &2.34(90)\\
& 1.632198  & 0.230(13) & 0.959(68) & 0.388(84) & 1.97(54) & 1.17(11) &2.08(69)\\
& 1.758807  & 0.201(17) & 0.730(86) & 0.45(11)  & 1.84(64) & 1.05(14) &1.95(72)\\
& 1.882824  & 0.211(16) & 0.769(76) & 0.439(69) & 1.69(41) & 0.92(13) &1.52(57)\\
& 2.004400  & 0.175(14) & 0.604(65) & 0.327(58) & 1.24(31) & 0.73(10) &1.18(43)\\
& 2.240774  & 0.124(22) & 0.43(11)  & 3.249(98) & 4.66(48) & 0.40(16) &0.48(60)\\
\hline
\hline
\end{tabular}
\caption{Delta form factors from quenched Wilson fermions.}
\label{quenched-W-k0.1554-results}
\end{table}

\begin{table}[H]
\small
\begin{tabular}{c|c|cccc|cc}
\hline\hline 
& & \multicolumn{4}{c}{Axial} &\multicolumn{2}{c}{Pseudoscalar} \\
& $Q^2$ (GeV$^2$) & $g_1$ & $g_3$ & $h_1$ & $h_3$ & $\tilde{g}$ &$\tilde{h}$\\
\hline
& 0.000000  & 0.578(13) &  ---      &  ---     &  ---    & ---      & --- \\
& 0.1728690 & 0.695(72) & 14.0(2.9) & 20.0(8.3)&390(330) &10.85(62) & 120(130)\\
& 0.3389351 & 0.550(36) & 7.03(78)  & 5.1(2.0) &97(41)   & 7.69(27) &59(23)\\
& 0.4989433 & 0.474(28) & 4.52(45)  & 2.4(1.1) &33(17)   & 5.78(23) &26(11)\\
& 0.6535119 & 0.467(23) & 3.99(31)  & 2.45(60) &26.8(9.0)& 4.23(22) &10.3(6.8)\\
& 0.8031605 & 0.408(17) & 2.88(18)  & 1.53(33) &13.1(3.6)& 3.54(17) &11.6(3.3)\\
& 0.9483310 & 0.363(16) & 2.18(14)  & 1.14(25) &8.1(2.4) & 2.98(16) &8.9(2.3)\\
& 1.226705  & 0.326(17) & 1.58(12)  & 1.02(18) &5.1(1.4) & 2.21(17) &5.9(1.5)\\
$m_\pi = 490(4)$ MeV
& 1.360524  & 0.278(16) & 1.253(95) & 0.67(14) &3.2(1.0) & 1.75(14) &4.1(1.1)\\
& 1.491113  & 0.252(17) & 1.14(10)  & 0.42(13) &2.59(93) & 1.30(14) &2.0(1.1)\\
& 1.618694  & 0.226(15) & 0.935(78) & 0.40(11) &2.02(63) & 1.15(13) &2.01(82)\\
& 1.743467  & 0.202(20) & 0.71(10)  & 0.54(14) &2.17(77) & 1.02(16) &1.50(89)\\
& 1.865609  & 0.216(19) & 0.751(87) & 0.501(87)&1.74(48) & 0.99(15) &1.79(70)\\
& 1.985279  & 0.173(16) & 0.580(71) & 0.348(73)&1.25(36) & 0.73(12) &1.10(50)\\
& 2.217769  & 0.110(24) & 0.41(12)  &-0.09(13) &-0.02(58)& 0.31(18) &0.21(72)\\
\hline\hline
\end{tabular}
\caption{Delta form factors from quenched Wilson fermions.}
\label{quenched-W-k0.1558-results}
\end{table}

\begin{table}[H]
\small
\begin{tabular}{c|c|cccc|cc}
\hline\hline 
& & \multicolumn{4}{c}{Axial} 
&   \multicolumn{2}{|c}{Pseudoscalar} \\
& $Q^2$ (GeV$^2$) & $g_1$ & $g_3$ & $h_1$ & $h_3$ & $\tilde{g}$ &$\tilde{h}$\\
\hline\hline
& 0.000000 & 0.571(18) &  ---     & ---      &    ---   &  ---    & ---  \\
& 0.1726468& 0.77(11)  & 17.4(4.5)& 27(12)   & 560(490) &13.1(1.1)&290(230)\\
& 0.3381149& 0.546(52) & 7.4(1.1) & 4.9(2.7) & 102(61)  &8.27(45) &59(40)\\
& 0.4972318& 0.458(44) & 4.53(66) & 2.0(1.6) & 33(24)   &6.17(36) &27(18)\\
& 0.6506769& 0.495(41) & 4.28(51) & 3.2(1.0) & 32(15)   &4.48(31) &15.4(9.8)\\
& 0.7990168& 0.419(26) & 2.96(25) & 1.81(49) & 13.4(5.2)&3.73(23) &14.4(4.4)\\
& 0.9427295& 0.361(23) & 2.14(19) & 1.21(37) & 7.5(3.4) &3.16(22) &11.2(3.2)\\
& 1.217848 & 0.354(25) & 1.68(16) & 1.42(26) & 6.4(2.0) &2.41(25) &7.5(2.3)\\
$m_\pi = 411(4)$ MeV
& 1.349909 & 0.276(22) & 1.23(12) & 0.72(21) & 3.4(1.4) &1.78(18) &4.4(1.4)\\
& 1.478675 & 0.251(24) & 1.16(13) & 0.44(19) & 3.1(1.3) &1.23(18) &1.8(1.5)\\
& 1.604379 & 0.224(20) & 0.93(10) & 0.43(16) & 2.26(87) &1.15(16) &2.4(1.1)\\
& 1.727230 & 0.203(28) & 0.68(13) & 0.65(21) & 2.5(1.1) &0.96(21) &1.0(1.3)\\
& 1.847415 & 0.230(24) & 0.75(11) & 0.63(13) & 1.87(64) &1.16(20) &2.9(1.0)\\
& 1.965099 & 0.171(18) & 0.565(85)& 0.39(10) & 1.32(47) &0.73(15) &1.13(68)\\
& 2.193551 & 9.100(30) & 0.37(14) &-0.18(19) & 8.70(83) &0.16(21) &-0.2(1.1)\\
\end{tabular}
\caption{Delta form factors from quenched Wilson fermions. Results for $h_1$ and $h_3$ are plagued by statistical noise.}
\label{quenched-W-k0.1562-results}
\end{table}

\begin{table}[H]
\small
\begin{tabular}{c|c|cccc|cc}
\hline\hline 
& & \multicolumn{4}{c}{Axial} 
&   \multicolumn{2}{|c}{Pseudoscalar} \\
& $Q^2$ (GeV$^2$) & $g_1$ & $g_3$ & $h_1$ & $h_3$ & $\tilde{g}$ &$\tilde{h}$\\
\hline\hline
&0.000000  & 0.640(26) & ---     & ---      & ---      &  ---    &  --- \\
&0.1240682 & 0.62(39)  & 15(26)  &8(74)     &-100(4500)&14.4(4.8)&230(1800)\\
&0.2450231 & 0.51(17)  & 9.8(6.0)&-2(16)    &120(540) & 11.7(1.8)&440(320)\\
&0.3630880 & 0.41(12)  & 6.0(3.1)&-4.4(7.7) &-25(200) & 9.1(1.2) &260(140)\\
&0.4784607 & 0.28(10)  & 1.9(2.1)&-8.1(4.9) &-160(104)& 7.0(1.1) &87(83)\\
&0.5913172 & 0.385(69) & 4.0(1.1)&-0.9(2.5) &-7(40)   & 5.88(67) &85(37)\\
&0.7018154 & 0.349(60) &3.42(78) &-1.0(1.9) & 2(25)   & 5.41(61) &87(27)\\
&0.9162911 & 0.368(57) &3.22(60) & 1.1(1.3) &21(15)   & 3.46(56) &29(17)\\
&1.020513  & 0.283(47) &2.07(43) &-0.6(1.0) &-0.3(9.3)& 3.43(51) &35(12)\\
&1.122869  & 0.222(50) &1.39(43) &-1.15(98) &-9.6(8.5)& 2.22(52) &17(11)\\
&1.223456  & 0.234(45) &1.43(36) &-0.50(77) &-3.2(6.3)& 2.26(48) &13.3(8.5)\\
$m_\pi = 353(4)$ MeV
&1.322363  & 0.212(66) &1.31(51) &-1.1(1.0) &-5.6(8.1)& 2.34(77) &18(13)\\
&1.419670  & 0.262(46) &1.50(31) &0.48(60)  &2.5(4.2) & 1.70(50) &7.0(7.1)\\
&1.515454  & 0.237(48) &1.32(30) &0.12(53)  &2.0(3.6) & 2.08(54) &12.3(6.3)\\
&1.702723  & 9.586(67) &0.41(38) &-1.15(83) &-6.4(4.8)& 1.25(99) &14(11)\\
&1.794332  & 0.148(44) &0.70(24) &-0.35(44) &-1.8(2.5)& 1.07(56) &8.4(5.2)\\
&1.884666  & 0.151(39) &0.65(20) &0.00(66)  &-0.3(2.0)& 0.90(48) &3.1(4.0)\\
&1.973777  & 0.190(66) &0.82(32) &0.34(44)  &1.3(2.3) & 0.65(67) &-6.1(5.8)\\
&2.061714  & 0.141(42) &0.66(20) &0.05(35)  &0.5(1.8) & 0.37(52) &5.4(4.0)\\
&2.148521  & 0.173(56) &0.77(26) &0.38(35)  &2.1(1.7) & 1.07(59) &4.3(4.1)\\
&2.234241  & 0.06(31)  &0.4(1.6) &-0.1(2.6) &0.0(11.0)&-0.5(3.5) &-5(32)\\
&2.402577  & 0.14(39)  &0.5(1.5) &0.7(2.3)  &2.5(9.2) & 1.0(3.4) &2(18) \\
\end{tabular}
\caption{Delta form factors from mixed-action fermions.}
\label{milc-mixed_FF-results}
\end{table}

\begin{table}[H]
\small
\begin{tabular}{c|c|cccc|cc}
\hline\hline 
& & \multicolumn{4}{c}{Axial} 
&   \multicolumn{2}{|c}{Pseudoscalar} \\
& $Q^2$ (GeV$^2$) & $g_1$ & $g_3$ & $h_1$ & $h_3$ & $\tilde{g}$ &$\tilde{h}$\\
\hline\hline
&0.0        & 0.604(38) &      ---      &     ---      &    ---      & ---       &  ---        \\
&0.2060893  & 0.08(45)  &  -5.6(18.0)   &  -44(45)     & -1500(2000) &  9.4(9.3) & -1100(1800) \\
&0.4030337  & 0.66(20)  &    9.3(4.3)   &   11(10)     &   130(240)  &  6.6(3.0) &   -300(280) \\
&0.5919523  & 0.79(19)  &   10.3(2.9)   &   13(6)      &   210(110)  &  4.5(2.4) &    -82(140) \\
&0.7737530  & 0.41(20)  &    3.5(2.3)   &    2.2(5.2)  &     8(65)   &  7.2(2.2) &     125(90) \\
&0.9491845  & 0.21(12)  &    0.6(1.7)   &   -1.0(2.5)  &   -34(25)   &  4.5(1.3) &      42(39) \\
&1.118873   & 0.51(14)  &    3.9(1.1)   &    4.8(2.2)  &    45(19)   &  1.3(1.4) &     -20(32) \\
&1.443061   & 0.47(21)  &    2.3(1.3)   &    4.3(2.6)  &    21(17)   &  3.2(1.9) &      27(30) \\
&1.598405   & 0.31(14)  &    1.25(76)   &    2.1(1.6)  &    4.7(8.8) &  0.1(1.2) &      -1(16) \\
&1.749719   & 0.14(13)  &    0.91(69)   &    0.0(1.3)  &    2.2(7.2) &  0.2(1.4) &     -12(16) \\
&1.897302   & -0.09(29) &   -0.3(1.4)   &   -2.2(3.2)  &   -11(16)   & -4.6(4.6) &     -62(57) \\
$m_\pi = 297(5)$ MeV & 2.041417 &-0.03(33) &  0.1(1.6)   & -0.4(2.9)  & 3(14)& 0.2(3.1) & 6(33) \\
&2.182297   &  0.17(23) &    0.51(91)   &    0.8(1.8)  &    4.6(7.7) &  2.8(2.7) &       9(18) \\
&2.320150   & -0.06(14) &   -0.41(57)   &   -0.7(1.1)  &   -3.6(4.4) &  0.4(1.4) &       2(11) \\
\end{tabular}
\caption{Delta form factors from Domain-Wall fermions.}
\label{dwf_FF-results}
\end{table}

\bibliography{paper_ref}

\begin{thebibliography}{41}
\expandafter\ifx\csname natexlab\endcsname\relax\def\natexlab#1{#1}\fi
\expandafter\ifx\csname bibnamefont\endcsname\relax
  \def\bibnamefont#1{#1}\fi
\expandafter\ifx\csname bibfnamefont\endcsname\relax
  \def\bibfnamefont#1{#1}\fi
\expandafter\ifx\csname citenamefont\endcsname\relax
  \def\citenamefont#1{#1}\fi
\expandafter\ifx\csname url\endcsname\relax
  \def\url#1{\texttt{#1}}\fi
\expandafter\ifx\csname urlprefix\endcsname\relax\def\urlprefix{URL }\fi
\providecommand{\bibinfo}[2]{#2}
\providecommand{\eprint}[2][]{\url{#2}}

\bibitem[{\citenamefont{Jansen}(2008)}]{Jansen:2008vs}
\bibinfo{author}{\bibfnamefont{K.}~\bibnamefont{Jansen}}
  (\bibinfo{year}{2008}), \eprint{0810.5634}.

\bibitem[{\citenamefont{Durr et~al.}(2008)}]{Durr:2008zz}
\bibinfo{author}{\bibfnamefont{S.}~\bibnamefont{Durr}} \bibnamefont{et~al.},
  \bibinfo{journal}{Science} \textbf{\bibinfo{volume}{322}},
  \bibinfo{pages}{1224} (\bibinfo{year}{2008}).

\bibitem[{\citenamefont{Aoki et~al.}(2009)}]{Aoki:2008sm}
\bibinfo{author}{\bibfnamefont{S.}~\bibnamefont{Aoki}} \bibnamefont{et~al.}
  (\bibinfo{collaboration}{PACS-CS}), \bibinfo{journal}{Phys. Rev. D}
  \textbf{\bibinfo{volume}{79}}, \bibinfo{pages}{034503}
  (\bibinfo{year}{2009}), \eprint{0807.1661}.

\bibitem[{\citenamefont{Durr et~al.}(2011)\citenamefont{Durr, Fodor, Hoelbling,
  Katz, Krieg et~al.}}]{Durr:2010aw}
\bibinfo{author}{\bibfnamefont{S.}~\bibnamefont{Durr}},
  \bibinfo{author}{\bibfnamefont{Z.}~\bibnamefont{Fodor}},
  \bibinfo{author}{\bibfnamefont{C.}~\bibnamefont{Hoelbling}},
  \bibinfo{author}{\bibfnamefont{S.}~\bibnamefont{Katz}},
  \bibinfo{author}{\bibfnamefont{S.}~\bibnamefont{Krieg}},
  \bibnamefont{et~al.}, \bibinfo{journal}{JHEP}
  \textbf{\bibinfo{volume}{1108}}, \bibinfo{pages}{148} (\bibinfo{year}{2011}),
  \eprint{1011.2711}.

\bibitem[{\citenamefont{Alexandrou
  et~al.}(2009{\natexlab{a}})}]{Alexandrou:2009qu}
\bibinfo{author}{\bibfnamefont{C.}~\bibnamefont{Alexandrou}}
  \bibnamefont{et~al.} (\bibinfo{collaboration}{ETM Collaboration}),
  \bibinfo{journal}{Phys.Rev.} \textbf{\bibinfo{volume}{D80}},
  \bibinfo{pages}{114503} (\bibinfo{year}{2009}{\natexlab{a}}),
  \eprint{0910.2419}.

\bibitem[{\citenamefont{Aoki et~al.}(2010)}]{Aoki:2009ix}
\bibinfo{author}{\bibfnamefont{S.}~\bibnamefont{Aoki}} \bibnamefont{et~al.}
  (\bibinfo{collaboration}{PACS-CS Collaboration}),
  \bibinfo{journal}{Phys.Rev.} \textbf{\bibinfo{volume}{D81}},
  \bibinfo{pages}{074503} (\bibinfo{year}{2010}), \eprint{0911.2561}.

\bibitem[{\citenamefont{Beane et~al.}(2012)}]{Beane:2011sc}
\bibinfo{author}{\bibfnamefont{S.}~\bibnamefont{Beane}} \bibnamefont{et~al.}
  (\bibinfo{collaboration}{NPLQCD Collaboration}), \bibinfo{journal}{Phys.Rev.}
  \textbf{\bibinfo{volume}{D85}}, \bibinfo{pages}{034505}
  (\bibinfo{year}{2012}), \eprint{1107.5023}.

\bibitem[{\citenamefont{Dudek et~al.}(2012)\citenamefont{Dudek, Edwards, and
  Thomas}}]{Dudek:2012gj}
\bibinfo{author}{\bibfnamefont{J.~J.} \bibnamefont{Dudek}},
  \bibinfo{author}{\bibfnamefont{R.~G.} \bibnamefont{Edwards}},
  \bibnamefont{and} \bibinfo{author}{\bibfnamefont{C.~E.}
  \bibnamefont{Thomas}}, \bibinfo{journal}{Phys.Rev.}
  \textbf{\bibinfo{volume}{D86}}, \bibinfo{pages}{034031}
  (\bibinfo{year}{2012}), \eprint{1203.6041}.

\bibitem[{\citenamefont{Feng et~al.}(2010)\citenamefont{Feng, Jansen, and
  Renner}}]{Feng:2009ij}
\bibinfo{author}{\bibfnamefont{X.}~\bibnamefont{Feng}},
  \bibinfo{author}{\bibfnamefont{K.}~\bibnamefont{Jansen}}, \bibnamefont{and}
  \bibinfo{author}{\bibfnamefont{D.~B.} \bibnamefont{Renner}},
  \bibinfo{journal}{Phys.Lett.} \textbf{\bibinfo{volume}{B684}},
  \bibinfo{pages}{268} (\bibinfo{year}{2010}), \eprint{0909.3255}.

\bibitem[{\citenamefont{Yamazaki et~al.}(2004)}]{Yamazaki:2004qb}
\bibinfo{author}{\bibfnamefont{T.}~\bibnamefont{Yamazaki}} \bibnamefont{et~al.}
  (\bibinfo{collaboration}{CP-PACS Collaboration}),
  \bibinfo{journal}{Phys.Rev.} \textbf{\bibinfo{volume}{D70}},
  \bibinfo{pages}{074513} (\bibinfo{year}{2004}), \eprint{hep-lat/0402025}.

\bibitem[{\citenamefont{Kotulla et~al.}(2002)\citenamefont{Kotulla, Ahrens,
  Annand, Beck, Caselotti et~al.}}]{Kotulla:2002cg}
\bibinfo{author}{\bibfnamefont{M.}~\bibnamefont{Kotulla}},
  \bibinfo{author}{\bibfnamefont{J.}~\bibnamefont{Ahrens}},
  \bibinfo{author}{\bibfnamefont{J.}~\bibnamefont{Annand}},
  \bibinfo{author}{\bibfnamefont{R.}~\bibnamefont{Beck}},
  \bibinfo{author}{\bibfnamefont{G.}~\bibnamefont{Caselotti}},
  \bibnamefont{et~al.}, \bibinfo{journal}{Phys.Rev.Lett.}
  \textbf{\bibinfo{volume}{89}}, \bibinfo{pages}{272001}
  (\bibinfo{year}{2002}), \eprint{nucl-ex/0210040}.

\bibitem[{\citenamefont{Lopez~Castro and Mariano}(2001)}]{LopezCastro:2000cv}
\bibinfo{author}{\bibfnamefont{G.}~\bibnamefont{Lopez~Castro}}
  \bibnamefont{and} \bibinfo{author}{\bibfnamefont{A.}~\bibnamefont{Mariano}},
  \bibinfo{journal}{Phys.Lett.} \textbf{\bibinfo{volume}{B517}},
  \bibinfo{pages}{339} (\bibinfo{year}{2001}), \eprint{nucl-th/0006031}.

\bibitem[{\citenamefont{Bernard et~al.}(2005)\citenamefont{Bernard, Hemmert,
  and Meissner}}]{Bernard:2005fy}
\bibinfo{author}{\bibfnamefont{V.}~\bibnamefont{Bernard}},
  \bibinfo{author}{\bibfnamefont{T.~R.} \bibnamefont{Hemmert}},
  \bibnamefont{and} \bibinfo{author}{\bibfnamefont{U.-G.}
  \bibnamefont{Meissner}}, \bibinfo{journal}{Phys.Lett.}
  \textbf{\bibinfo{volume}{B622}}, \bibinfo{pages}{141} (\bibinfo{year}{2005}),
  \eprint{hep-lat/0503022}.

\bibitem[{\citenamefont{Hemmert et~al.}(1998)\citenamefont{Hemmert, Holstein,
  and Kambor}}]{Hemmert:1997ye}
\bibinfo{author}{\bibfnamefont{T.~R.} \bibnamefont{Hemmert}},
  \bibinfo{author}{\bibfnamefont{B.~R.} \bibnamefont{Holstein}},
  \bibnamefont{and} \bibinfo{author}{\bibfnamefont{J.}~\bibnamefont{Kambor}},
  \bibinfo{journal}{J.Phys.} \textbf{\bibinfo{volume}{G24}},
  \bibinfo{pages}{1831} (\bibinfo{year}{1998}), \eprint{hep-ph/9712496}.

\bibitem[{\citenamefont{Jenkins and Manohar}(1991)}]{Jenkins:1991es}
\bibinfo{author}{\bibfnamefont{E.~E.} \bibnamefont{Jenkins}} \bibnamefont{and}
  \bibinfo{author}{\bibfnamefont{A.~V.} \bibnamefont{Manohar}},
  \bibinfo{journal}{Phys.Lett.} \textbf{\bibinfo{volume}{B259}},
  \bibinfo{pages}{353} (\bibinfo{year}{1991}).

\bibitem[{\citenamefont{Fettes and Meissner}(2001)}]{Fettes:2000bb}
\bibinfo{author}{\bibfnamefont{N.}~\bibnamefont{Fettes}} \bibnamefont{and}
  \bibinfo{author}{\bibfnamefont{U.~G.} \bibnamefont{Meissner}},
  \bibinfo{journal}{Nucl.Phys.} \textbf{\bibinfo{volume}{A679}},
  \bibinfo{pages}{629} (\bibinfo{year}{2001}), \eprint{hep-ph/0006299}.

\bibitem[{\citenamefont{Bernard}(2008)}]{Bernard:2007zu}
\bibinfo{author}{\bibfnamefont{V.}~\bibnamefont{Bernard}},
  \bibinfo{journal}{Prog.Part.Nucl.Phys.} \textbf{\bibinfo{volume}{60}},
  \bibinfo{pages}{82} (\bibinfo{year}{2008}), \eprint{0706.0312}.

\bibitem[{\citenamefont{Dashen et~al.}(1994)\citenamefont{Dashen, Jenkins, and
  Manohar}}]{Dashen:1993jt}
\bibinfo{author}{\bibfnamefont{R.~F.} \bibnamefont{Dashen}},
  \bibinfo{author}{\bibfnamefont{E.~E.} \bibnamefont{Jenkins}},
  \bibnamefont{and} \bibinfo{author}{\bibfnamefont{A.~V.}
  \bibnamefont{Manohar}}, \bibinfo{journal}{Phys.Rev.}
  \textbf{\bibinfo{volume}{D49}}, \bibinfo{pages}{4713} (\bibinfo{year}{1994}),
  \eprint{hep-ph/9310379}.

\bibitem[{\citenamefont{Brown and Weise}(1975)}]{Brown:1975di}
\bibinfo{author}{\bibfnamefont{G.}~\bibnamefont{Brown}} \bibnamefont{and}
  \bibinfo{author}{\bibfnamefont{W.}~\bibnamefont{Weise}},
  \bibinfo{journal}{Phys.Rept.} \textbf{\bibinfo{volume}{22}},
  \bibinfo{pages}{279} (\bibinfo{year}{1975}).

\bibitem[{\citenamefont{Choi et~al.}(2010)\citenamefont{Choi, Plessas, and
  Wagenbrunn}}]{Choi:2010ty}
\bibinfo{author}{\bibfnamefont{K.-S.} \bibnamefont{Choi}},
  \bibinfo{author}{\bibfnamefont{W.}~\bibnamefont{Plessas}}, \bibnamefont{and}
  \bibinfo{author}{\bibfnamefont{R.}~\bibnamefont{Wagenbrunn}},
  \bibinfo{journal}{Phys.Rev.} \textbf{\bibinfo{volume}{D82}},
  \bibinfo{pages}{014007} (\bibinfo{year}{2010}), \eprint{1005.0337}.

\bibitem[{\citenamefont{Alexandrou et~al.}(2007)\citenamefont{Alexandrou,
  Koutsou, Leontiou, Negele, and Tsapalis}}]{Alexandrou:2007xj}
\bibinfo{author}{\bibfnamefont{C.}~\bibnamefont{Alexandrou}},
  \bibinfo{author}{\bibfnamefont{G.}~\bibnamefont{Koutsou}},
  \bibinfo{author}{\bibfnamefont{T.}~\bibnamefont{Leontiou}},
  \bibinfo{author}{\bibfnamefont{J.~W.} \bibnamefont{Negele}},
  \bibnamefont{and} \bibinfo{author}{\bibfnamefont{A.}~\bibnamefont{Tsapalis}},
  \bibinfo{journal}{Phys.Rev.} \textbf{\bibinfo{volume}{D76}},
  \bibinfo{pages}{094511} (\bibinfo{year}{2007}), \eprint{0912.0394}.

\bibitem[{\citenamefont{Alexandrou
  et~al.}(2011{\natexlab{a}})\citenamefont{Alexandrou, Koutsou, Negele,
  Proestos, and Tsapalis}}]{Alexandrou:2010uk}
\bibinfo{author}{\bibfnamefont{C.}~\bibnamefont{Alexandrou}},
  \bibinfo{author}{\bibfnamefont{G.}~\bibnamefont{Koutsou}},
  \bibinfo{author}{\bibfnamefont{J.}~\bibnamefont{Negele}},
  \bibinfo{author}{\bibfnamefont{Y.}~\bibnamefont{Proestos}}, \bibnamefont{and}
  \bibinfo{author}{\bibfnamefont{A.}~\bibnamefont{Tsapalis}},
  \bibinfo{journal}{Phys.Rev.} \textbf{\bibinfo{volume}{D83}},
  \bibinfo{pages}{014501} (\bibinfo{year}{2011}{\natexlab{a}}),
  \eprint{1011.3233}.

\bibitem[{\citenamefont{Alexandrou
  et~al.}(2009{\natexlab{b}})\citenamefont{Alexandrou, Korzec, Koutsou, Lorce,
  Negele et~al.}}]{Alexandrou:2009hs}
\bibinfo{author}{\bibfnamefont{C.}~\bibnamefont{Alexandrou}},
  \bibinfo{author}{\bibfnamefont{T.}~\bibnamefont{Korzec}},
  \bibinfo{author}{\bibfnamefont{G.}~\bibnamefont{Koutsou}},
  \bibinfo{author}{\bibfnamefont{C.}~\bibnamefont{Lorce}},
  \bibinfo{author}{\bibfnamefont{J.~W.} \bibnamefont{Negele}},
  \bibnamefont{et~al.}, \bibinfo{journal}{Nucl.Phys.}
  \textbf{\bibinfo{volume}{A825}}, \bibinfo{pages}{115}
  (\bibinfo{year}{2009}{\natexlab{b}}), \eprint{0901.3457}.

\bibitem[{\citenamefont{Alexandrou et~al.}(2012)\citenamefont{Alexandrou,
  Papanicolas, and Vanderhaeghen}}]{Alexandrou:2012da}
\bibinfo{author}{\bibfnamefont{C.}~\bibnamefont{Alexandrou}},
  \bibinfo{author}{\bibfnamefont{C.}~\bibnamefont{Papanicolas}},
  \bibnamefont{and}
  \bibinfo{author}{\bibfnamefont{M.}~\bibnamefont{Vanderhaeghen}}
  (\bibinfo{year}{2012}), \eprint{1201.4511}.

\bibitem[{\citenamefont{Alexandrou
  et~al.}(2011{\natexlab{b}})\citenamefont{Alexandrou, Gregory, Korzec,
  Koutsou, Negele et~al.}}]{Alexandrou:2011py}
\bibinfo{author}{\bibfnamefont{C.}~\bibnamefont{Alexandrou}},
  \bibinfo{author}{\bibfnamefont{E.~B.} \bibnamefont{Gregory}},
  \bibinfo{author}{\bibfnamefont{T.}~\bibnamefont{Korzec}},
  \bibinfo{author}{\bibfnamefont{G.}~\bibnamefont{Koutsou}},
  \bibinfo{author}{\bibfnamefont{J.~W.} \bibnamefont{Negele}},
  \bibnamefont{et~al.}, \bibinfo{journal}{Phys.Rev.Lett.}
  \textbf{\bibinfo{volume}{107}}, \bibinfo{pages}{141601}
  (\bibinfo{year}{2011}{\natexlab{b}}), \eprint{1106.6000}.

\bibitem[{\citenamefont{Alexandrou
  et~al.}(2011{\natexlab{c}})}]{Alexandrou:2010hf}
\bibinfo{author}{\bibfnamefont{C.}~\bibnamefont{Alexandrou}}
  \bibnamefont{et~al.} (\bibinfo{collaboration}{ETM Collaboration}),
  \bibinfo{journal}{Phys.Rev.} \textbf{\bibinfo{volume}{D83}},
  \bibinfo{pages}{045010} (\bibinfo{year}{2011}{\natexlab{c}}),
  \eprint{1012.0857}.

\bibitem[{\citenamefont{Jiang and Tiburzi}(2008{\natexlab{a}})}]{Jiang:2008we}
\bibinfo{author}{\bibfnamefont{F.-J.} \bibnamefont{Jiang}} \bibnamefont{and}
  \bibinfo{author}{\bibfnamefont{B.~C.} \bibnamefont{Tiburzi}},
  \bibinfo{journal}{Phys.Rev.} \textbf{\bibinfo{volume}{D78}},
  \bibinfo{pages}{017504} (\bibinfo{year}{2008}{\natexlab{a}}),
  \eprint{0803.3316}.

\bibitem[{for()}]{form_ref}
\bibinfo{note}{Http://www.nikhef.nl/$\sim$form}.

\bibitem[{\citenamefont{Tsapalis}(2006)}]{Tsapalis:2006kn}
\bibinfo{author}{\bibfnamefont{A.}~\bibnamefont{Tsapalis}},
  \bibinfo{journal}{Nucl.Phys.Proc.Suppl.} \textbf{\bibinfo{volume}{153}},
  \bibinfo{pages}{320} (\bibinfo{year}{2006}).

\bibitem[{\citenamefont{Bernard et~al.}(2001)\citenamefont{Bernard, Burch,
  Orginos, Toussaint, DeGrand et~al.}}]{Bernard:2001av}
\bibinfo{author}{\bibfnamefont{C.~W.} \bibnamefont{Bernard}},
  \bibinfo{author}{\bibfnamefont{T.}~\bibnamefont{Burch}},
  \bibinfo{author}{\bibfnamefont{K.}~\bibnamefont{Orginos}},
  \bibinfo{author}{\bibfnamefont{D.}~\bibnamefont{Toussaint}},
  \bibinfo{author}{\bibfnamefont{T.~A.} \bibnamefont{DeGrand}},
  \bibnamefont{et~al.}, \bibinfo{journal}{Phys.Rev.}
  \textbf{\bibinfo{volume}{D64}}, \bibinfo{pages}{054506}
  (\bibinfo{year}{2001}), \eprint{hep-lat/0104002}.

\bibitem[{\citenamefont{Aoki et~al.}(2011)}]{Aoki:2010dy}
\bibinfo{author}{\bibfnamefont{Y.}~\bibnamefont{Aoki}} \bibnamefont{et~al.}
  (\bibinfo{collaboration}{RBC Collaboration, UKQCD Collaboration}),
  \bibinfo{journal}{Phys.Rev.} \textbf{\bibinfo{volume}{D83}},
  \bibinfo{pages}{074508} (\bibinfo{year}{2011}), \eprint{1011.0892}.

\bibitem[{\citenamefont{Belyaev et~al.}(1985)\citenamefont{Belyaev, Blok, and
  Kogan}}]{Belyaev:1984ib}
\bibinfo{author}{\bibfnamefont{V.}~\bibnamefont{Belyaev}},
  \bibinfo{author}{\bibfnamefont{B.~Y.} \bibnamefont{Blok}}, \bibnamefont{and}
  \bibinfo{author}{\bibfnamefont{Y.}~\bibnamefont{Kogan}},
  \bibinfo{journal}{Sov.J.Nucl.Phys.} \textbf{\bibinfo{volume}{41}},
  \bibinfo{pages}{280} (\bibinfo{year}{1985}).

\bibitem[{\citenamefont{Zhu}(2001)}]{Zhu:2000zd}
\bibinfo{author}{\bibfnamefont{S.-L.} \bibnamefont{Zhu}},
  \bibinfo{journal}{Phys.Rev.} \textbf{\bibinfo{volume}{C63}},
  \bibinfo{pages}{018201} (\bibinfo{year}{2001}), \eprint{nucl-th/0009062}.

\bibitem[{\citenamefont{Erkol}()}]{erkol_thesis}
\bibinfo{author}{\bibfnamefont{G.}~\bibnamefont{Erkol}},
  \bibinfo{note}{http://irs.ub.rug.nl/ppn/297396218}.

\bibitem[{\citenamefont{Jido et~al.}(2000)\citenamefont{Jido, Hatsuda, and
  Kunihiro}}]{Jido:1999hd}
\bibinfo{author}{\bibfnamefont{D.}~\bibnamefont{Jido}},
  \bibinfo{author}{\bibfnamefont{T.}~\bibnamefont{Hatsuda}}, \bibnamefont{and}
  \bibinfo{author}{\bibfnamefont{T.}~\bibnamefont{Kunihiro}},
  \bibinfo{journal}{Phys.Rev.Lett.} \textbf{\bibinfo{volume}{84}},
  \bibinfo{pages}{3252} (\bibinfo{year}{2000}), \eprint{hep-ph/9910375}.

\bibitem[{\citenamefont{Hemmert et~al.}(2003)\citenamefont{Hemmert, Procura,
  and Weise}}]{Hemmert:2003cb}
\bibinfo{author}{\bibfnamefont{T.~R.} \bibnamefont{Hemmert}},
  \bibinfo{author}{\bibfnamefont{M.}~\bibnamefont{Procura}}, \bibnamefont{and}
  \bibinfo{author}{\bibfnamefont{W.}~\bibnamefont{Weise}},
  \bibinfo{journal}{Phys.Rev.} \textbf{\bibinfo{volume}{D68}},
  \bibinfo{pages}{075009} (\bibinfo{year}{2003}), \eprint{hep-lat/0303002}.

\bibitem[{\citenamefont{Procura}(2008)}]{Procura:2008ze}
\bibinfo{author}{\bibfnamefont{M.}~\bibnamefont{Procura}},
  \bibinfo{journal}{Phys.Rev.} \textbf{\bibinfo{volume}{D78}},
  \bibinfo{pages}{094021} (\bibinfo{year}{2008}), \eprint{0803.4291}.

\bibitem[{\citenamefont{Procura et~al.}(2007)\citenamefont{Procura, Musch,
  Hemmert, and Weise}}]{Procura:2006gq}
\bibinfo{author}{\bibfnamefont{M.}~\bibnamefont{Procura}},
  \bibinfo{author}{\bibfnamefont{B.}~\bibnamefont{Musch}},
  \bibinfo{author}{\bibfnamefont{T.}~\bibnamefont{Hemmert}}, \bibnamefont{and}
  \bibinfo{author}{\bibfnamefont{W.}~\bibnamefont{Weise}},
  \bibinfo{journal}{Phys.Rev.} \textbf{\bibinfo{volume}{D75}},
  \bibinfo{pages}{014503} (\bibinfo{year}{2007}), \eprint{hep-lat/0610105}.

\bibitem[{\citenamefont{Jiang and Tiburzi}(2008{\natexlab{b}})}]{Jiang:2008aqa}
\bibinfo{author}{\bibfnamefont{F.-J.} \bibnamefont{Jiang}} \bibnamefont{and}
  \bibinfo{author}{\bibfnamefont{B.~C.} \bibnamefont{Tiburzi}},
  \bibinfo{journal}{Phys.Rev.} \textbf{\bibinfo{volume}{D77}},
  \bibinfo{pages}{094506} (\bibinfo{year}{2008}{\natexlab{b}}),
  \eprint{0801.2535}.

\bibitem[{\citenamefont{Dinter et~al.}(2011)\citenamefont{Dinter, Alexandrou,
  Constantinou, Drach, Jansen et~al.}}]{Dinter:2011sg}
\bibinfo{author}{\bibfnamefont{S.}~\bibnamefont{Dinter}},
  \bibinfo{author}{\bibfnamefont{C.}~\bibnamefont{Alexandrou}},
  \bibinfo{author}{\bibfnamefont{M.}~\bibnamefont{Constantinou}},
  \bibinfo{author}{\bibfnamefont{V.}~\bibnamefont{Drach}},
  \bibinfo{author}{\bibfnamefont{K.}~\bibnamefont{Jansen}},
  \bibnamefont{et~al.}, \bibinfo{journal}{Phys.Lett.}
  \textbf{\bibinfo{volume}{B704}}, \bibinfo{pages}{89} (\bibinfo{year}{2011}),
  \eprint{1108.1076}.

\bibitem[{\citenamefont{Alexandrou
  et~al.}(2011{\natexlab{d}})\citenamefont{Alexandrou, Constantinou, Dinter,
  Drach, Jansen et~al.}}]{Alexandrou:2011aa}
\bibinfo{author}{\bibfnamefont{C.}~\bibnamefont{Alexandrou}},
  \bibinfo{author}{\bibfnamefont{M.}~\bibnamefont{Constantinou}},
  \bibinfo{author}{\bibfnamefont{S.}~\bibnamefont{Dinter}},
  \bibinfo{author}{\bibfnamefont{V.}~\bibnamefont{Drach}},
  \bibinfo{author}{\bibfnamefont{K.}~\bibnamefont{Jansen}},
  \bibnamefont{et~al.}, \bibinfo{journal}{PoS}
  \textbf{\bibinfo{volume}{LATTICE2011}}, \bibinfo{pages}{150}
  (\bibinfo{year}{2011}{\natexlab{d}}), \eprint{1112.2931}.

\end{thebibliography}
\end{document}